\documentclass[longauth]{aa}

\usepackage{euclid}
\usepackage{graphicx}
\usepackage{natbib}
\usepackage{scalerel}

\usepackage[table]{xcolor}

\usepackage{txfonts}
\usepackage[pdfencoding=auto,psdextra]{hyperref}
\hypersetup{
    colorlinks=true,
    linkcolor=blue,
    filecolor=magenta,      
    urlcolor=blue,
    citecolor=blue
}
\makeatletter
\renewcommand*\aa@pageof{, page \thepage{} of \pageref*{LastPage}}
\makeatother

\usepackage[utf8]{inputenc}

\usepackage[switch, modulo]{lineno}

\begin{document}
   \title{\Euclid preparation. XVIII. The NISP photometric system}

\newcommand{\orcid}[1]{} 
\author{\normalsize Euclid Collaboration: M.~Schirmer\orcid{0000-0003-2568-9994}$^{1}$\thanks{\email{schirmer@mpia.de}}, K.~Jahnke\orcid{0000-0003-3804-2137}$^{1}$,
 G.~Seidel\orcid{0000-0003-2907-353X}$^{1}$,
 H.~Aussel\orcid{0000-0002-1371-5705}$^{2}$,
 C.~Bodendorf$^{3}$,
 F.~Grupp$^{3,4}$,
 F.~Hormuth$^{1}$,
 S.~Wachter$^{5}$,
 P.N.~Appleton$^{6}$,
 R.~Barbier$^{7}$
% ...
}

\institute{$^{1}$ Max-Planck-Institut f\"ur Astronomie, K\"onigstuhl 17, D-69117 Heidelberg, Germany\\
$^{2}$ AIM, CEA, CNRS, Universit\'{e} Paris-Saclay, Universit\'{e} de Paris, F-91191 Gif-sur-Yvette, France\\
$^{3}$ Max Planck Institute for Extraterrestrial Physics, Giessenbachstr. 1, D-85748 Garching, Germany\\
$^{4}$ Universit\"ats-Sternwarte M\"unchen, Fakult\"at f\"ur Physik, Ludwig-Maximilians-Universit\"at M\"unchen, Scheinerstrasse 1, 81679 M\"unchen, Germany\\
$^{5}$ Carnegie Observatories, Pasadena, CA 91101, USA\\
$^{6}$ Infrared Processing and Analysis Center, California Institute of Technology, Pasadena, CA 91125, USA\\
$^{7}$ Univ Lyon, Univ Claude Bernard Lyon 1, CNRS/IN2P3, IP2I Lyon, UMR 5822, F-69622, Villeurbanne, France\\
$^{8}$ Centro de Astrof\'{\i}sica da Universidade do Porto, Rua das Estrelas, 4150-762 Porto, Portugal\\
$^{9}$ Instituto de Astrof\'isica e Ci\^encias do Espa\c{c}o, Universidade do Porto, CAUP, Rua das Estrelas, PT4150-762 Porto, Portugal\\
$^{10}$ Institut de Ci\`{e}ncies del Cosmos (ICCUB), Universitat de Barcelona (IEEC-UB), Mart\'{i} i Franqu\`{e}s 1, 08028 Barcelona, Spain\\
$^{11}$ Institut d’Estudis Espacials de Catalunya (IEEC), Carrer Gran Capit\'a 2-4, 08034 Barcelona, Spain\\
$^{12}$ Institute of Space Sciences (ICE, CSIC), Campus UAB, Carrer de Can Magrans, s/n, 08193 Barcelona, Spain\\
$^{13}$ Department of Astronomy, University of Geneva, ch. d\'Ecogia 16, CH-1290 Versoix, Switzerland\\
$^{14}$ Department of Mathematics and Physics E. De Giorgi, University of Salento, Via per Arnesano, CP-I93, I-73100, Lecce, Italy\\
$^{15}$ INFN, Sezione di Lecce, Via per Arnesano, CP-193, I-73100, Lecce, Italy\\
$^{16}$ Universit\'e de Paris, CNRS, Astroparticule et Cosmologie, F-75013 Paris, France\\
$^{17}$ Institut d'Astrophysique de Paris, 98bis Boulevard Arago, F-75014, Paris, France\\
$^{18}$ Aix-Marseille Univ, CNRS, CNES, LAM, Marseille, France\\
$^{19}$ Observatoire Astronomique de Strasbourg (ObAS), Universit\'e de Strasbourg - CNRS, UMR 7550, Strasbourg, France\\
$^{20}$ European Space Agency/ESTEC, Keplerlaan 1, 2201 AZ Noordwijk, The Netherlands\\
$^{21}$ INAF-Sezione di Lecce, c/o Dipartimento Matematica e Fisica, Via per Arnesano, I-73100, Lecce, Italy\\
$^{22}$ Institute of Cosmology and Gravitation, University of Portsmouth, Portsmouth PO1 3FX, UK\\
$^{23}$ Aix-Marseille Univ, CNRS/IN2P3, CPPM, Marseille, France\\
$^{24}$ INAF-Osservatorio di Astrofisica e Scienza dello Spazio di Bologna, Via Piero Gobetti 93/3, I-40129 Bologna, Italy\\
$^{25}$ INAF-Osservatorio Astronomico di Padova, Via dell'Osservatorio 5, I-35122 Padova, Italy\\
$^{26}$ INAF-Osservatorio Astrofisico di Torino, Via Osservatorio 20, I-10025 Pino Torinese (TO), Italy\\
$^{27}$ Department of Mathematics and Physics, Roma Tre University, Via della Vasca Navale 84, I-00146 Rome, Italy\\
$^{28}$ INFN-Sezione di Roma Tre, Via della Vasca Navale 84, I-00146, Roma, Italy\\
$^{29}$ INAF-Osservatorio Astronomico di Capodimonte, Via Moiariello 16, I-80131 Napoli, Italy\\
$^{30}$ INAF-IASF Milano, Via Alfonso Corti 12, I-20133 Milano, Italy\\
$^{31}$ Institut de F\'{i}sica d’Altes Energies (IFAE), The Barcelona Institute of Science and Technology, Campus UAB, 08193 Bellaterra (Barcelona), Spain\\
$^{32}$ Port d'Informaci\'{o} Cient\'{i}fica, Campus UAB, C. Albareda s/n, 08193 Bellaterra (Barcelona), Spain\\
$^{33}$ INAF-Osservatorio Astronomico di Roma, Via Frascati 33, I-00078 Monteporzio Catone, Italy\\
$^{34}$ Department of Physics "E. Pancini", University Federico II, Via Cinthia 6, I-80126, Napoli, Italy\\
$^{35}$ INFN section of Naples, Via Cinthia 6, I-80126, Napoli, Italy\\
$^{36}$ Dipartimento di Fisica e Astronomia ''Augusto Righi'' - Alma Mater Studiorum Universit\'a di Bologna, Viale Berti Pichat 6/2, I-40127 Bologna, Italy\\
$^{37}$ INAF-Osservatorio Astrofisico di Arcetri, Largo E. Fermi 5, I-50125, Firenze, Italy\\
$^{38}$ Centre National d'Etudes Spatiales, Toulouse, France\\
$^{39}$ Institut national de physique nucl\'eaire et de physique des particules, 3 rue Michel-Ange, 75794 Paris C\'edex 16, France\\
$^{40}$ Institute for Astronomy, University of Edinburgh, Royal Observatory, Blackford Hill, Edinburgh EH9 3HJ, UK\\
$^{41}$ Jodrell Bank Centre for Astrophysics, Department of Physics and Astronomy, University of Manchester, Oxford Road, Manchester M13 9PL, UK\\
$^{42}$ ESAC/ESA, Camino Bajo del Castillo, s/n., Urb. Villafranca del Castillo, 28692 Villanueva de la Ca\~nada, Madrid, Spain\\
$^{43}$ European Space Agency/ESRIN, Largo Galileo Galilei 1, 00044 Frascati, Roma, Italy\\
$^{44}$ Institute of Physics, Laboratory of Astrophysics, Ecole Polytechnique F\'{e}d\'{e}rale de Lausanne (EPFL), Observatoire de Sauverny, 1290 Versoix, Switzerland\\
$^{45}$ Departamento de F\'isica, Faculdade de Ci\^encias, Universidade de Lisboa, Edif\'icio C8, Campo Grande, PT1749-016 Lisboa, Portugal\\
$^{46}$ Instituto de Astrof\'isica e Ci\^encias do Espa\c{c}o, Faculdade de Ci\^encias, Universidade de Lisboa, Campo Grande, PT-1749-016 Lisboa, Portugal\\
$^{47}$ Universit\'e Paris-Saclay, CNRS, Institut d'astrophysique spatiale, 91405, Orsay, France\\
$^{48}$ INFN-Padova, Via Marzolo 8, I-35131 Padova, Italy\\
$^{49}$ INAF-Osservatorio Astronomico di Trieste, Via G. B. Tiepolo 11, I-34131 Trieste, Italy\\
$^{50}$ Istituto Nazionale di Astrofisica (INAF) - Osservatorio di Astrofisica e Scienza dello Spazio (OAS), Via Gobetti 93/3, I-40127 Bologna, Italy\\
$^{51}$ Istituto Nazionale di Fisica Nucleare, Sezione di Bologna, Via Irnerio 46, I-40126 Bologna, Italy\\
$^{52}$ Dipartimento di Fisica "Aldo Pontremoli", Universit\'a degli Studi di Milano, Via Celoria 16, I-20133 Milano, Italy\\
$^{53}$ INAF-Osservatorio Astronomico di Brera, Via Brera 28, I-20122 Milano, Italy\\
$^{54}$ INFN-Sezione di Milano, Via Celoria 16, I-20133 Milano, Italy\\
$^{55}$ Institute of Theoretical Astrophysics, University of Oslo, P.O. Box 1029 Blindern, N-0315 Oslo, Norway\\
$^{56}$ Leiden Observatory, Leiden University, Niels Bohrweg 2, 2333 CA Leiden, The Netherlands\\
$^{57}$ Jet Propulsion Laboratory, California Institute of Technology, 4800 Oak Grove Drive, Pasadena, CA, 91109, USA\\
$^{58}$ Technical University of Denmark, Elektrovej 327, 2800 Kgs. Lyngby, Denmark\\
$^{59}$ Mullard Space Science Laboratory, University College London, Holmbury St Mary, Dorking, Surrey RH5 6NT, UK\\
$^{60}$ Universit\'e de Gen\`eve, D\'epartement de Physique Th\'eorique and Centre for Astroparticle Physics, 24 quai Ernest-Ansermet, CH-1211 Gen\`eve 4, Switzerland\\
$^{61}$ Department of Physics and Helsinki Institute of Physics, Gustaf H\"allstr\"omin katu 2, 00014 University of Helsinki, Finland\\
$^{62}$ NOVA optical infrared instrumentation group at ASTRON, Oude Hoogeveensedijk 4, 7991PD, Dwingeloo, The Netherlands\\
$^{63}$ Argelander-Institut f\"ur Astronomie, Universit\"at Bonn, Auf dem H\"ugel 71, 53121 Bonn, Germany\\
$^{64}$ Dipartimento di Fisica e Astronomia “Augusto Righi” - Alma Mater Studiorum Università di Bologna, via Piero Gobetti 93/2, I-40129 Bologna, Italy\\
$^{65}$ INFN-Sezione di Bologna, Viale Berti Pichat 6/2, I-40127 Bologna, Italy\\
$^{66}$ Institute for Computational Cosmology, Department of Physics, Durham University, South Road, Durham, DH1 3LE, UK\\
$^{67}$ Universit\'e C\^{o}te d'Azur, Observatoire de la C\^{o}te d'Azur, CNRS, Laboratoire Lagrange, Bd de l'Observatoire, CS 34229, 06304 Nice cedex 4, France\\
$^{68}$ CEA Saclay, DFR/IRFU, Service d'Astrophysique, Bat. 709, 91191 Gif-sur-Yvette, France\\
$^{69}$ Sorbonne Universit{\'e}s, UPMC Univ Paris 6 et CNRS, UMR 7095, Institut d'Astrophysique de Paris, 98 bis bd Arago, 75014 Paris, France\\
$^{70}$ Department of Physics and Astronomy, University of Aarhus, Ny Munkegade 120, DK–8000 Aarhus C, Denmark\\
$^{71}$ Centre for Astrophysics, University of Waterloo, Waterloo, Ontario N2L 3G1, Canada\\
$^{72}$ Department of Physics and Astronomy, University of Waterloo, Waterloo, Ontario N2L 3G1, Canada\\
$^{73}$ Perimeter Institute for Theoretical Physics, Waterloo, Ontario N2L 2Y5, Canada\\
$^{74}$ Institute of Space Science, Bucharest, Ro-077125, Romania\\
$^{75}$ Max-Planck-Institut f\"ur Astrophysik, Karl-Schwarzschild Str. 1, 85741 Garching, Germany\\
$^{76}$ IFPU, Institute for Fundamental Physics of the Universe, via Beirut 2, 34151 Trieste, Italy\\
$^{77}$ INFN-Sezione di Roma, Piazzale Aldo Moro, 2 - c/o Dipartimento di Fisica, Edificio G. Marconi, I-00185 Roma, Italy\\
$^{78}$ Dipartimento di Fisica e Astronomia “G.Galilei", Universit\'a di Padova, Via Marzolo 8, I-35131 Padova, Italy\\
$^{79}$ Centro de Investigaciones Energ\'eticas, Medioambientales y Tecnol\'ogicas (CIEMAT), Avenida Complutense 40, 28040 Madrid, Spain\\
$^{80}$ Instituto de Astrof\'isica e Ci\^encias do Espa\c{c}o, Faculdade de Ci\^encias, Universidade de Lisboa, Tapada da Ajuda, PT-1349-018 Lisboa, Portugal\\
$^{81}$ Universidad Polit\'ecnica de Cartagena, Departamento de Electr\'onica y Tecnolog\'ia de Computadoras, 30202 Cartagena, Spain\\
$^{82}$ INAF-IASF Bologna, Via Piero Gobetti 101, I-40129 Bologna, Italy\\
$^{83}$ Kapteyn Astronomical Institute, University of Groningen, PO Box 800, 9700 AV Groningen, The Netherlands\\
$^{84}$ Dipartimento di Fisica, Universit\'a degli Studi di Torino, Via P. Giuria 1, I-10125 Torino, Italy\\
$^{85}$ INFN-Sezione di Torino, Via P. Giuria 1, I-10125 Torino, Italy\\
$^{86}$ Space Science Data Center, Italian Space Agency, via del Politecnico snc, 00133 Roma, Italy\\
$^{87}$ INFN, Sezione di Trieste, Via Valerio 2, I-34127 Trieste TS, Italy\\
$^{88}$ SISSA, International School for Advanced Studies, Via Bonomea 265, I-34136 Trieste TS, Italy\\
$^{89}$ Departamento de Astrof\'{i}sica, Universidad de La Laguna, E-38206, La Laguna, Tenerife, Spain\\
$^{90}$ Instituto de Astrof\'isica de Canarias, Calle V\'ia L\'actea s/n, E-38204, San Crist\'obal de La Laguna, Tenerife, Spain\\
$^{91}$ Institut de Recherche en Astrophysique et Plan\'etologie (IRAP), Universit\'e de Toulouse, CNRS, UPS, CNES, 14 Av. Edouard Belin, F-31400 Toulouse, France\\
$^{92}$ Dipartimento di Fisica - Sezione di Astronomia, Universit\'a di Trieste, Via Tiepolo 11, I-34131 Trieste, Italy\\
$^{93}$ Dipartimento di Fisica e Scienze della Terra, Universit\'a degli Studi di Ferrara, Via Giuseppe Saragat 1, I-44122 Ferrara, Italy\\
$^{94}$ INAF, Istituto di Radioastronomia, Via Piero Gobetti 101, I-40129 Bologna, Italy\\
$^{95}$ INFN-Bologna, Via Irnerio 46, I-40126 Bologna, Italy\\
$^{96}$ Department of Physics and Astronomy, University of California, Davis, CA 95616, USA\\
$^{97}$ University of Lyon, UCB Lyon 1, CNRS/IN2P3, IUF, IP2I Lyon, France\\
$^{98}$ INFN-Sezione di Genova, Via Dodecaneso 33, I-16146, Genova, Italy\\
$^{99}$ INAF-Istituto di Astrofisica e Planetologia Spaziali, via del Fosso del Cavaliere, 100, I-00100 Roma, Italy\\
$^{100}$ Department of Physics, Oxford University, Keble Road, Oxford OX1 3RH, UK\\
$^{101}$ School of Physics, HH Wills Physics Laboratory, University of Bristol, Tyndall Avenue, Bristol, BS8 1TL, UK\\
$^{102}$ Instituto de F\'isica Te\'orica UAM-CSIC, Campus de Cantoblanco, E-28049 Madrid, Spain\\
$^{103}$ Institut de Ciencies de l'Espai (IEEC-CSIC), Campus UAB, Carrer de Can Magrans, s/n Cerdanyola del Vall\'es, 08193 Barcelona, Spain\\
$^{104}$ Department of Physics, P.O. Box 64, 00014 University of Helsinki, Finland\\
$^{105}$ Department of Physics, Lancaster University, Lancaster, LA1 4YB, UK\\
$^{106}$ Universit\'e PSL, Observatoire de Paris, Sorbonne Universit\'e, CNRS, LERMA, F-75014, Paris, France\\
$^{107}$ Code 665, NASA Goddard Space Flight Center, Greenbelt, MD 20771 and SSAI, Lanham, MD 20770, USA\\
$^{108}$ Helsinki Institute of Physics, Gustaf H{\"a}llstr{\"o}min katu 2, University of Helsinki, Helsinki, Finland\\
$^{109}$ Centre de Calcul de l'IN2P3, 21 avenue Pierre de Coubertin F-69627 Villeurbanne Cedex, France\\
$^{110}$ Dipartimento di Fisica, Sapienza Universit\`a di Roma, Piazzale Aldo Moro 2, I-00185 Roma, Italy\\
$^{111}$ Zentrum f\"ur Astronomie, Universit\"at Heidelberg, Philosophenweg 12, D- 69120 Heidelberg, Germany\\
$^{112}$ Institut f\"ur Theoretische Physik, University of Heidelberg, Philosophenweg 16, 69120 Heidelberg, Germany\\
$^{113}$ Ruhr University Bochum, Faculty of Physics and Astronomy, Astronomical Institute (AIRUB), German Centre for Cosmological Lensing (GCCL), 44780 Bochum, Germany\\
$^{114}$ Institute for Computational Science, University of Zurich, Winterthurerstrasse 190, 8057 Zurich, Switzerland\\
$^{115}$ Departamento de F\'isica, FCFM, Universidad de Chile, Blanco Encalada 2008, Santiago, Chile\\
$^{116}$ Department of Astrophysical Sciences, Peyton Hall, Princeton University, Princeton, NJ 08544, USA\\
$^{117}$ Department of Physics, P.O.Box 35 (YFL), 40014 University of Jyv\"askyl\"a, Finland\\
}

   \date{Received; accepted}

 

  \abstract{\Euclid will be the first space mission to survey most of the extragalactic sky in the $0.95$--$2.02$\,\micron\ range, to a $5\,\sigma$ point-source median depth of $24.4$\,AB\,mag. This unique photometric data set will find wide use beyond \Euclid's core science. In this paper, we present accurate computations of the Euclid \ymagm, \jmagm, and \hmagm passbands used by the Near-Infrared Spectrometer and Photometer (NISP), and the associated photometric system. We pay particular attention to passband variations in the field of view, accounting among others for spatially variable filter transmission, and variations of the angle of incidence on the filter substrate using optical ray tracing. The response curves' cut-on and cut-off wavelengths -- and their variation in the field of view -- are determined with $\sim$0.8\,nm accuracy, essential for the photometric redshift accuracy required by \Euclid. 
  
  After computing the photometric zeropoints in the AB mag system, we present linear transformations from and to common ground-based near-infrared photometric systems, for normal stars, red and brown dwarfs, and galaxies separately. A {\tt Python} tool to compute accurate magnitudes for arbitrary passbands and spectral energy distributions is provided. We discuss various factors from space weathering to material outgassing that may slowly alter \Euclid's spectral response. At the absolute flux scale, the \Euclid in-flight calibration program connects the NISP photometric system to \textit{Hubble} Space Telescope spectrophotometric white dwarf standards; at the relative flux scale, the chromatic evolution of the response is tracked at the milli-mag level. In this way, we establish an accurate photometric system that is fully controlled throughout \Euclid's lifetime.}

   \titlerunning{\Euclid Preparation: The NISP photometric system}
   \authorrunning{Euclid Collaboration}
   
   \maketitle
%
%
\section{Introduction}
\begin{figure*}[t]
\centering
\includegraphics[angle=0,width=1.0\hsize]{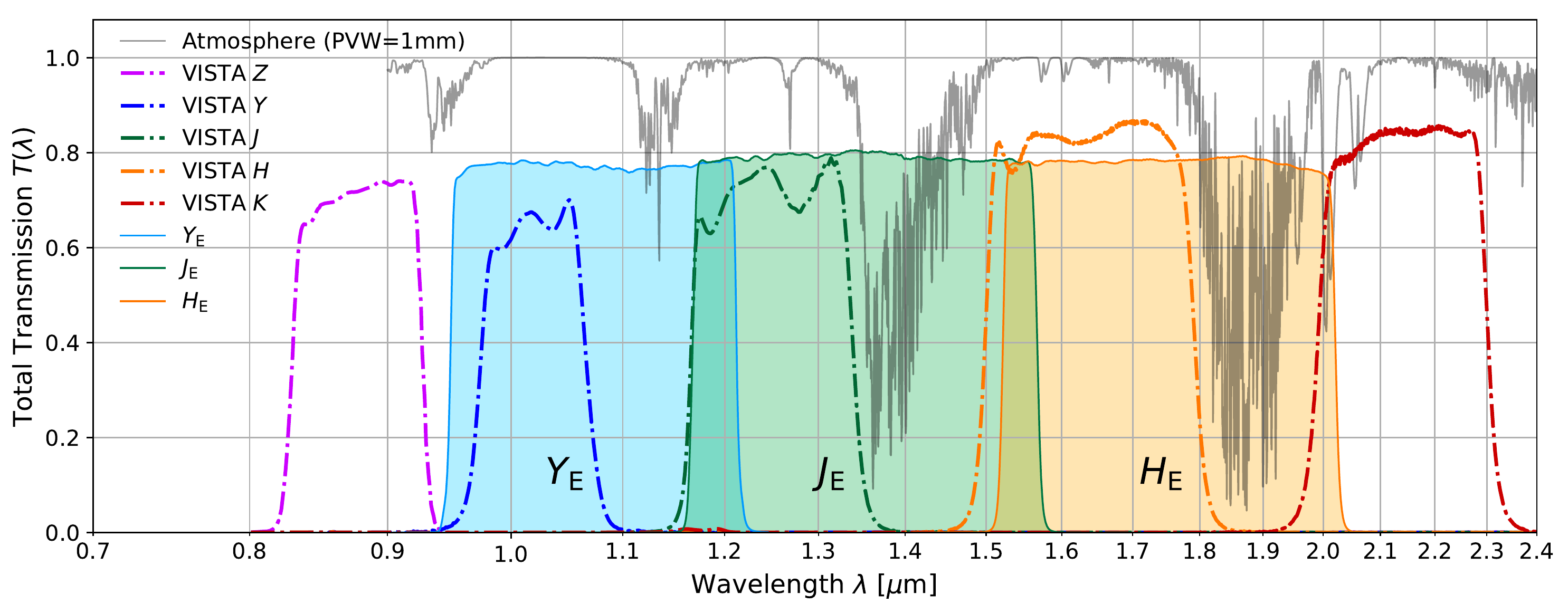}
\caption{Comparison of the NISP spectral response (shaded) with typical ground-based NIR passbands, in this case VIRCAM@VISTA \citep{sutherland2015}. Both sets of response curves account for mirrors, filters, and detector quantum efficiency (QE). The gray line displays the atmospheric transmission for a precipitable water vapor of $1$\,mm and at zenith, taken from the ESO VISTA instrument description. The ground-based $Z$ and $K$ bands lie outside the NISP wavelength range, whereas $Y$, $J$, and $H$ cover approximately half of the corresponding NISP passbands.}
\label{fig:passband_comparison}
\end{figure*}

The \Euclid mission will observe 15\,000\,deg$^2$ of extragalactic sky \citep{scaramella2021} from the Sun--Earth Lagrange Point L2. It will employ weak gravitational lensing and galaxy clustering -- including baryonic acoustic oscillations and redshift space distortions -- as cosmological probes, to determine the expansion history and growth rate of cosmic structures over the last 10 billion years. The measurements will be performed in several tomographic redshift bins, covering the time when the acceleration of the Universe became important \citep{laureijs2011,blanchard2020}. In this way, \Euclid addresses the nature and properties of dark energy, dark matter, gravitation, and the Universe's initial conditions. The results should be decisive for the validity of the $\Lambda$CDM concordance model and General Relativity on cosmic scales.

Among many other data, \Euclid must determine the near-infrared (NIR) photometry of at least one billion galaxies to a relative accuracy of better than 1.5\%.
This will establish a NIR photometric reference system that will be in wide use for arguably the next few decades. To achieve these observations with \Euclid's 1.2\,m telescope and within the planned mission duration of six years, the NISP instrument \citep{prieto2012,maciaszek2016} carries a large focal plane array (FPA) of 16 Teledyne HAWAII-2RG detectors.
The NISP wide-field optical system uses filters with a diameter of $130$\,mm, the largest NIR filters flown in a civilian spacecraft to date.

\Euclid's core cosmology science requires the $0.95$--$2.02$\,\micron\ range to be covered in three passbands (\ymagm, \jmagm, \hmagm; see Fig.~\ref{fig:passband_comparison}), with rectangular shape, equal relative spectral width $\Delta\lambda/\lambda$, and without inter-passband gaps. While overlapping with the common ground-based $Y$, $J$ and $H$ passbands, the Euclid passbands are about twice as wide as they are not constrained by atmospheric absorption. The passbands' flanks -- i.e. the transition regions from out-of-band blocking to full in-band transmission -- must be defined by the filters alone. Together with complementary ground-based photometry, these passbands enable the calculation of the mean photometric redshifts (photo-$z$) of the tomographic redshift bins with an accuracy of $0.002\,(1+z)$ \citep{laureijs2011,ilbert2021}. 

The photo-$z$ accuracy required for \Euclid implies that the edges of the NISP passbands must be known to better than $1.0$\,nm. In this paper, we show that the passbands are blueshifted by up to 6\,nm when going from the center of the focal plane toward its corners. This effect is mostly due to variations in the angle of incidence (AOI) on the filter surface, and to a lesser degree to a systematic blueshift of the filter's transmission toward its edges. Using a careful assessment of the transmission measurements conducted by the filter coating manufacturer, and accurate ray tracing methods, we determine the passbands with sub-nm accuracy anywhere in the field of view; this will also serve all purposes of legacy science.

This paper is organized as follows. In Sect.~\ref{sec:measurements}, we introduce the filter substrates, the various transmission measurements and their limitations. In particular, we focus on intrinsic, local variations of the filter transmission. In Sect.~\ref{sec:rootcausespbvar}, we use optical ray tracing to quantify the AOI on the filter surface, which blueshifts the passband for non-zero AOIs. In Sect.~\ref{sec:computation}, we use these results to compute the effective filter transmission and its blueshift toward the corners of the focal plane. In Sect.~\ref{sec:totaltransmission}, we compute the spectral response, i.e. the total system transmission including telescope optics, filters and detectors. We introduce the NISP photometric system in Sect.~\ref{sec:photometricsystem}, together with transformations to common ground-based NIR photometric systems. We summarize the main results and data products in Sect.~\ref{sec:summary}.

\section{Filter characteristics and transmission data\label{sec:measurements}}
\subsection{Terminology\label{sec:definitions}}
\subsubsection{Beam footprint}
With `beam footprint' we refer to the intersection of the optical beam on the filter surface. The beam footprint is asymmetric in shape and has a decentered obstruction. It's geometry is important for the transmission calculations. More details about the footprint can be found in Sect.~\ref{sec:footprintgeometry}.

\subsubsection{Passband, transmission and response}
With `passband' we refer to the \ymagm, \jmagm, and \hmagm wavelength intervals transmitted by the NISP filters. The filter passbands are characterized by their wavelength-dependent `local transmission', $t(\lambda)$, which varies as a function of position on the filter substrate. The filters `effective transmission', $T(\lambda)$, is obtained by integrating $t(\lambda)$ over the beam footprint on the filter substrate (Sect.~\ref{sec:computation}). The `response' -- i.e. the total system throughput -- is computed by multiplying $T(\lambda)$ with the transmission of other optical elements and the detector quantum efficiency (QE) (Sect.~\ref{sec:totaltransmission}).

The `mean peak transmission' is computed over the wavelength interval where the transmission exceeds 97\% of its maximum value. 

\begin{figure}[t]
\centering
\includegraphics[angle=0,width=0.86\hsize]{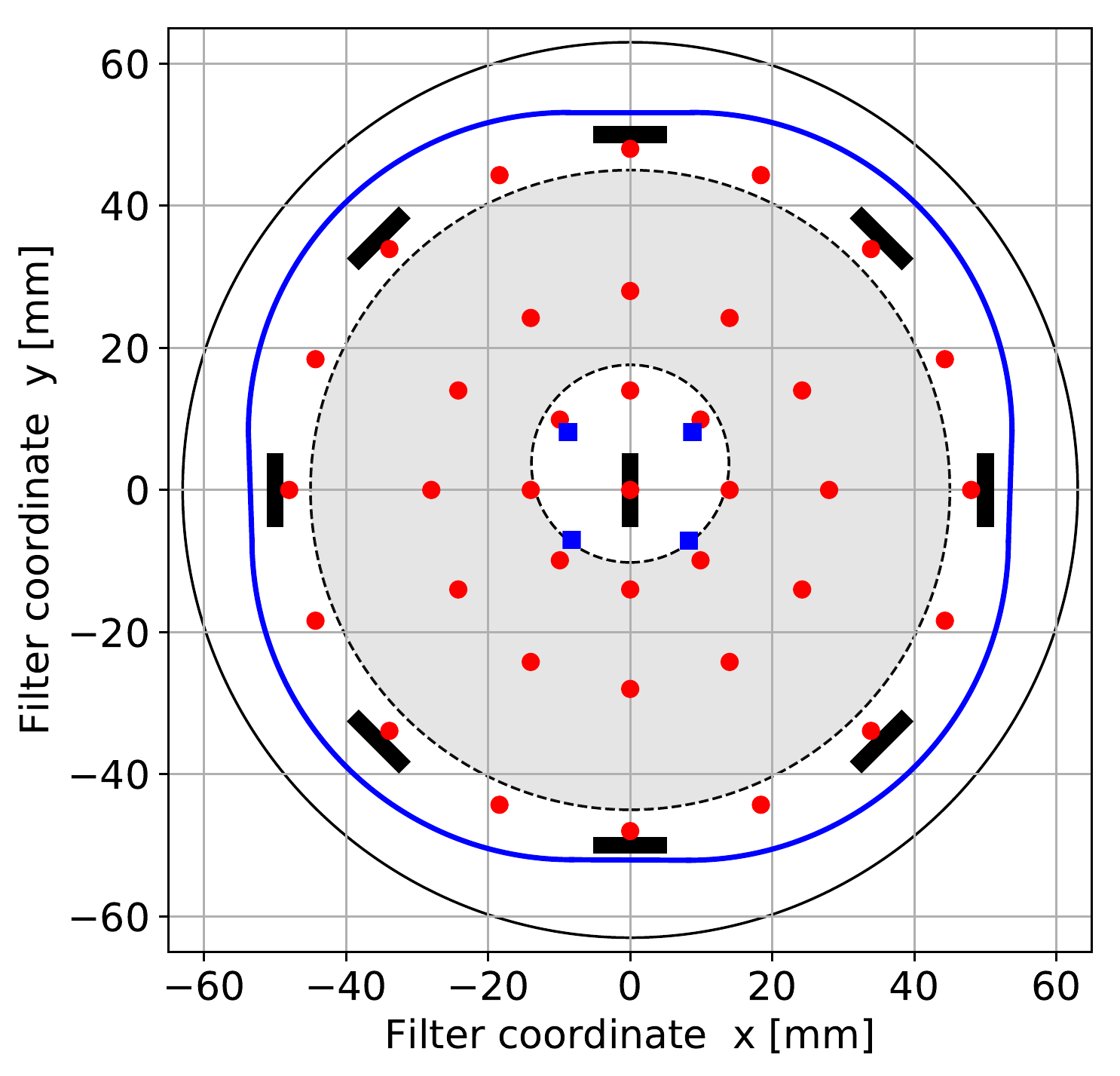}
\caption{Filter geometry and measurement points. The filter has a diameter of 130\,mm (black circle). The blue line shows the baffle, blocking light paths outside without causing vignetting. The 9-point original measurement apertures are shown by the black rectangles (to scale). Our refined 37-point measurements are marked by red dots instead of black rectangles, to avoid cluttering. The gray annulus displays the 90\,mm obstructed beam footprint, i.e. the part of the filter seen by a source at the center of the FPA. The obstruction is off-centered due to the telescope's off-axis design. \textit{The response curves made available online are integrated over this central footprint}. The light of a source mapped onto any of the four corners of the FPA intersects different parts of the filter; the centers of the correspondingly shifted footprints are shown by the blue squares.}
\label{fig:filtergeometry}
\end{figure}

\subsubsection{Wavelength-related parameters: Cut-on and cut-off}
At the `50\% cut-on wavelength' (hereafter: cut-on), the near-rectangular transmission curve has transitioned halfway from out-of-band blocking to the mean peak transmission. The 50\% cut-off is found correspondingly, at the long-wavelength end of the passband; specifically, we use cubic spline interpolation to locate the cut-on and cut-off with a precision better than $0.1$\,nm. We refer to the transition regions as the `passband flanks', which are approximately centered at the cut-on and cut-off. These quantities can be computed for both the transmission and the response curves.

We calculate the `passband width' as the wavelength interval between the cut-on and cut-off. Lastly, we define the passband's `central wavelength' as 
\begin{equation}
  \lambda_{\rm cen} = \frac{\int\lambda\,T(\lambda)\,{\rm d}\lambda} 
                      {\int T(\lambda)\,{\rm d}\lambda}\,.
\end{equation}
Because of the nearly rectangular transmission curve, the central wavelength computed in this way is within less than 0.5\,nm of the mid-point between the cut-on and cut-off.


\subsection{Filter substrate, dielectric coatings, and blocking\label{sec:coatings}}
The 130\,mm-diameter NISP filter substrates were made by Heraeus (Germany) from their proprietary {`Suprasil 3001'} type fused silica, and shaped by Winlight Optics (France). 
%
%
The filters have a center thickness of $11$--$12$\,mm, depending on the specific filter, with the sky-facing side being slightly convex with a curvature radius of about 10\,000\,mm.
%
%

The filters carry quarterwave stacks of dielectric layers with alternating high and low refractive index, defining the filter's passband through interference. The coatings were performed by Optics Balzers Jena (OBJ, Germany; now Materion Balzers Optics). Coating stacks of up to 200 alternating interference layers of SiO$_2$ and Nb$_2$O$_5$ were deposited, to a total stack height of up to $20$\,\micron\ per filter side, using a plasma assisted magnetron sputtering process (PARMS). The coatings block photons outside the nominal passbands very efficiently (Sect.~\ref{sec:blueleak}). The coating layers defining a single passband flank are deposited on both sides of the filter, which becomes relevant in Sect.~\ref{sec:raytracing}. A characteristic effect of interference filters is the blueshift of the passband when the filter is tilted in a collimated beam against the surface normal vector. This is because the phase factor of an individual dielectric layer decreases with increasing AOI \citep{rienstra1998,smith2008}.

%
%

\begin{figure}[t]
\centering
\includegraphics[angle=0,width=1.0\hsize]{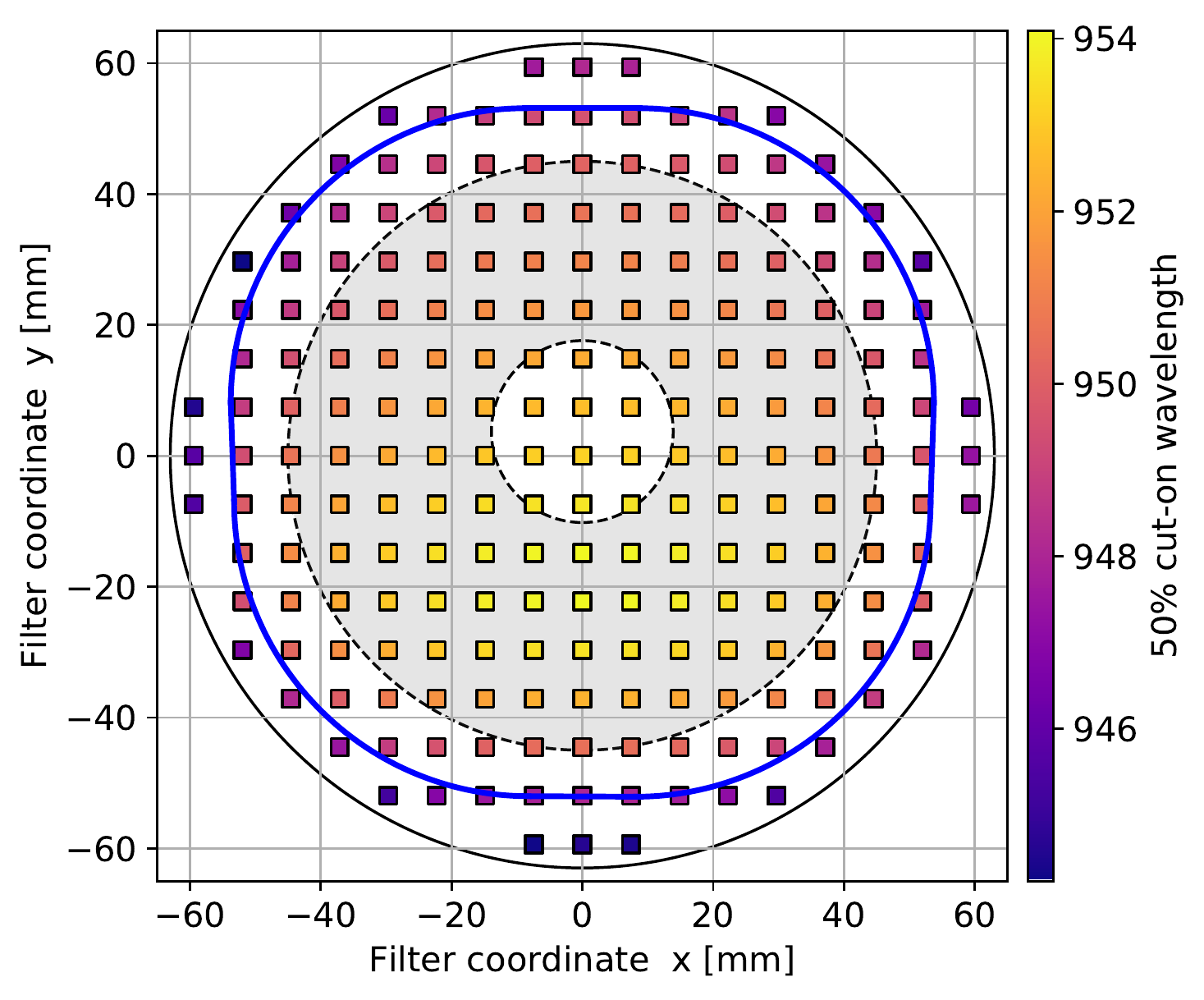}
\caption{Variation of the 50\% cut-on wavelength of the \yband engineering qualification filter. Measurements were done with a single-beam photometer in 213 apertures $2\times2$\,mm in size (shown to scale), at AOI $\theta=\ang{0}$. The cut-on varies by 8\,nm over the unbaffled area, and is a consequence of the coating layers getting systematically thinner by about 0.4\% toward the outer edge. When the gray annulus moves across this surface (different object positions in the field of view), the effective transmission changes by $0.5$--$1.0$\,nm. Lines are the same as in Fig.~\ref{fig:filtergeometry}.}
\label{fig:EQM}
\end{figure}

\subsection{Measurement strategy with the PE950\label{sec:PE950}}
The transmission of all filters -- one flight model and two flight spares for each band -- was measured with a Perkin Elmer Lambda 950 (PE950). This is a double beam, double monochromator, intensity ratio recording spectrophotometer working in the 175--3\,300\,nm range. In the NIR regime, the PE950 used by OBJ has a nominal wavelength reproducibility of $0.05$\,nm, an absolute wavelength accuracy of $\leq0.32$\,nm, and is certified periodically by Perkin Elmer Corp.

Measurements were taken at 1\,nm wavelength steps at nine positions (Fig.~\ref{fig:filtergeometry}). They were repeated for two AOIs, $\theta=\ang{0}$ and $\theta=\ang{7}$, the latter approximating the largest AOI realized on the filters. This calibrates the blueshift of the passband for oblique AOIs. Hereafter, we refer to these data as the `9-point data'.

Higher spatial sampling was achieved for the single \yband `engineering quali\-fi\-cation model' filter -- not used in NISP -- using a single-beam photometer with lower spectral resolution, to determine the cut-on at over 200 positions (Fig.~\ref{fig:EQM}).

The assumption behind this dual approach was that the coating process would be repeatable, replicating the topography of the cut-on and cut-off `surfaces' for all filters. The sparsely sampled 9-point PE950 data taken over the full wavelength range would be complemented by the high spatial resolution measurements of the cut-on surface. This would be sufficient to create accurate interpolation models, to compute the effective passbands by integration over the 90\,mm obstructed beam footprints (Fig.~\ref{fig:filtergeometry}). We show in Sect.~\ref{sec:nonrepeatability} that this assumption is invalid.

\subsection{Non-repeatability of the coating process and wavelength accuracy of the PE950
\label{sec:nonrepeatability}}
The flight model filters passed all acceptance tests and were integrated in NISP. Our analysis of the local transmission data, however, shows that the topography of the cut-on and cut-off surfaces varies considerably, even between flight models and flight spares of the same passband. Consequently, the assumption of a repeatable coating process is invalid, and the high spatial resolution mapping data (Fig.~\ref{fig:EQM}) is not suitable to improve the interpolation models. A more detailed analysis of this will be presented elsewhere, based on physical models of the coating layer stacks.

Furthermore, we show in the following that the PE950 operated at least partially outside its wavelength specifications summarized at the beginning of Sect.~\ref{sec:PE950}. 

\begin{figure}[t]
\centering
\includegraphics[angle=0,width=1.0\hsize]{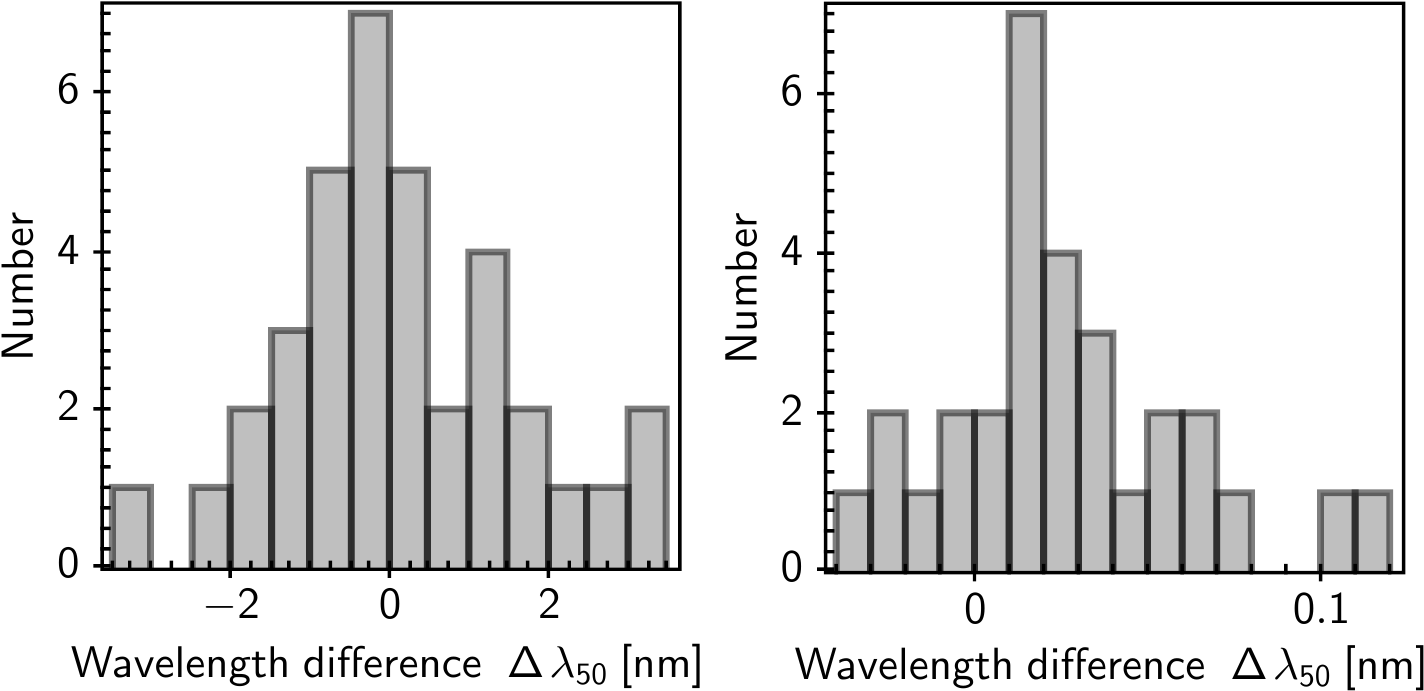}
\caption{Wavelength accuracy of the PE950 spectrophotometer. \textit{Left panel}: Joint long-term reproducibility of the cut-on and cut-off wavelengths \textit{between} the 9-point measurement runs on the same substrates. The RMS is $\sigma_{\rm wave}^{\rm long}=1.47$\,nm. \textit{Right panel}: Short-term reproducibility \textit{within} single 37-point measurement runs, with 
$\sigma_{\rm wave}^{\rm short}=0.035$\,nm.}
\label{fig:PE950_reproducibility}
\end{figure}

\subsubsection{Inconsistencies in the 9-point transmission data\label{sec:9pointinconsistencies}}
During the first measurement run of the flight spare filters, not all $\theta=\ang{7}$ data were taken for the \ymagm- and \hband models. A complete set of the 9-point data including the $\theta=\ang{0}$ setting was repeated at a later time. Having the same positions at $\theta=0^\circ$ measured twice, we can check the wavelength reproducibility and accuracy of the PE950 (Sect.~\ref{sec:PE950}). To this end, we compute the normalized wavelength difference $\Delta\lambda_{50}=(\lambda_1-\lambda_2)/\sqrt{2}$ for each measurement position, where $\lambda_1$ and $\lambda_2$ are the cut-ons in the two measurement runs; the same is done for the cut-off. Assuming normally distributed errors, the RMS of $\Delta\lambda_{50}$ is then an estimator of the PE950's wavelength accuracy.

We do not detect a systematic wavelength offset between the runs. However, we find that  the long-term RMS, $\sigma_{\rm wave}^{\rm long}$, of the $\Delta\lambda_{50}$ data from the 9-point measurements exceeds the PE950's specification; we refer to this RMS as 'long-term', because many months have passed between the measurements (see Sect.~\ref{sec:37point} for the short-term RMS). In particular, $\sigma_{\rm wave}^{\rm long}=1.47$\,nm (left panel of Fig.~\ref{fig:PE950_reproducibility}) is higher than the specification of $\leq0.32$\,nm. This implies that both the PE950's internal wavelength reproducibility of 0.05\,nm as well as the wavelength accuracy were not met for the flight spare runs. Aging effects are ruled out since the filter coatings are very durable. We do not have repeated measurements for the flight model filters, and hence assume that the same uncertainties apply to them.

\subsubsection{37-point transmission data on flight spare filters\label{sec:37point}}
To better understand these inconsistencies, we designed a refined test protocol for the second set of flight spare filters that were still accessible at that time. Transmission was measured at 37 points (red dots in Fig.~\ref{fig:filtergeometry}), including the central position of the original 9-point pattern. The remaining positions were placed in three rings with diameters of 14, 28, and 48\,mm, respectively. Five of the positions were measured repeatedly, to check for hysteresis, wavelength reproducibility, and systematic drifts.

We compute the short-term RMS of the $\Delta\lambda_{50}$ data, 
$\sigma_{\rm wave}^{\rm short}=0.035$\,nm, from the repeatedly visited points (right panel in Fig.~\ref{fig:PE950_reproducibility}). This is within the PE950's nominal wavelength reproducibility of 0.05\,nm, for all three flight spares. Meaningful systematic effects are not found.

Comparing the passband flanks of the common central position between the 37-point and the 9-point measurements, we find $\sigma_{\rm wave}^{\rm long}=0.52$\,nm ($0.16$\,nm) for the first (second) run of 9-point data. This is above (below) the PE950 specification of $\leq0.32$\,nm for wavelength accuracy.

\subsubsection{Conclusion about wavelength accuracy\label{sec:wavelength_accuracy}}
The origin of these inconsistencies, in particular between the original 9-point measurements (Sect.~\ref{sec:9pointinconsistencies}), remains unclear. It could be due to temporal instabilities and uncalibrated long-term drifts of the PE950, or tolerances in the preparation and execution of the measurements. While the 37-point data show that the PE950 can deliver data with high internal consistency, it is not clear whether this applies to the flight model data, for which only a single measurement epoch is available. Therefore, we assume that the 9-point flight model data have the same uncertainties as the 9-point flight spare data.

The spatial interpolation of the local transmission is based on 9 data points; in these runs, the wavelength has a statistical uncertainty of $\sigma_{\rm wave}^{\rm long}=1.47$\,nm (Sect.~\ref{sec:9pointinconsistencies}). Integration over the interpolating function (Sect.~\ref{sec:computation}) has an averaging effect on these uncertainties, plausibly by a factor of $\sqrt{9}$. Here, we are more conservative and set $\sigma_{50}=0.8$\,nm for the cut-on and cut-off wavelengths of the integrated transmission in Sect.~\ref{sec:computation}. The passband's central wavelength, when approximated as the midpoint between the cut-on and cut-off, has an uncertainty of $\sigma_{\rm cen}=\sigma_{50}\,/\sqrt{2}=0.6$\,nm.

\begin{figure*}[t]
\centering
\includegraphics[angle=0,width=1.0\hsize]{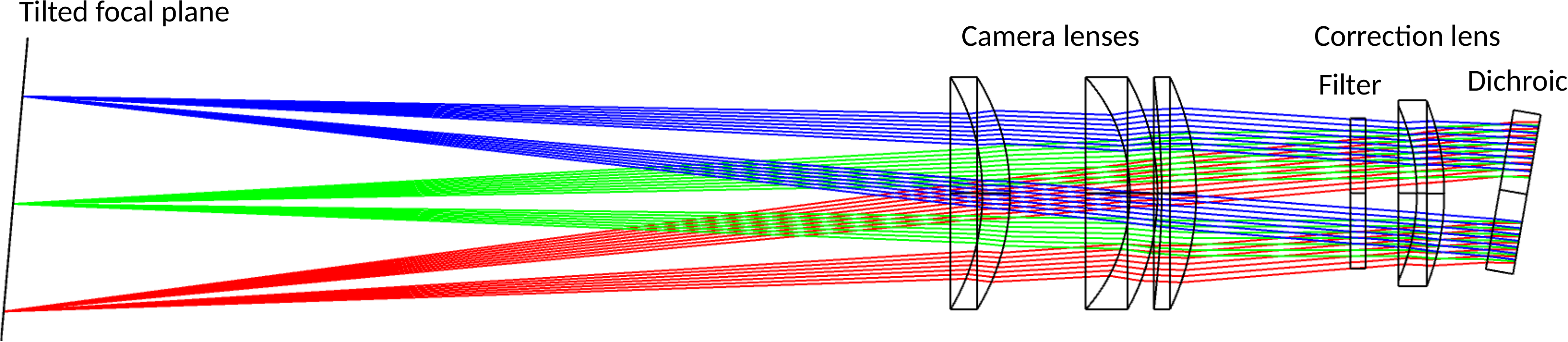}
\caption{Ray tracing of the NISP optical assembly (NI-OA). Shown are the obstructed beams for three sources located at the center of the focal plane and at two opposing edges. The NI-OA consists of four lenses; the filter also acts as a lens, albeit only weakly. The dichroic element is located in the pupil plane of the telescope, and is not part of NISP. Notice how the beam paths cover different parts of the filter surface, because -- contrary to the dichroic -- the filter is located outside the pupil plane. This leads to $10$--$23$\% of the passband variations in the focal plane due to local variations of the filter transmission. Notice how the AOI on the filter changes for the three sources; this is the dominant cause for passband variations.}
\label{fig:nioa_fields}
\end{figure*}

\section{Root causes of the NISP passband variations\label{sec:rootcausespbvar}}
Passband variations, i.e. changes of the cut-on and cut-off as a function of position in the focal plane, must be accurately known for \Euclid to meet its requirements on photo-$z$ accuracy. Two factors contribute to passband variations: 
\begin{enumerate}
\itemsep0em
\item increasingly bluer transmission toward the filter edge, in combination with a moving beam footprint across the filter surface (up to 23\% of the total blueshift for NISP);
\item variations of the AOI on the filter surface, in combination with the filter coating's effective refractive index, $n_{\rm eff}$ (up to 90\% of the total blueshift).
\end{enumerate}
\Euclid's telescope is based on a Korsch design \citep{korsch1977} in an off-axis configuration \citep{laureijs2011}. This leads to a number of effects we need to account for when quantifying the two factors introduced above:
\begin{itemize}
    \itemsep0em 
    \item a decentered central obstruction of the beam by the secondary mirror  (Fig.~\ref{fig:filtergeometry});
    \item a tilted FPA, and slightly tilted NISP optics;
    \item elliptical beam footprint shape;
    \item AOI asymmetries on the planar and curved filter surfaces.
\end{itemize}

The results of our analysis in this Section are used in Sect.~\ref{sec:computation}, where we compute the effective filter transmission and its variation across the field of view.

\subsection{Beam footprint geometry on the filter surface\label{sec:footprintgeometry}}
Consider Fig.~\ref{fig:nioa_fields}, displaying the optical paths for three different point sources at two edges and in the center of the FPA. The corresponding footprints on the filter surface have an annular shape with outer and inner diameters of about 91.0\,mm and 27.8\,mm, respectively (Fig.~\ref{fig:filter_incident_angle}). The off-axis configuration leads to a slight ellipticity and off-centered central obstruction. A single footprint covers $\sim$56\% of the filter area, and its location on the filter surface depends on the source position in the field of view (Figs.~\ref{fig:nioa_fields}, \ref{fig:filter_incident_angle}); the footprint can move away from the filter center by up to 11.9\,mm. In combination with the spatially variable filter transmission (Fig.~\ref{fig:EQM}), this causes a dependence of the effective passband on the image field. 

Depending on the position on the filter, the major axis of the footprint changes between 89.8\,mm and 92.9\,mm. The footprint shapes are nearly circular, with differences from 0.0 to 1.3\,mm between the minor and major axes. The circular obstruction has a diameter of 27.8\,mm for the central footprint, and scales correspondingly for footprints at the corners. The obstruction is spatially offset in the beam by 3.7\,mm (Figs.~\ref{fig:filtergeometry}, \ref{fig:EQM}, \ref{fig:filter_incident_angle}, \ref{fig:filter_incident_angle_differential}).


\subsection{Transmission blueshift due to oblique AOI\label{sec:blueshift}}
Like all optical interference filters, the transmitted passband is blueshifted with increasing AOI, $\theta$, due to the changing phase-difference of reflected and incident rays. The blueshift also depends on the coating's effective refractive index, $n_{\rm eff}(\lambda)$ \citep[see e.g.][]{amra2021}. In vacuum, and with the small angle approximation applicable here, 
a particular passband feature seen at wavelength $\lambda_0$ for $\theta=0^\circ$ is blueshifted to 
\begin{equation}
  \lambda=\lambda_0\;\sqrt{1 -\left( \frac{ \sin\theta} 
  {n_{\rm eff}(\lambda_0)}\right)^2}\;;
  \label{eq:neffdlambda}
\end{equation}
for details see \cite{smith2008} and \cite{loefdahl2011}. The maximum AOI realized on the NISP filters is $\theta=7.5^\circ$; polarization splitting due to polarization-dependent Fresnel reflections is ignored, as it becomes important for larger angles, only.

Thus, to compute the passband blueshift for a given footprint, we have to determine two quantities: $n_{\rm eff}$ from measurements and / or the filter coating design (Sect.~\ref{sec:neff}), and the spatial dependence of $\theta(x,y)$, with $x$ and $y$ specifying the location where the individual rays intersect the filter's surface (Sect.~\ref{sec:raytracing}).

\begin{table}[ht]
\caption{Effective refractive index. Columns 2 and 3 show the design values of the coating layer stacks for the cut-on and cut-off wavelengths, and columns 4 and 5 the measured values for the flight model filters.}
\smallskip
\label{table:neff}
\smallskip
\begin{tabular}{|l|rr|rr|}
\hline
  & \multicolumn{2}{c|}{Design values} & \multicolumn{2}{c|}{Measured values} \\
\hline
Filter & $n_{\rm eff}^{\rm cut-on}$ & $n_{\rm eff}^{\rm cut-off}$ & $n_{\rm eff}^{\rm cut-on}$ & $n_{\rm eff}^{\rm cut-off}$\\[3pt]
\hline
\ymagm & 1.764 & 1.759 & $1.75\pm0.07$ & $1.79\pm0.12$\\
\jmagm & 1.769 & 1.776 & $1.66\pm0.09$ & $1.78\pm0.10$\\
\hmagm & 1.773 & 1.773 & $1.74\pm0.05$ & $1.86\pm0.05$\\
\hline
\end{tabular}
\end{table}

\subsection{Effective refractive index \texorpdfstring{$n_{\rm eff}$}{n\_eff}\label{sec:neff}}
While $n_{\rm eff}$ is known from the coating design, imperfections in the coating process cause thickness variations of the layers, and thus modulate $n_{\rm eff}$. In Table~\ref{table:neff} we list the expected and measured values of $n_{\rm eff}$, obtained using Eq.~(\ref{eq:neffdlambda}) by comparing the cut-on and cut-off determined from the OBJ measurements at $\theta=\ang{0}$ and $\theta=\ang{7}$ (Sect.~\ref{sec:PE950}).

The 9-point measurements of $n_{\rm eff}$ for the flight model filters show a RMS of up to 6\%. This can be caused by local thickness variations of the layers, and by the inherent wavelength uncertainty of the PE950 data. The 37-point measurements -- having higher internal consistency -- show more subtle variations of $n_{\rm eff}$ across the flight spare substrates, decorrelating for spatial distances above $20$--$30$\,mm. Since we integrate the effective passband over a $\sim$90\,mm beam footprint, the variations in $n_{\rm eff}$ are smoothed accordingly. We thus adopt the mean of all design values listed in Table~\ref{table:neff}, $\langle n_{\rm eff}\rangle=1.769$, similar to the mean value ($1.763$) of the measured data. The error made by this simplification -- i.e. constant $n
_{\rm eff}$ -- is found negligible (see Sect.~\ref{sec:passband_uncertainties}).

\begin{figure}[t]
\centering
\includegraphics[angle=0,width=0.96\hsize]{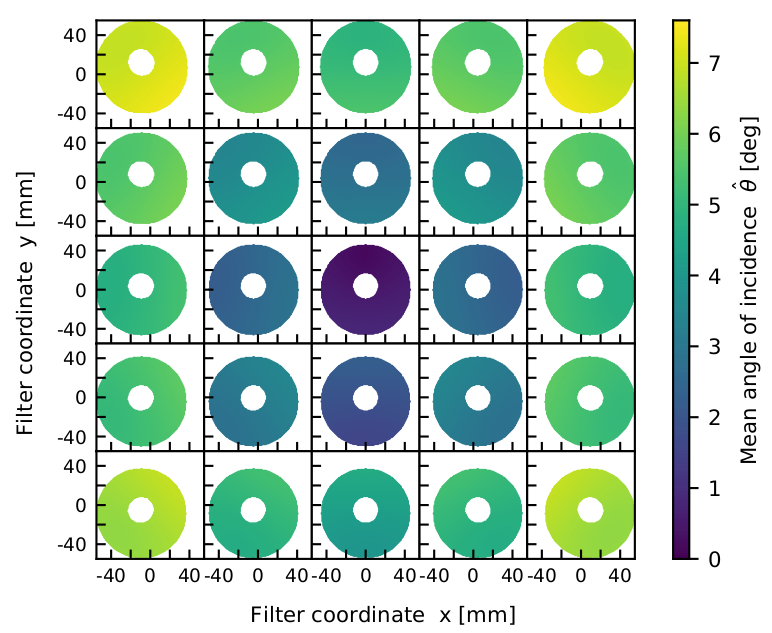}
\caption{The mean AOI, $\hat\theta$, computed from the AOI on the front and back filter side, for different footprint positions on the filter surface. The coordinate squares represent the largest area on the filter within which the footprint can move: footprints displayed at the grid corners correspond to the corners of the FPA. The linear obstructions by the three spider arms  -- holding the secondary mirror -- are too thin to be seen in this representation. Note that we actually work with a $9\times9$ grid; for this figure, every 2nd row and column has been omitted for clarity.}
\label{fig:filter_incident_angle}
\end{figure}

\subsection{AOI from {\tt Zemax} optical ray tracing\label{sec:raytracing}}
As stated above, the sky-facing side of the NISP filters is slightly convex with a curvature radius of $\sim$10\,000\,mm, and their rear side is flat; thus the angles of incidence and exitance are different. As mentioned in Sect.~\ref{sec:coatings}, a \textit{single} passband flank is defined by coatings on \textit{both} surfaces, with more details undisclosed to us. Since the angles of incidence and exitance of a ray differ at most by $[-0.32^\circ;+0.25^\circ]$, we simply use their mean value, $\hat{\theta}$. The error made by this simplification is estimated in Sect.~\ref{sec:computation}, by repeating the computation for both angles or filter surfaces.

\begin{figure}[t]
\centering
\includegraphics[angle=0,width=1.0\hsize]{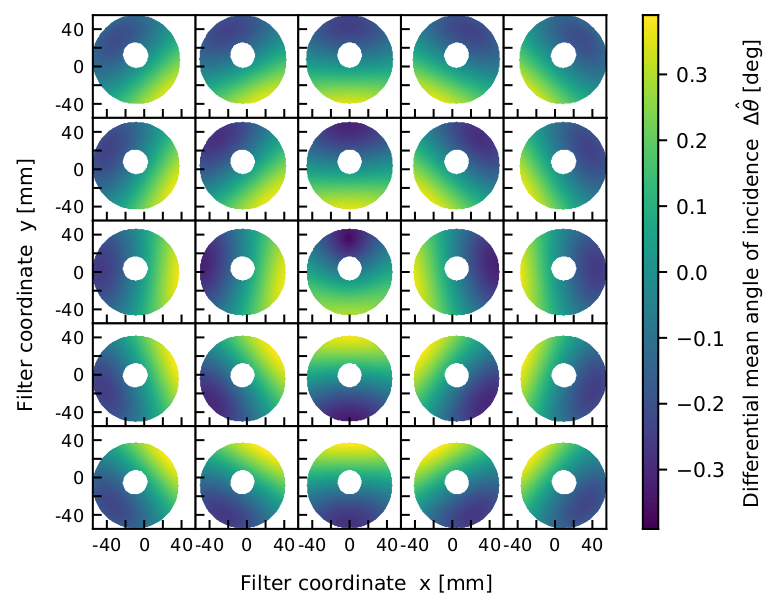}
\caption{Same as Fig.~\ref{fig:filter_incident_angle}, after subtracting  $\langle\hat\theta\rangle$ of each footprint. This highlights the variation of $\hat\theta$ within the footprints. Note the different color scale compared to Fig.~\ref{fig:filter_incident_angle}.}
\label{fig:filter_incident_angle_differential}
\end{figure}

Using {\tt Zemax} ray tracing, we determine the distribution of $\theta(x,y)$ on both filter surfaces for a grid of $9\times9$ positions in the field of view. For each position, the pupil plane of the optical system is sampled with a regular grid of $100\times100$ rays, some of which are vignetted\footnote{The coordinate system used by {\tt Zemax} for the filter plane must be rotated by \ang{180} to match the $(x,y)$ filter substrate coordinate system, in which OBJ performed the transmission measurements. Figures~\ref{fig:filter_incident_angle} and \ref{fig:filter_incident_angle_differential} use the filter substrate coordinate system.}.

The resulting values of $\hat\theta(x,y)$ are shown in Fig.~\ref{fig:filter_incident_angle}, where every second footprint from the $9\times9$ grid is omitted for clarity. In the computation of the effective transmission (Sect.~\ref{sec:computation}), we use a 2D spline interpolation for $\hat\theta(x,y)$ for each node in the $9\times9$ grid. Averaging $\hat\theta$ for each footprint, we find $\langle\hat\theta\rangle=0.5^\circ$ for the central position; this fairly large value is caused by a deliberate small tilt of the NISP optics to optimize the image quality in the off-axis system. For the footprints in the bottom and top corners, we find $\langle\hat\theta\rangle=6.5^\circ$ and $7.1^\circ$, respectively. Within a footprint, $\hat\theta$ varies by about $\pm0.3^\circ$; this is displayed in Fig.~\ref{fig:filter_incident_angle_differential}, where we plot $\Delta\hat\theta=\hat\theta-\langle\hat\theta\rangle$. 

The conclusion of this Section is that $n_{\rm eff}$ and $\hat\theta(x,y)$ are well understood and accurately known for the computation of the effective filter transmission, $T(\lambda)$, in Sect.~\ref{sec:computation}.

\section{Computing the filters' effective transmission\label{sec:computation}}
\subsection{Numerical integration with blueshift\label{sec:beamintegration}}
\subsubsection{Computational principle\label{sec:comp_principle}}
Let $t(\lambda,x,y)$ be the \textit{local} filter transmission for $\theta(x,y)=0^\circ$, where $x$ and $y$ are the Cartesian filter coordinates defined in Fig.~\ref{fig:filtergeometry}. For an oblique angle $\theta$, we find the blueshifted transmission, $t^\prime$, by looking up the original transmission, $t$, at the correspondingly redshifted wavelength $\lambda^\prime$, 
\begin{equation}
t^\prime(\lambda,x,y,\theta) = t\left(\lambda^\prime(\theta),x,y\right),
\label{eq:tprimet}
\end{equation}
with
\begin{equation}
\lambda^\prime(\theta) = \lambda\left[1 -\left( \frac{\sin\theta} 
  {n_{\rm eff}(\lambda)}\right)^2\right]^{-1/2}\;.
  \label{eq:neff_scaling}
\end{equation}
The scaling factor is just the inverse of the one in Eq.~(\ref{eq:neffdlambda}), used to compute the blueshift. 

Note that in general $t^\prime(\lambda)\neq t(\lambda^\prime)$, as the transmission at an oblique angle decreases with respect to normal incidence. However, for small values of $\theta$, and for angle-tuned coating designs, this effect can be neglected: For NISP we measure a reduction in mean peak transmission of $0.03$\%, $0.17$\% and $0.04$\% for \ymagm, \jmagm, and \hmagm, respectively, for $\theta=\ang{7}$ with respect to $\theta=\ang{0}$, and thus Eq.~(\ref{eq:tprimet}) remains valid.

Now consider a source image anywhere in the focal plane. The corresponding chief ray intersects the filter substrate at a point $P$. The effective transmission, $T(\lambda)$, for the source is given by integrating the local blueshifted transmission, $t(\lambda^\prime)$, in the filter plane over the annular footprint $A$ centered on $P$,
\begin{equation}
    T(\lambda) = \int\displaylimits_A t\left(\lambda^\prime(\theta),x,y\right)\,{\rm d}^2 A\,.
    \label{eq:bandpassintegration}
\end{equation}
To numerically compute this integral, we construct a spline interpolating function\footnote{Using {\tt Mathematica}'s {\tt Interpolation[]} routine.} to the 9-point data that sample $t\left(\lambda^\prime(\theta),x,y\right)$. The interpolation is quadratic in the two spatial dimensions and cubic in the spectral dimension. And we replace $\theta(x,y)$ by a spline interpolation of the mean AOI, $\hat\theta(x,y)$ (see Sect.~\ref{sec:raytracing}). 

\subsubsection{Simplifications to the beam footprint geometry}
To facilitate the numeric integration in Eq.~(\ref{eq:bandpassintegration}), we simplify the footprint geometry. The outer edge of the footprint is nearly circular, with the largest ellipticity of $91.8\times93.0$\,mm realized for a source in one of the corners of the FPA. We approximate the edge with a circle, taking the mean of the minor and major axis as the radius. The circle's diameter varies between 89.8\,mm and 92.4\,mm, depending on position on the filter substrate.

Likewise, the decentered obstruction is considered circular. We approximate it with a fixed diameter of 27.8\,mm, and a fixed spatial offset of $y=+3.7$\,mm in the filter coordinate system, independent of footprint location.

These simplifications alter the cut-on and cut-off wavelengths by less than 0.01\,nm, and are ignored in our error budget.

\subsubsection{Simplifications in the transmission integration\label{sec:anglesimplifications}}
In Sect.~\ref{sec:raytracing}, we show that due to the unknown coating design we must use the mean AOI, $\hat\theta(x,y)$, from the curved and planar filter sides. To estimate the error made by this approximation, we recompute Eq.~(\ref{eq:bandpassintegration}) for the central footprint position twice, once using $\theta(x,y)$ for the curved side, and once for the planar side. The resulting wavelength shifts of the passbands' cut-on and cut-off are within $[-0.014;0.017]$\,nm with respect to the computation for $\hat\theta(x,y)$. Hence this simplification is valid, and the error small enough to be ignored in the error budget.

\begin{figure*}[t]
\centering
\includegraphics[angle=0,width=0.78\hsize]{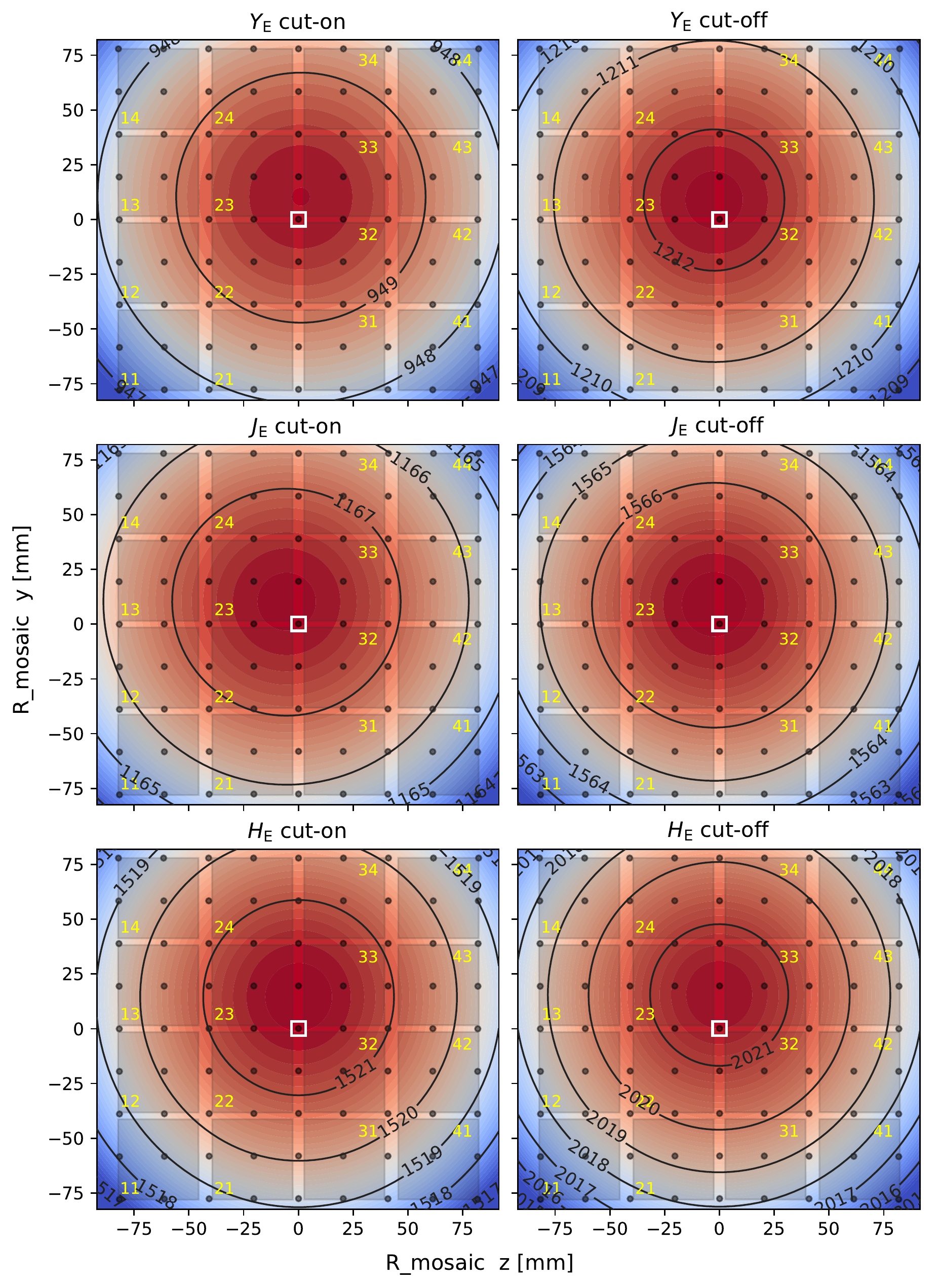}
\caption{Cut-on and cut-off wavelengths in nm as a function of focal plane position, cold and in vacuum. The blueshift toward the focal plane corners is evident. The black lines and colored background show the polynomial fits introduced in Sect.~\ref{sec:passband_variations}; the fit parameters are listed in Table~\ref{table:pbvar_coeffs}, and the joint residuals are shown in Fig.~\ref{fig:bpvar_residuals}. The fits are based on the $9\times9$ object positions (small dots) for which we inferred the AOI distribution on the filter surface using {\tt Zemax} ray tracing. The shaded squares display the 16 NISP detectors, numbered in yellow from 11 to 44. The position of the number indicates the location of the (1|1) pixel of a detector. The published response curves (Sect.~\ref{sec:data_products}) were computed for the central dot marked with the white square. {\tt R\_mosaic} $(z,y)$ is a physical coordinate system in the focal plane, to describe -- among others -- the detector positions.}
\label{fig:pbvar_YJH}
\end{figure*}

\subsection{Passband variations\label{sec:passband_variations}}
\subsubsection{Mathematical model for the total passband shift}
To determine the variation of the cut-on and cut-off across the focal plane, we evaluate $T(\lambda)$ 
for each node of the $9\times9$ grid introduced in Sect.~\ref{sec:raytracing}. The results are displayed in Fig.~\ref{fig:pbvar_YJH}, for cold conditions and in vacuum (Sect.~\ref{sec:tempvacshifts}), in the physical {\tt R\_mosaic} $(z,y)$ coordinate system in which the positions of the NISP detectors are accurately known. The transformation of the filter coordinate system used in Fig.~\ref{fig:filter_incident_angle} to the 
{\tt R\_mosaic} system is
\begin{alignat}{2}
z_{\rm R\_mosaic} &= -x_{\rm Filter}\\
y_{\rm R\_mosaic} &= -y_{\rm Filter}\;,
\end{alignat}
i.e. the panels in Fig.~\ref{fig:pbvar_YJH} are rotated by \ang{180} with respect to the panels in Fig.~\ref{fig:filter_incident_angle}.

For accurate and unbiased photo-$z$ measurements, we need a mathematical prescription that provides the cut-on and cut-off wavelengths for any position in the focal plane. To this end, we fit 2D 3rd degree polynomials of the following form to the cut-on and cut-off surfaces in the focal plane for each filter,
\begin{equation}
    \lambda_{\rm 50}^{\rm on/off}(z,y) = a_0 + \sum_{i=1}^3 b_i\, z^{\;i} + 
    \sum_{i=1}^3 c_i\, y^{\;i}\;.
    \label{eq:pbvar_polynomials}
\end{equation}
These fits have residuals with an RMS of $\sigma=0.011$\,nm, with a maximum value of $0.06$\,nm (Fig.~\ref{fig:bpvar_residuals}). The coefficients are listed in Table~\ref{table:pbvar_coeffs} in the appendix. The inclusion of crossterms did not improve the quality of the fits.

\begin{figure*}[t]
\centering
\includegraphics[angle=0,width=1.0\hsize]{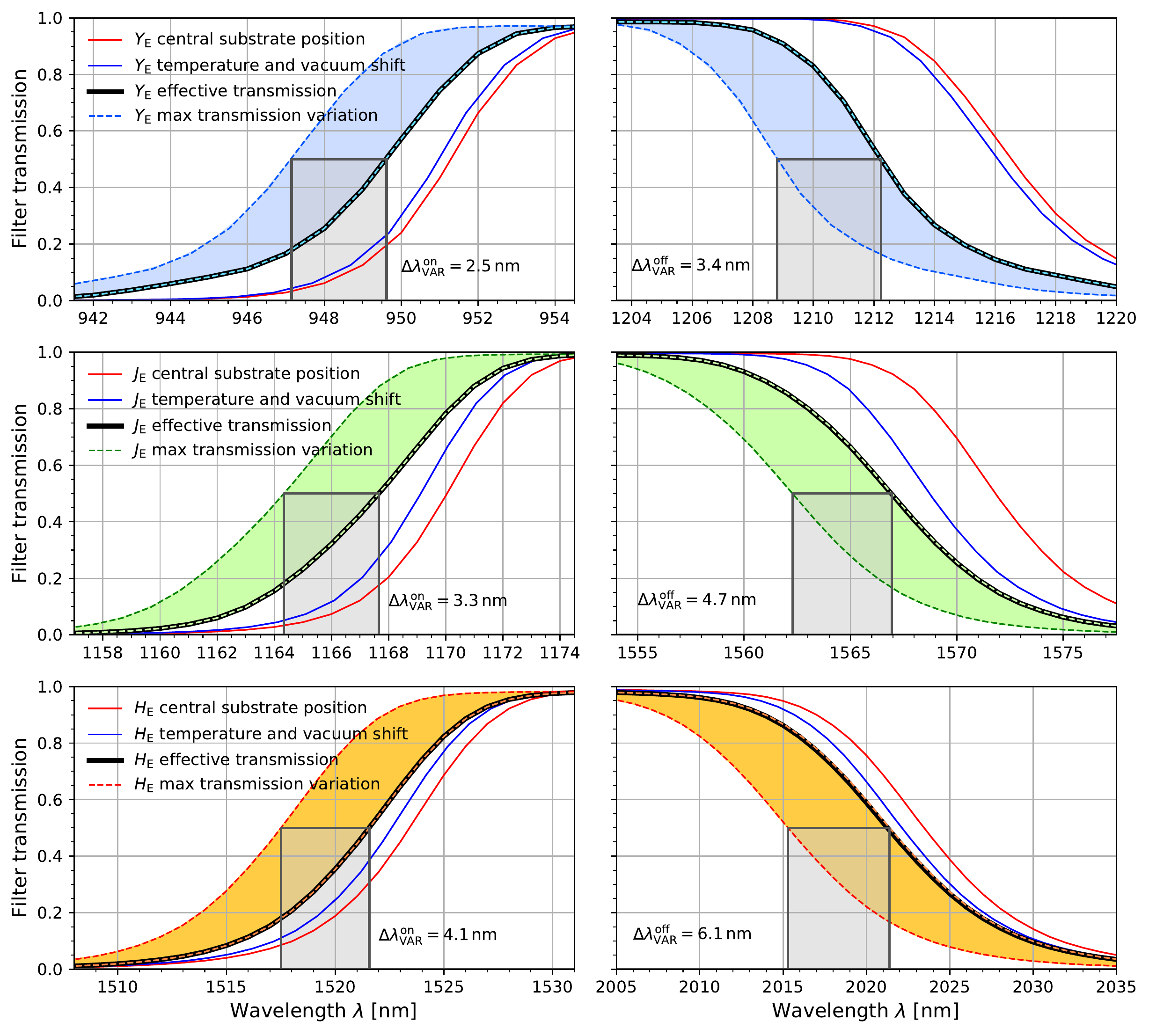}
\caption{NISP passband flanks and variations. The solid red lines display the local transmission at the filter center (see Fig.~\ref{fig:filtergeometry}). The solid blue line shows the combined vacuum and temperature shifts on the red curve. The thick black line shows the effective transmission after integrating over the central beam footprint (white squares in Fig.~\ref{fig:pbvar_YJH}), including the temperature and vacuum shifts. The colored areas enclosed by dashed lines show the range of blueshifts seen by the FPA: the left-most dashed line corresponds to the bluest FPA corners, and the right-most to the reddest passband near -- but not necessarily coincident with -- the FPA center. The gray rectangles show the maximum spread, $\Delta\lambda_{\rm VAR}^{\rm on/off}$, of the cut-on and cut-off.}
\label{fig:filterprofiles}
\end{figure*}

The maximum blueshift, $\Delta\lambda_{\rm VAR}$, seen by the FPA increases about linearly with wavelength, from $2.7$\,nm for the \yband cut-on, to $5.8$\,nm for the \hband cut-off. These values are also given in Table~\ref{table:passbandproperties} and shown in Fig.~\ref{fig:filterprofiles}. 

By setting $\theta=0$ for all rays in the computation, we switch off the dependence on the AOI and compute the contribution of the intrinsic filter variations, only. We find that the latter contribute between $10$--$23$\% to the total focal plane variations.


Note that our published NISP response curves (Sect.~\ref{sec:data_products}) were computed for an object at the center of the FPA, near the location where the cut-on and cut-off have their longest wavelengths (Fig.~\ref{fig:pbvar_YJH}). 

\begin{figure}[t]
\centering
\includegraphics[angle=0,width=0.9\hsize]{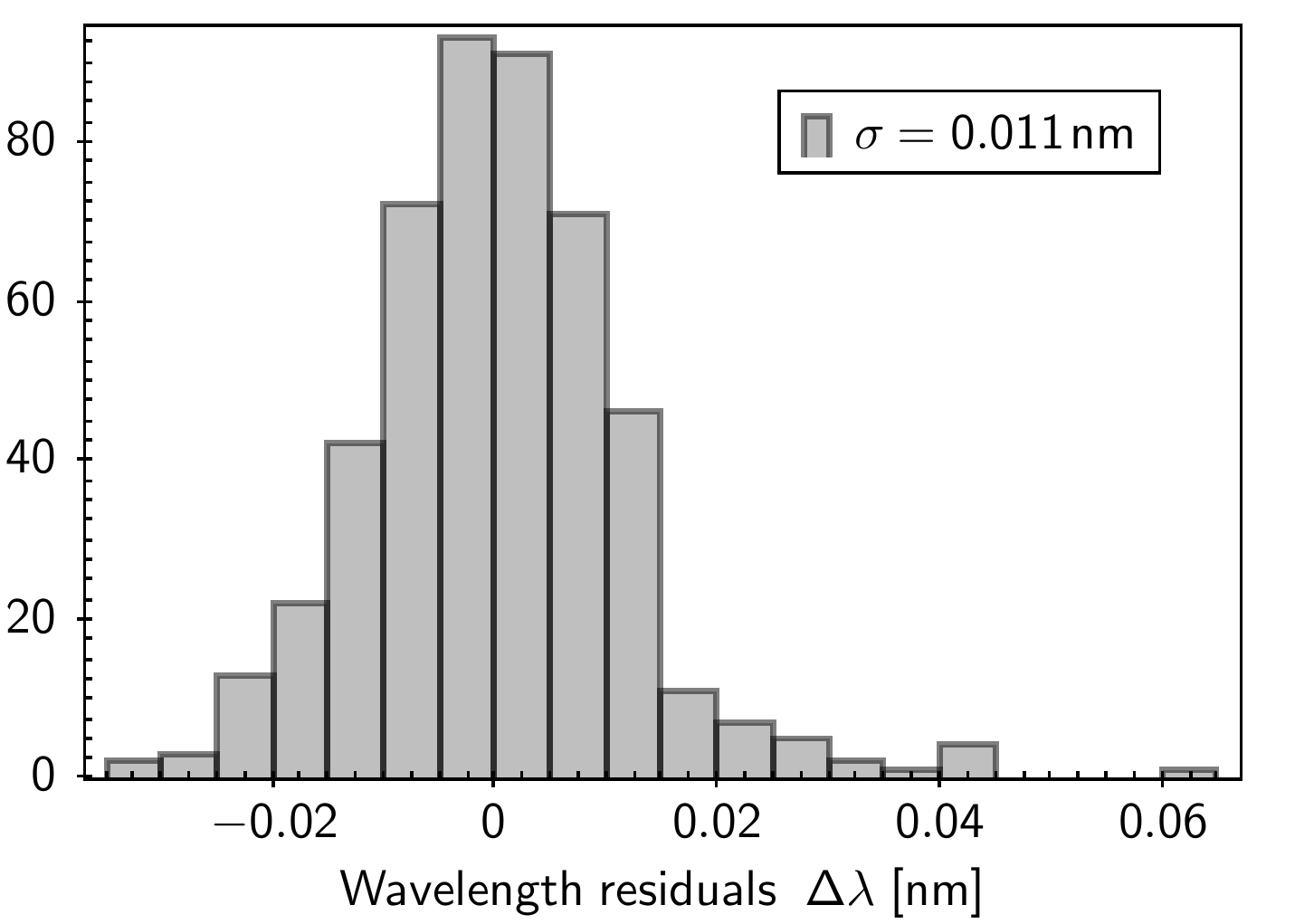}
\caption{Joint wavelength residuals of the polynomial fits to the passband variations, for all filters (their cut-on and cut-off; see also Fig.~\ref{fig:pbvar_YJH}). The RMS is $\sigma=0.011$\,nm; 92\% of the residuals are smaller than 0.02\,nm.}
\label{fig:bpvar_residuals}
\end{figure}

\subsubsection{Passband variations for the field corners due to uncertainties in \texorpdfstring{$n_{\rm eff}$}{n\_eff}\label{sec:neff_uncertainty}}
In Sect.~\ref{sec:neff}, we reviewed the uncertainties and variations in $n_{\rm eff}$. They enter $T(\lambda)$ through Eq.~(\ref{eq:neff_scaling}), increasing or decreasing the blueshift in the field corners with respect to the field center. For a conservative estimate, we recompute $T(\lambda)$ twice, increasing and decreasing $n_{\rm eff}$ by 5\%. We find the following:

At the center of the FPA, where the typical AOI is $\theta=0.5^\circ$, the passbands always shift by 0.002\,nm or less; the published response curves (Sect.~\ref{sec:data_products}) are computed for this footprint.

At the corners of the FPA with typical values of $\theta=7^\circ$, however, significant shifts occur. In case $n_{\rm eff}$ is larger (smaller) by 5\%, the blueshift $\Delta\lambda_{\rm neff}$ for \ymagm, \jmagm, and \hmagm increases (decreases) by 0.2, 0.3 and 0.4\,nm, respectively. In relative terms, the difference of the cut-on wavelength between the center of the FPA and its corners increases (decreases) by $7$--$10$\%; same for the cut-off. 

\subsubsection{Passband variations and effect on source photometry\label{sec:pbvar_source_photom}}
Without correcting for passband variations, the measured flux for an individual source depends on its position in the field of view, for two reasons. First, the blueshifted passband selects a different part of the source's spectral energy distribution (SED), increasing or decreasing the measured flux. Second, the cut-off experiences a larger blueshift than the cut-on when moving from the center of the FPA to a corner. This `passband compression' reduces the widths of the \ymagm-, \jmagm, and \hmagm-bands by 0.7, 0.8 and 1.2\,nm, respectively, typically resulting in a SED-dependent flux reduction.

To estimate the joint amplitude of both effects, we use the catalogs from the Euclid SC8 (`science challenge 8'), simulating 100\,deg$^2$ of sky with realistic SEDs as they will be seen by the Euclid Wide Survey \citep{scaramella2021}. We randomly select 1.8 million stars and 1.1 million redshifted galaxies from the SC8 catalogs (see Sect.\ref{sec:euclid_sc8} for details), and compute the difference in source photometry for a position at the center of the FPA and another one in the most blue-shifted corner. The results are shown in Fig.~\ref{fig:photometry_difference_galaxies} for galaxies and Fig.~\ref{fig:photometry_difference_stars} for stars. The effect on individual source photometry is on the order of $1$--$5$\,mmag. The differential photometry for galaxies shows a very low but also broad tail up to 0.6\,mag. This tail is comprised of galaxies with bright emission lines that shift in -- or out -- of a passband as the source moves across the field of view. In principle, if such emission line galaxies were identified in the survey data, they could be used to improve the knowledge error of the passbands' cut-on and cut-off as tabulated in Table~\ref{table:passbandproperties}.

We note that due to the slightly reduced transmission at oblique angle of incidence, generic flux losses of up to 0.2\% can occur in the corners of the focal plane (Sect.~\ref{sec:comp_principle}). This effect is systematic and automatically removed by the illumination correction procedure in the data processing pipeline.

\subsection{Temperature and vacuum wavelength shifts\label{sec:tempvacshifts}}
The transmission measured at ambient temperature\footnote{In order to not mix up transmission and temperature, we denote temperature by $\tau$ throughout this paper.}, $\tau$, and pressure will change in vacuum and at near cryogenic temperatures. The temperature dependence is caused by the thermal contraction of the dielectric coating layers, as well as a change in $n_{\rm eff}$ \citep[e.g.][for SiO$_2$]{tan2000}. 

To quantify these effects, \ymagm-, \jmagm-, and \hmagm-band coating samples were cryocycled in a vacuum chamber at Martin-Luther-University Halle-Wittenberg (Germany), while simultaneously recording their transmission. We determine the cut-on and cut-off 
in the same manner as for the actual filters. As expected, the passbands redshift slightly when going into vacuum, and experience a larger blueshift when cold; the combined effect on the passband flanks is shown in Fig.~\ref{fig:filterprofiles}.

\begin{table}[t]
\caption{Fit parameters to compute the blueshift for cooldown, Eq.~(\ref{eq:tempdep}), and the redshift for the transition to vacuum, Eq.~(\ref{eq:vacdep}).}
\smallskip
\label{table:tempvacshiftcoeffs}
\smallskip
\begin{tabular}{|l|cc|cc|}
\hline
  & \multicolumn{2}{c|}{Cold effect} & \multicolumn{2}{c|}{Vacuum effect} \\
\hline
Filter & $p_1$ & $p_2$ & $q_1$ & $q_2$ \\
\hline
\ymagm & $0.155$ & $-7.321\times 10^{-4}$ & $0.0376$ & $1.591\times 10^{-4}$ \\
\jmagm & $5.603$ & $-5.680\times 10^{-3}$ & $0.2243$ & $-1.464\times 10^{-4}$ \\
\hmagm & $0.707$ & $-1.083\times 10^{-3}$ & $0.0788$ & $-1.443\times 10^{-5}$ \\
\hline
\end{tabular}
\end{table}

\subsubsection{Temperature dependence}
The temperature-related blueshift is normally computed with the Sellmeier equation \citep[see][for a review]{fang2019}. Here, however, since the transmission was measured at two temperatures only, we linearly fit the cooldown blueshift as
\begin{equation}
   \Delta\lambda_{\rm Temp}(\tau,\lambda) = \frac{\tau_1-\tau}{\tau_1-\tau_0}\,\left(p_1+p_2\,\frac{\lambda}{1\,{\rm nm}}\right)\;,
\label{eq:tempdep}
\end{equation}
with $\tau_1=295$\,K and $\tau_0=120$\,K given by the setup, and $p_1$ and $p_2$ are the filter-dependent fit coefficients in Table~\ref{table:tempvacshiftcoeffs}.

The expected in-flight filter temperature is $\tau=132$\,K. The central wavelengths of the \ymagm-, \jmagm- and \hband filters shift by $-0.60$\,nm, $-2.04$\,nm, and $-1.13$\,nm, respectively. We correct all wavelengths for their individual temperature dependence, $\Delta\lambda_{\rm Temp}(\tau,\lambda)$.

Note that $\tau=132$\,K is the temperature we use for the passbands characteristics reported in Table~\ref{table:passbandproperties}. It is unlikely that the passbands need to be corrected for the \textit{actual} in-flight temperature that will only be known after launch; the expected temperature bracket for the NISP optics is $130$--$135$\,K. For reference, a change by $+1$\,K in filter temperature would cause a redshift by $0.004$--$0.012$\,nm, only. Furthermore, the error we make by using the linear fit in Eq.~(\ref{eq:tempdep}) instead of the nonlinear Sellmeier equation is less than $0.1$\,nm, and therefore neglected.

\subsubsection{Vacuum dependence}
In analogy to the temperature dependence, we compute a linear fit of the passband redshift when going into vacuum, 
\begin{equation}
   \Delta\lambda_{\rm Vac}(\lambda) = q_1+q_2\,\frac{\lambda}{1\,{\rm nm}}\;.
   \label{eq:vacdep}
\end{equation}
The coefficients are collected in Table~\ref{table:tempvacshiftcoeffs}. The corrections are minor, amounting to $0.21$\,nm, $0.023$\,nm, and $0.053$\,nm at the central wavelengths of the \ymagm-, \jmagm-, and \hmagm-band filters, respectively. Like for the temperature dependence, we correct all wavelengths for their individual vacuum dependence. 

\subsection{Summary of the relevant passband uncertainties\label{sec:passband_uncertainties}}
In Sect.~\ref{sec:neff}, we discussed the uncertainties of $n_{\rm eff}$ and use a constant $\langle n_{\rm eff}\rangle = 1.769$ for all filters. In Sect.~\ref{sec:neff_uncertainty}, we estimate corresponding errors of the cut-on and cut-off of $0.2$, $0.3$ and $0.4$\,nm for filters \ymagm, \jmagm, and \hmagm, respectively, for the corners of the FPA. This error is negligible at the center of the FPA.

In Sect.~\ref{sec:wavelength_accuracy}, we placed a conservative measurement accuracy for the cut-on and cut-off wavelengths of $0.8$\,nm, applicable to all filters. Adding in quadrature the errors for $n_{\rm eff}$ as given in the previous paragraph, we obtain increased total uncertainties \textit{in the FPA corners} of $\sigma_{50}=0.8$, 0.9 and 0.9\,nm, respectively, for the three bands. The estimated uncertainty for the passbands' central wavelengths is $\sigma_{\rm cen}=0.6$\,nm for any point in the field of view, for all bands. The uncertainties are listed in Table~\ref{table:passbandproperties}.

All other uncertainties, and errors made due to simplifications, are below 0.1\,nm and neglected.

\subsection{Effect on photometric redshifts}
Since photo-$z$ estimates are one of NISP's principal purposes, we want to briefly describe the effect of the derived passband know\-ledge uncertainties on the photo-$z$ estimates. Obviously, the effect is highly dependent on the sources' redshifted SEDs, and on the uncertainties of complementary optical passbands used. It is beyond this paper to quantify this accurately. Nonetheless, a simple example illustrates the overall suitability of the computed NISP passbands for photo-$z$ purposes:

Consider the $0.95$--$1.21$\,\micron\ range of the \yband (Table~\ref{table:passbandproperties}). The characteristic 4000\,\AA\;break falls within the \yband for redshifts $z\in[1.375,2.025]$. If the 4000\,\AA\;break was the sole spectral feature determining the photo-$z$ of a source, then the uncertainty in the central wavelength, $\sigma_{\rm cen}=0.6$\,nm, would introduce a redshift bias of $\Delta z=\sigma_{\rm cen}/400\,{\rm nm}=0.0015$, compared to the requirement of $0.002\,(1+z)=0.0048$ for $z=1.375$. Hence, in this example, the knowledge error of the passband flanks is sufficient. In practice, there are many sources of bias, which need to be quantified with realistic simulations for a wide range of redshifted SEDs, folding in complementary ground-based observations and their uncertainties \citep[e.g.][]{ilbert2021}. We remark that the bias $\Delta z$ can be calibrated, as well as the effects of the passband variations across the focal plane.

\section{Total system response\label{sec:totaltransmission}}
\Euclid observes simultaneously with NISP and its Visible Instrument \citep[VIS;][]{cropper2012} using a dichroic beam-splitter. To prepare well-defined input passbands for both instruments, \Euclid relies on a finely orchestrated balance of mirror coatings, the dichroic, filters, the NISP optical assembly (NI-OA, see Sect.~\ref{sec:NIOA}), and detector QE, as illustrated in Fig.~\ref{fig:chromatic_selection}, and  more quantitatively in Fig.~\ref{fig:transmission_elements}. While the NISP filters alone define the cut-on and cut-off wavelengths, it is only in conjunction with the other elements that excellent out-of-band blocking is achieved. The total NISP transmission is therefore given by
\begin{equation}
  T_{\rm tot}(\lambda) = T_{\rm Tel}(\lambda) \;\, T_{\rm NI\text -OA}(\lambda) \;\,T_{\rm Filter}(\lambda)\;\,T_{\rm QE}(\lambda),
  \label{throughput2}
\end{equation}
where $T_{\rm Tel}(\lambda)$ is the telescope contribution, accounting for the mirrors and the dichroic. Dependencies of the individual factors $T$ on the AOI are negligible. We explain \Euclid's chromatic selection function 
(Figs.~\ref{fig:chromatic_selection} and \ref{fig:transmission_elements}) in the following.

\subsection{\Euclid's chromatic selection function\label{sec:chromaticselection}}
\subsubsection{Telescope transmission}
The telescope optics common to VIS and NISP consists of -- in this order -- the primary mirror M1, the secondary mirror M2, folding mirrors FoM1 and FoM2, the tertiary mirror M3, and the dichroic element. M1, M2 and M3 are coated with protective silver and provide a very broad wavelength coverage longward of $0.34$\,\micron. After the dichroic and in the VIS optical path only, folding mirror FoM3 is also coated with protected silver.

At wavelengths of 0.5\,\micron\ and below, VIS and NISP have considerable QE, which is undesirable for both instruments: At these wavelengths, galaxy images are increasingly dominated by intrinsic substructures, adding noise to the VIS weak lensing shape measurement; and for NISP, unfiltered UV/blue photons would contaminate the NIR photometry. Therefore, folding mirrors FoM1 and FoM2 have dielectric coatings including three layers of gold, each. By means of destructive interference and absorption, these layers block light shortward of $0.42$\,\micron, reach maximum reflectivity at 0.50\,\micron, and maintain it beyond $2$\,\micron.

\begin{figure}[t]
\centering
\includegraphics[angle=0,width=1.0\hsize]{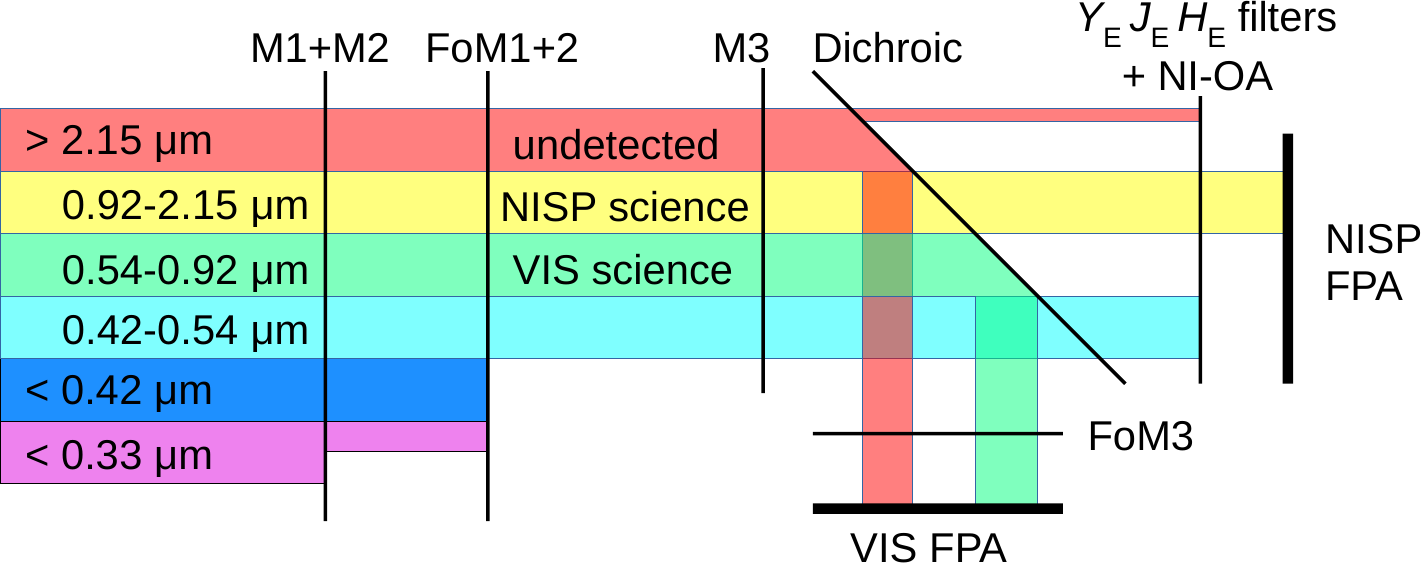}
\caption{Chromatic selection function of the \Euclid optical elements. Note that the VIS detectors have zero QE for photons with $\lambda>2.15$\,\micron. The exact behavior of the dichroic above $2.2$\,\micron\ is unknown: longer wavelengths could enter NISP and would be blocked by the filters. See Sect.~\ref{sec:chromaticselection} and Fig.~\ref{fig:transmission_elements} for more details.}
\label{fig:chromatic_selection}
\end{figure}

\begin{figure*}[t]
\centering
\includegraphics[angle=0,width=1.0\hsize]{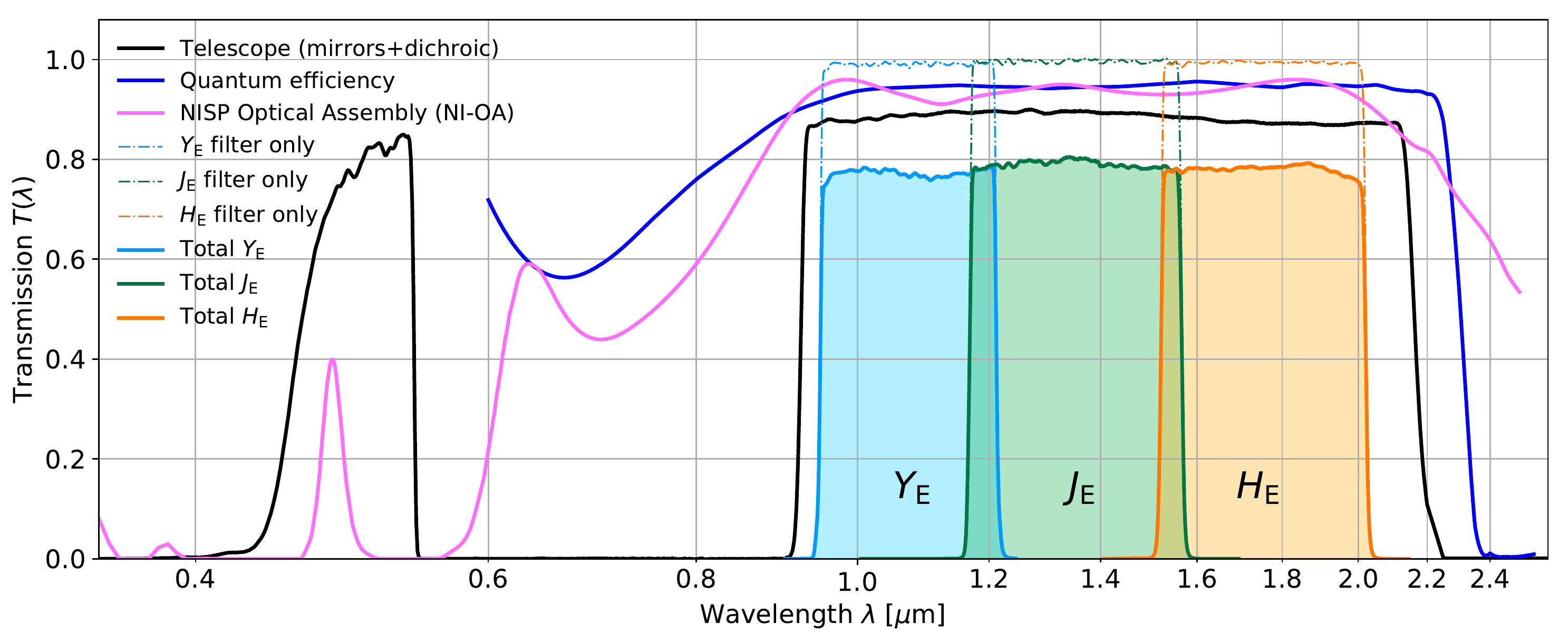}
\caption{Transmission of elements in the NISP optical path. The telescope component (black line) contains all mirrors and the dichroic; the latter redirects the $0.54$--$0.93$\,\micron\ range to VIS, and cuts off at $2.15$\,\micron. The pink line shows the transmission of the NI-OA. The blue line shows the interpolated mean detector QE, originally measured at 50\,nm intervals. The thin lines show the effective filter transmission integrated over the beam footprint, and the shaded areas the total response (accounting for filter, telescope, NI-OA and QE). The residual optical transmission from the dichroic and the NI-OA between $0.45$ and $0.54\,\micron$ is fully suppressed by the filters' out-of-band blocking (see Fig.~\ref{fig:blocking}). These curves do not include effects from particulate contamination.}
\label{fig:transmission_elements}
\end{figure*}

The dichroic has a complex wavelength selection function (Figs.~\ref{fig:chromatic_selection} and \ref{fig:transmission_elements}). It cuts out a wavelength range from $0.54$--$0.93$\,\micron\ 
and reflects it to VIS. Photons within $0.93$--$2.15$\,\micron\ are transmitted to NISP. Photons below $0.54$\,\micron\ are also sent to NISP, and are blocked by the filters and -- partially -- by the NI-OA (Sect.~\ref{sec:blueleak}). Photons above $2.15$\,\micron\ are mostly reflected to VIS again. This does not matter for VIS, as the CCDs do not detect photons longward of $\sim$1.1\,\micron. Note that the dichroic's reflectance was not determined nor specified beyond $2.2$\,\micron. The NISP filters block out to $2.9$\,\micron\ (Sect.~\ref{sec:blueleak}), and the detectors have a QE cut-off at $2.3$\,\micron\ (Sect.~\ref{sec:NISPqe}).

The combined transmission, $T_{\rm Tel}(\lambda)$, of this arrangement entering the NISP instrument is shown as the black line in Fig.~\ref{fig:transmission_elements}. For more details, see also \cite{venancio2020}. 
\begin{table*}[t]
\caption{Summary of the \Euclid NISP AB mag photometric system. Wavelengths and their uncertainties are computed from higher precision data and rounded to one digit. The first half of the table lists the global passband characteristics; `Width' refers to the distance between the 50\% cut-on and cut-off wavelengths, $\langle T_{\rm peak}\rangle$ to the mean in-band response, and ZP to the photometric zeropoint in the AB system. The second half of the table lists the \textit{total} uncertainties of the cut-on and cut-off wavelengths, $\sigma_{50}$, and that of the central wavelength, $\sigma_{\rm cen}$.}
\smallskip
\label{table:passbandproperties}
\smallskip
\begin{tabular}{|lrrrrrrrr|}
\hline
\rowcolor{gray!20}
&\multicolumn{8}{c|}{NISP spectral response characteristics}\\
\hline
Filter  & $0.1$\% cut-on & $50$\% cut-on & $50$\% cut-off & 0.1\% cut-off & $\lambda_{\rm cen}$ & Width & $\langle T_{\rm peak}\rangle$ & ZP\\
  & [nm] & [nm] & [nm] & [nm] & [nm] & [nm] & & [AB mag]\\
\hline
\ymagm &  937.5 &  949.6 & 1212.3 & 1243.2 & 1080.9 & 262.7 & 0.772 & 25.04\\
\jmagm & 1151.1 & 1167.6 & 1567.0 & 1595.0 & 1367.3 & 399.4 & 0.790 & 25.26\\
\hmagm & 1495.6 & 1521.5 & 2021.4 & 2056.8 & 1771.4 & 499.9 & 0.782 & 25.21\\
\hline
\rowcolor{gray!20}
& \multicolumn{3}{c}{Wavelength uncertainty (center of FPA)} & \multicolumn{4}{c}{Wavelength uncertainty (corner of FPA)} & \\
\hline
Filter & $\sigma_{50}$ & $\sigma_{\rm cen}$ & & $\sigma_{50}$ & $\sigma_{\rm cen}$ & & & \\
 & [nm] & [nm] & & [nm] & [nm] & & & \\[2pt]
\hline
\ymagm & 0.8 & 0.6 & & 0.8 & 0.6 & & & \\
\jmagm & 0.8 & 0.6 & & 0.9 & 0.6 & & & \\
\hmagm & 0.8 & 0.6 & & 0.9 & 0.6 & & & \\
\hline
\end{tabular}
\end{table*}

\subsubsection{NISP optical assembly (NI-OA) transmission}\label{sec:NIOA}
NISP carries four lenses, collectively known as the NI-OA \citep{bodendorf2019,grupp2019}; their transmission was not measured. However, we have the transmission from witness samples that were  present in the coating chamber, used by industry to demonstrate that a minimum transmission requirement has been exceeded. The joint transmission of these witness samples is shown as the pink curve in Fig.~\ref{fig:transmission_elements}. 

Note that this transmission \textit{estimate} varies between $0.93$--$0.96$ over the $0.95$--$2.02$\,\micron\ range. It is unclear whether these variations are also replicated in the flight model optics that may deviate by several percent. Steep gradients are not expected across this wavelength range, since only few coating layers were used. The true transmission of NI-OA will be constrained further in-flight, once observations of white dwarf (WD) spectrophotome\-tric standard stars are available (Sect.~\ref{sec:inflightcalibrationabs}).

The NI-OA blocks photons below $0.58$\,\micron, apart from a narrow transmission window between $0.46$--$0.52$\,\micron\ that overlaps with wavelengths transmitted by the dichroic. The residual optical transmission into NISP is suppressed by the filters (Sect.~\ref{sec:blueleak}).

\subsubsection{NISP detector quantum efficiency\label{sec:NISPqe}}
The NISP detectors have a sharp QE cut-off at $2.3$\,\micron, controlled by the Cd concentration in the HgCdTe alloy \citep{rogalski2005}. This provides further rejection of long wavelengths, in addition to the dichroic sending most photons above $2.15$\,\micron\ to VIS. 

The QE was measured for each pixel in steps of 50\,nm from $0.60$ to $2.55$\,\micron\ \citep{waczynski2016}. The relative accuracy -- i.e. the knowledge error of the QE curve shape -- across this wavelength range is 1\%. The absolute scaling is uncertain to 5\%, since the pure QE is difficult to disentangle from the detector gain \citep[][and A.~Waczynski, priv.~comm.]{secroun2018}. For the computation of the out-of-band blocking, only, we linearly extrapolate the QE to zero from $0.60$\,\micron\ to $0.30$\,\micron.

The mean QE curves of the detectors are very uniform across the $0.95$--$2.2$\,\micron\ range \citep{bai2018}, and within just a few percent of each other. Computed for individual detectors at monochromatic wavelengths, the pixel-to-pixel RMS is $1.2$--$2.4$\% globally, and $1.0$\% locally. Their impact on the passband flanks is entirely negligible. 

For this paper, we use the mean QE (blue line in Fig.~\ref{fig:transmission_elements}) computed from all pixels of 14 out of the 16 detectors. We exclude one detector because of its slightly lower QE below 1.0\,\micron, and another because of its increasing QE towards 2.0\,\micron. 
The original QE values between $1.65$--$1.75$\,\micron\ are excluded from all detectors due to a measurement artifact, and replaced by a local, linear interpolation.

\subsection{NISP out-of-band blocking and flux contamination\label{sec:blueleak}}
\subsubsection{Out-of-band blocking}
By requirement, the NISP filters must reject light to better than $10^{-3}$ long- and shortward of their passband; the requirement over the VIS range from $0.54$--$0.93$\,\micron\ is relaxed to $10^{-2}$, since these photons are directed to VIS already. The blocking of the flight model filters was measured locally at the center of the substrates, in steps of 2\,nm from $0.3$ to $2.9$\,\micron. 

With a transmission of $10^{-5}$ to $10^{-7}$ below $0.55$\,\micron, the filters eliminate the residual optical transmission from the dichroic and the NI-OA. Between $0.55$\,\micron\ and the filters' cut-on, and longward of the filters' cut-off, the blocking is better than $10^{-3}$--$10^{-4}$.

The NISP \textit{total out-of-band blocking}, including telescope, NI-OA, filters, and QE, is shown in Fig.~\ref{fig:blocking}. Below $0.9$\,\micron, the blocking is $10^{-7}$ or better. Above $2.2$\,\micron, the blocking is $10^{-5}$, improving to $10^{-7}$ at $2.4$\,\micron, albeit there are some uncertainties from extrapolating the telescope transmission to longer wavelengths. Between $0.9$ and $2.2$\,\micron, the blocking is at least $10^{-4}$.

\subsubsection{Flux contamination from out-of-band photons}
To estimate the contribution from out-of-band photons to the `measured flux' -- i.e. the total flux reaching the detector including out-of-band light --, we consider sources with a power-law SED $f_\nu\propto\nu^{\,\alpha}$ for frequency $\nu$. For $\alpha=1$ and $\alpha=-3$ we have a blue and red source with $\jmag-\hmag$ colors of $-0.27$\,mag and $+0.84$\,mag, respectively, in the AB mag system. We also consider $\alpha=0$, i.e. sources whose AB mag colors are zero.

\begin{figure*}[t]
\centering
\includegraphics[angle=0,width=1.0\hsize]{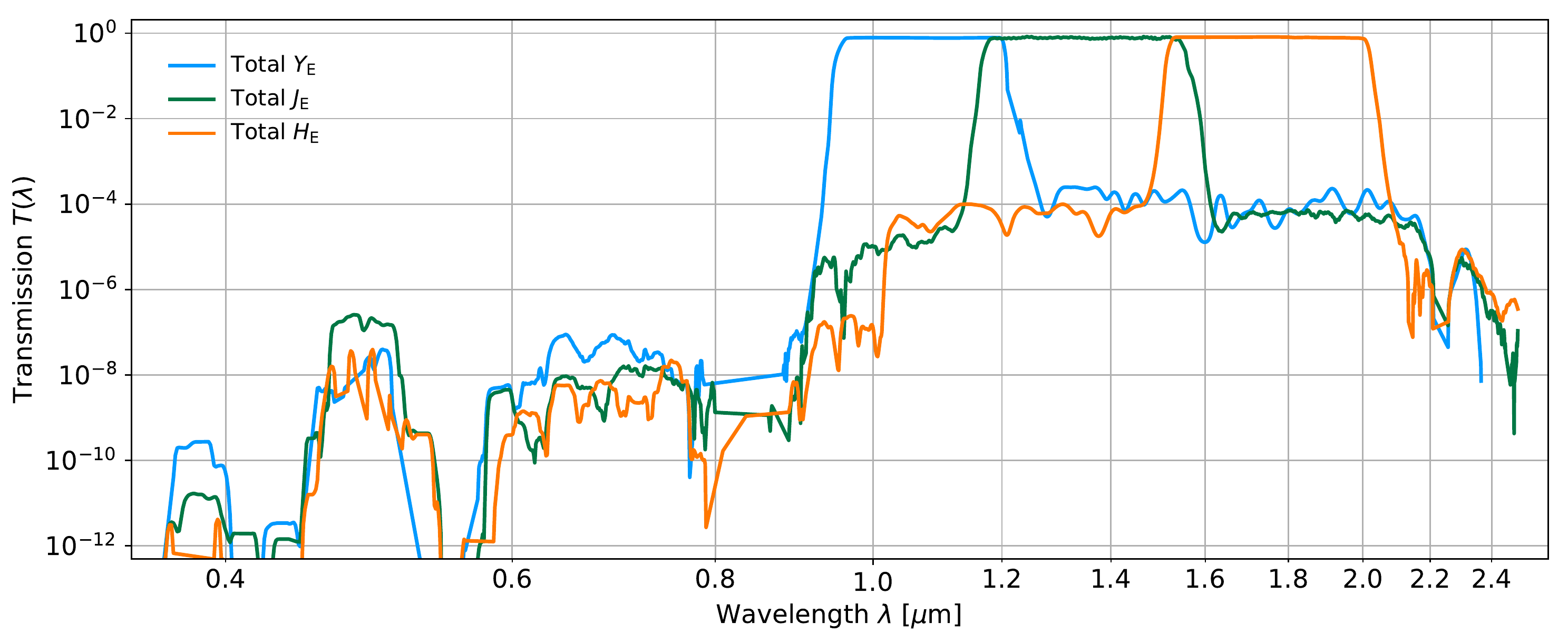}
\caption{Joint NISP out-of-band blocking, including flight model filters, telescope, NI-OA and QE. The curves are the same as the shaded curves in Fig.~\ref{fig:transmission_elements}, with the difference that the blocking was measured at the filter center only, in wavelength steps of 2\,nm. We smoothed the data on a 20\,nm baseline to reduce the noise at the lowest transmission levels. The residual optical transmission from the dichroic and the NI-OA at $0.45$--$0.54$\,\micron\ (Fig.~\ref{fig:transmission_elements}) -- shortward of the VIS bandpass -- is suppressed by a factor $\geq 4\times10^{-7}$. Overall, blocking is excellent across the full wavelength range.}
\label{fig:blocking}
\end{figure*}

We then compute the `in-band flux' by integrating $f_{\nu}\,T(\nu)$  within a passband's 0.1\% cut-on and cut-off wavelengths (Table~\ref{table:passbandproperties}), with $T(\nu)$ as shown in Fig.~\ref{fig:blocking}. The predicted measured flux is obtained by integrating over the frequency range corresponding to the $0.35$--$2.50$\,\micron\ interval.  

Accordingly, the \textit{maximum} relative contribution from out-of-band flux -- outside the 0.1\% cut-on and cut-off -- to the total flux in the \ymagm- and \jmagm-bands is $1.4\times10^{-3}$ and $2.3\times10^{-4}$, respectively, for red SEDs with $\alpha=-3$. For \hband, we find a maximum contribution of $1.9\times10^{-4}$ for the blue SED with $\alpha=1$. Out-of-band contamination of the NISP photometry is therefore at most at the level of 2\,mmag, and more typically $0.2$\,mmag, negligible for virtually all practical purposes. This is not restricted to wavelengths within $0.35$--$2.50$\,\micron, since there is essentially no sensitivity or response outside this range.

\subsection{In-flight changes of the total transmission\label{sec:inflightchanges}}
The transmission of the elements presented in this Section will slowly degrade with time. In practice, Eq.~(\ref{throughput2}) becomes
\begin{equation}
  T_{\rm tot}(\lambda,t_{\rm m}) = T_{\rm Tel}(\lambda) \;\, T_{\rm NI\text -OA}(\lambda) \;\,T_{\rm Filter}(\lambda)\;\,T_{\rm QE}(\lambda)\;\,T_{\rm Evol}(\lambda,t_{\rm m})\;.
  \label{throughput3}
\end{equation}
Here, $T_{\rm Evol}(\lambda,t_{\rm m})$ absorbs the combined evolution of the other four factors with progressing mission time, $t_{\rm m}$. At the end of the mission, six years after launch, we expect an unrecoverable reduction of 5\% in the response, i.e. $T_{\rm Evol}(\lambda)\sim0.95$. Several factors contribute:

During launch, particulates in the spacecraft and in the rocket fairing will be redistributed due to the acoustic and mechanical vibrations. This leads to an increased contamination of \Euclid's primary mirror, and therefore increased optical scattering.

Once in orbit at L2, \Euclid will be subject to space weathering from radiation damage and dust pitting that lower the system transmission. Radiation damage, for example, can darken dielectric coatings, an effect that is minimized by protective coating layers; see e.g.  \cite{sheik2008} for the \textit{Kepler} space telescope, and also \cite{pelizzo2021}.

Proton and electron radiation can degrade the QE of HgCdTe photodiodes such as in the NISP HAWAII-2RG detectors, as damage displacement in the alloy generates recombination centers. The expected total ionization dose at L2 for the NISP detectors at the end of the mission is 2.5\,krad; any reduction in QE, whether chromatic or achromatic, is expected to be minor. We note that QE degradation due to radiation damage is highly dependent on the photodiode architecture, see e.g. \cite{sun2020} and \cite{crouzet2020}, and can thus be very different for instruments other than NISP.

Surface pitting by dust and meteoroids \citep{rodmann2019} as well as by electrons \citep{simonetto2020} increases the micro-roughness of optical surfaces, and thus the wavelength-dependent scattering. The sky-facing primary mirror, in particular, will be subject to surface erosion. The level of dust pitting at L2 is fairly well known \citep{gruen1985,gaia2016}.

Material outgassing in vacuum \citep{green2001,chiggiato2020} will contaminate \Euclid's optical surfaces, mostly with water ice. This alters the total transmission by wavelength-dependent scattering, interference, and absorption. Contrary to radiation damage and surface erosion, this process is reversible by heating the spacecraft, restoring $T_{\rm Evol}(\lambda,t_{\rm m})$ closer to 1. We will discuss contamination effects in a forthcoming paper.

There are also indirect effects. Electron bombardment lowers the efficiency of a spacecraft's multi-layer insulation thermal blankets \citep{engelhart2017}, resulting in increasing internal temperatures, and thus a -- most likely negligible -- redshift of the passbands (Sect.~\ref{sec:tempvacshifts}).

Apart from the latter temperature effect, the cut-on and cut-off wavelengths are unaffected by these processes. Primarily, the system response -- in practice, $T_{\rm Evol}(\lambda,t_{\rm m})$ -- will gradually reduce over time, which we track with a tight in-flight calibration program (Sects.~\ref{sec:inflightcalibrationabs} and \ref{sec:inflightcalibrationrel}).

\section{The NISP photometric system\label{sec:photometricsystem}}
\subsection{Photometric zeropoints\label{sec:photozeropoints}}
In general, the number of photo-electrons created per second in a detector in the frequency interval $[\nu_{\rm min},\nu_{\rm max}]$ is 
\begin{equation}
    n_{\rm e} = A_{\rm eff}\,\int\displaylimits_{\nu_{\rm min}}^{\nu_{\rm max}}
    \frac{f_{\nu}(\nu)}{h \nu}\;T_{\rm tot}(\nu)\;{\rm d}\nu\;,
    \label{eq:numphotoelectrons}
\end{equation}
where $A_{\rm eff}$ is the effective collecting area, $f_\nu(\nu)$ the source's spectral flux density in frequency units, $h=6.602\times10^{-27}$\,erg\,Hz$^{-1}$ Planck's constant, $\nu$ the photon frequency, and $T_{\rm tot}(\nu)$ the total `transmission' -- i.e. the probability that a photon entering the telescope is converted to a photo-electron -- as given in Sects.~\ref{sec:computation} and \ref{sec:totaltransmission}. We assume a perfect detection and extraction chain, i.e.\ a photo-electron created in a pixel is registered by the readout electronics and caught by software processing.

\Euclid's photometric measurements will be given in the AB magnitude system \citep{oke1983}, where the relation between monochromatic -- i.e. per frequency interval -- AB magnitude and spectral flux density is\footnote{We use units of $1\,{\rm Jy} = 10^{-23}$\,erg\,s$^{-1}$\,cm$^{-2}$\,Hz$^{-1}$ for the spectral flux density, to avoid cluttering by units in the logarithms.}
\begin{equation}
m_{\rm AB} = -2.5\; \logten
    \left(\frac{f_\nu(\nu)}{1\,{\rm Jy}}\right)+8.90\;.
    \label{eq:ABdefinition}
\end{equation}
This equation is also valid for a broad-band observation in case of a frequency-flat source spectrum.

We define the photometric zeropoint, ZP, of a \Euclid broad-band observation as the magnitude of a source creating a single photo-electron per second, i.e.\ $n_{\rm e}=1$\,s$^{-1}$. Assuming a frequency-flat source spectrum, $f_{\nu}(\nu) = {\rm const}$, we solve Eq.~(\ref{eq:numphotoelectrons}) for $f_\nu$, replace it in Eq.~(\ref{eq:ABdefinition}), and find the ZP as
\begin{equation}
    {\rm ZP}=m_{\rm AB}=+2.5\,\logten\left(\frac{A_{\rm eff}}{n_{\rm e}\,h}\,
    \left(1\,{\rm Jy}\right)\,\int\displaylimits_{\nu_{\rm min}}^{\nu_{\rm max}}\frac{T_{\rm tot}(\nu)}{\nu} \,{\rm d}\nu
    \right)+8.90\;.
\end{equation}
Using $A_{\rm eff}=9.926\times10^3$\,cm$^{2}$ (L.~Venancio, priv.\ comm.) for \Euclid, 
\begin{equation}
    {\rm ZP}=26.84+2.5\,{\rm log}_{10}\,
    \int\displaylimits_{\nu_{\rm min}}^{\nu_{\rm max}}\frac{T_{\rm tot}(\nu)}{\nu} \,{\rm d}\nu\;.
    \label{eq:ZP}
\end{equation}

Setting $\nu_{\rm min}$ and $\nu_{\rm max}$ to the frequency interval where $T_{\rm tot}(\nu)$ exceeds $0.1\%$ of the mean peak total transmission (see Table~\ref{table:passbandproperties}), we derive the following AB mag photometric zeropoints,
\begin{eqnarray}
    {\rm ZP}_{Y_{\scriptscriptstyle\rm E}} = 25.04\pm0.05\,{\rm mag},\label{eq:ZPY}\\
    {\rm ZP}_{J_{\scriptscriptstyle\rm E}} = 25.26\pm0.05\,{\rm mag},\label{eq:ZPJ}\\
    {\rm ZP}_{H_{\scriptscriptstyle\rm E}} = 25.21\pm0.05\,{\rm mag}.\label{eq:ZPH}
\end{eqnarray}
When computed for the 50\% cut-on and cut-off wavelengths, these ZPs decrease by 0.01\,mag. 

The uncertainty of these ZPs is dominated by the $5\%$ uncertainty of the absolute QE measurement (Sect.~\ref{sec:NISPqe}). For comparison, the absolute accuracy of the spectrophotometers used to measure the mirror reflectances, and the NISP filter and dichroic transmissions, is $0.02$--$0.2$\%. An initial in-flight flux scaling will tie  the NISP photometric system to a WD spectrophotometric standard star with known absolute flux (Sect.~\ref{sec:inflightcalibrationabs}), constraining the ZPs much more accurately. Yet, throughout the mission, the ZPs will slowly evolve as explained in Sect.~\ref{sec:inflightchanges}.

\subsection{Transformations to other NIR systems\label{sec:transformations}}
The NISP photometric system, summarized in Table~\ref{table:passbandproperties}, is geared towards photo-$z$ measurements. It deviates considerably from common ground-based NIR systems, such as 2MASS \citep{carpenter2001}, Mauna Kea Observatories \citep[MKO;][]{leggett2006}, VISTA \citep{gonzalez2018}, and UKIRT \citep{hodkin2009}, the latter being built upon MKO \citep[see also][]{hewett2006}. All of these are primarily dictated by the atmospheric transmission windows (Fig.~\ref{fig:passband_comparison}).

\subsubsection{Linear transformations for stars and galaxies\label{sec:euclid_sc8}}
In Appendices \ref{apx:transformations_galaxies} to \ref{apx:transformations_stellar} we provide linear transformations between these external photometric systems and the NISP photometric system. The transformations are based on the Euclid SC8 catalogs, containing realistic SEDs for 
100\,deg$^2$ of sky as it will be seen by the Euclid Wide Survey \citep{scaramella2021}. We randomly selected 1.8 million stars and 1.1 million redshifted galaxies from the SC8 catalogs, and computed their colors in the NISP, 2MASS, MKO, and VISTA photometric systems. 

The stellar catalog is comprised of Besançon models at the faint end, and the observed stars of \cite{pickels2010} -- their Table 15 in the online edition -- for the bright end (1.8 million in total). Late M-type stars, and L and T dwarfs (hereafter `MLT types') were taken from \cite{barnett2019} (61\,000 in total). 

The galaxy catalog in SC8 is built on the Euclid flagship simulation\footnote{
\texttt{\url{https://www.euclid-ec.org/?page\_id=4133}}}, supported by CosmoHub \citep{tallada2020}. The templates are based on \cite{ilbert2009}, in turn relying on the templates by \cite{bruzual2003} and \cite{polletta2007}. Included are effects from two different internal extinction laws \citep{prevot1984,calzetti2000}. 

We then fit -- for the galaxies and stars separately -- linear relations including a constant offset, of the form
\begin{equation}
Y_{\rm A}=J_{\rm B}+c_0+c_1\,(Y_{\rm B}-J_{\rm B})\,.
\end{equation}
In this example, an instrument A $Y$-band magnitude is estimated from instrument B $Y$- and $J$-band magnitudes. Simple $\chi^2$ minimization yields the $c_0$ and $c_1$ coefficients including a 3\,$\sigma$ outlier rejection, and \textit{we choose the neighboring bands producing the best fit for each case}. 

Higher-order polynomial fits do not significantly reduce the fit residuals. We also find that when a constant offset $c_0$ is included in the fit as we do here, then there is no significant benefit in using more than two bands. For the majority -- but not all -- of the conversions even a two-band linear fit omitting the offset produces results with errors almost as low as the ones listed in Appendices \ref{apx:transformations_stellar} and \ref{apx:transformations_galaxies}.

The fits are performed for the following object classes:

{\bf Galaxies:} Equations~(\ref{eq:nisp_estimates_galaxies_a}) through (\ref{eq:nisp_estimates_galaxies_o}) convert 2MASS, MKO and VISTA NIR magnitudes to NISP magnitudes. Equations~(\ref{eq:ext_estimates_galaxies_a}) through (\ref{eq:ext_estimates_galaxies_o}) convert in the other direction, from NISP magnitudes to 2MASS, MKO and VISTA magnitudes.

{\bf Stars without MLT types:} Likewise, see Eqs.~(\ref{eq:nisp_estimates_stellarMAIN_a}) through (\ref{eq:nisp_estimates_stellarMAIN_o}), and Eqs.~(\ref{eq:ext_estimates_stellarMAIN_a}) through (\ref{eq:ext_estimates_stellarMAIN_o}). 

{\bf MLT types:} Likewise, see Eqs.~(\ref{eq:nisp_estimates_stellarMLT_a}) through (\ref{eq:nisp_estimates_stellarMLT_o}), and Eqs.~(\ref{eq:ext_estimates_stellarMLT_a}) through (\ref{eq:ext_estimates_stellarMLT_o}). 

{\bf Stars of all spectral types:} Likewise, see Eqs.~(\ref{eq:nisp_estimates_stellar_a}) through (\ref{eq:nisp_estimates_stellar_o}), and Eqs.~(\ref{eq:ext_estimates_stellar_a}) through (\ref{eq:ext_estimates_stellar_o}). 

\subsubsection{Limitations of the linear transformations\label{sec:limitations}}
The color range for which the transformations are applicable is given by the x-axes of the
plots in Appendices \ref{apx:transformations_galaxies} to \ref{apx:transformations_stellar}. The plots encompass 99\% of the sources in the SC8 catalogs, and therefore the validity range can be considered universal for these simple parameterizations.

The transformations for galaxies have residuals of $0.04$--$0.06$\,mag, and are therefore usable for many purposes needing a simple translation between the different photometric systems. Notable exceptions are the computed $Z$- and $K_{\rm s}$-band magnitudes, where residuals amount to $0.1$--$0.2$\,mag.

The transformations for stars of spectral types O to K have residuals of $0.02$--$0.04$\,mag. Notable exceptions are the computed $Z$-band magnitudes, with residuals as large as 0.11\,mag.

The transformations for MLT types have residuals of $0.04$--$0.12$\,mag. Notable exceptions are the $Z$-band magnitudes, with residuals up to 0.38\,mag. When working with such red types, the {\tt photometry} package (Sect.~\ref{sec:pythontool}) should be used.

The transformations for stars of all spectral types (O to K, MLT) have residuals of $0.03$--$0.08$\,mag. Again, notable exceptions are the $Z$-band magnitudes with residuals up to 0.25\,mag. 

As a caveat, we emphasize that the SC8 catalogs are representative for the Euclid Wide Survey area at galactic latitude $|b|\gtrsim30^\circ$. The stellar disk population is therefore underrepresented in these catalogs, as are lines of sight of appreciable dust extinction \citep{scaramella2021}. Our transformations must therefore be used with caution when working at low galactic latitudes. While the Euclid Wide Survey does not extend into the galactic plane, observations during e.g. a future mission extension could do so.

We emphasize that all transformations between filter systems given in this paper are computed in the AB mag system. The 2MASS magnitudes in \cite{cutri2003} are given in the Vega system. Before they can be used in any of our transformations, they must be converted to AB magnitudes, using e.g. the conversions by \cite{pons2019}.

\subsubsection{A {\tt Python} transformation tool\label{sec:pythontool}}
The linear transformations provided in this paper should be useful for many general purposes. However, individual objects may deviate by 0.1\,mag or more, in particular for galaxies and  very blue or red sources (see Sect.~\ref{sec:limitations}, and Figs.~\ref{fig:nisp_photo_colors_stars} to \ref{fig:ext_photo_colors_galaxies}). 

For accurate photometry and filter systems other than the ones presented here, we recommend to use the external, stand-alone {\tt Python} 
{\tt photometry}\footnote{\texttt{\url{https://github.com/haussel/photometry}}} package. The NISP response curves of this paper have been integrated in {\tt photometry}, which includes a notebook\footnote{See {\tt notebooks/Euclid\_TU\_checks.ipynb} in the {\tt photometry} source tree.} showing how to recompute magnitudes in other passbands for arbitrary stellar and non-stellar SEDs.  

\begin{figure*}[t]
\centering
\includegraphics[angle=0,width=1.0\hsize]{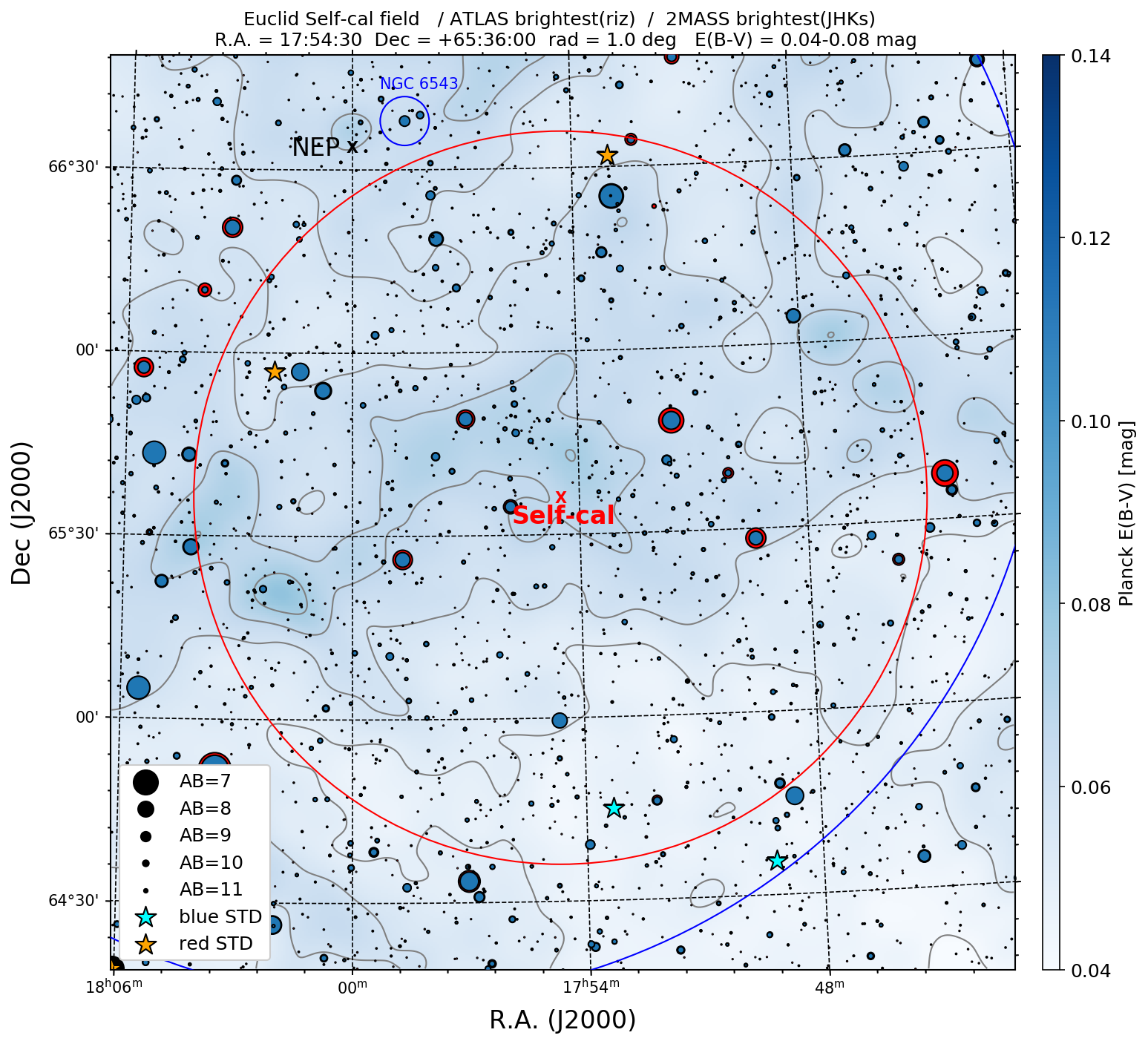}
\caption{Approximate outline of the \Euclid self-calibration field (red circle) near the North Ecliptic Pole (NEP), which will be observed with an irregular dither pattern. The star symbol sizes scale with the brightest AB magnitude in the ATLAS-Refcat2 $riz$ bands \citep[blue;][]{tonry2018} and the 2MASS $JHK_{\rm s}$ bands \citep[red, converted to AB mag;][]{cutri2003}. Cyan and orange stars show hot (blue) WD spectrophotometric standards, and red extragalactic spectrophotometric standards, respectively. The large blue circle indicates the boundary of the Euclid Deep Field North.}
\label{fig:selfcal}
\end{figure*}

\subsection{In-flight lifetime of the NISP photometric system}
\subsubsection{Validity}
At the time of this writing, \Euclid is still one year from launch. The transmission measurements presented here are not expected to change significantly until launch. The space-grade optical coatings are very durable, and all hardware is kept in clean conditions in a controlled environment up to and including launch. Only a small degree of particulate contamination is expected during this time, which is well under control, and whose additional contribution to the flux loss will be much smaller than the uncertainty of 5\% for the absolute detector QE.

Therefore, the cut-on and cut-off wavelengths presented in this paper are not expected to change, neither due to the remaining time on-ground, nor during flight. What will change over time is -- to first order --  the overall response as encoded in the ZPs of  Sect.~\ref{sec:photozeropoints}. This will be tracked using a tight calibration program outlined in Sects.~\ref{sec:inflightcalibrationabs} and \ref{sec:inflightcalibrationrel}. Hence, the AB magnitudes of sources reported in future \Euclid data releases are unaffected by -- i.e. corrected for -- any response losses.

Second order effects arise due to wavelength-dependent changes of the response.
For example, scattering by dust and water ice on optical surfaces is larger at shorter wavelengths, making objects appear redder than they are. In Eq.~(\ref{throughput3}), we have absorbed the time dependence of the response in the additional term 
$T_{\rm Evol}(\lambda,t_{\rm m})$. The \Euclid calibration program (Sects.~\ref{sec:inflightcalibrationabs} and \ref{sec:inflightcalibrationrel}) has been specifically developed to determine and monitor $T_{\rm Evol}(\lambda,t_{\rm m})$ with great accuracy. This will eventually even enable SED-dependent corrections of source fluxes, once our understanding of the system and $T_{\rm Evol}(\lambda,t_{\rm m})$ has matured at later data releases.

Conclusively, we expect that the NISP photometric system presented in this paper will maintain its validity throughout the mission. This should also hold for the linear transformation equations in Sect.~\ref{sec:transformations}, whose uncertainties are dominated by the large diversity of the redshifted source SEDs. In the following two Sections, we summarize the in-flight maintenance.

\subsubsection{Absolute calibration observations\label{sec:inflightcalibrationabs}}
For a number of science applications the absolute photometry of NISP observations must be known to within 5\% of established reference photometric systems. To this end, \Euclid will observe early one of several suitable WD spectrophotometric standard stars on a $5\times5$ grid per detector and per filter, before the beginning of the routine science observations; hot, stable WDs are excellent flux calibrators \citep{bohlin2014,bohlin2020}. \Euclid will also take spectra of a WD on five positions per detector. Furthermore, one of the \Euclid WDs is located in \Euclid's self-calibration field (Sect.~\ref{sec:inflightcalibrationrel}), which is visited regularly on a monthly basis.

Our approved \textit{Hubble} Space Telescope (HST) program (ID \#16702) has begun obtaining NIR spectra of a total of six WDs in \Euclid's survey area. These data will tightly connect the NISP photometric system to the well-known HST photometric system \citep{bohlin2019}. Other spectrophotometric standards that happen to be in the survey area will be used for additional spot checks of the absolute calibration.

\subsubsection{Relative calibration observations\label{sec:inflightcalibrationrel}}
\Euclid's relative photometric accuracy must be better than 1.5\%. To fulfill this requirement -- independently of the WD observations -- \Euclid will observe a self-calibration field every $25$--$35$ days \citep{scaramella2021}. 
Figure~\ref{fig:selfcal} shows the location of the `Self-cal field' near the North Ecliptic Pole (NEP), within \Euclid's continuous viewing zone. Each visit will reach a depth of about $25.0$\,AB\,mag, one magnitude deeper than the Euclid Wide Survey. At each dither point, VIS and NISP images in all bands will be taken, including a NISP spectroscopic exposure. Over time, the Self-cal field will become the deepest Euclid field.

The Self-cal field contains about $40\,000$ unsaturated stars in the  $J=16.5$--$24.0$\,AB\,mag interval, with a signal-to-noise ratio of $S/N\gtrsim10$. \cite{gaia2019} have shown that about 9\% of all Gaia stars are variable down to a precision of $5$--$10$\,mmag. Even if half of all stars detected by \Euclid were variable, the statistical basis for ZP monitoring is excellent, also because variability between stars is uncorrelated. We expect to detect ZP changes down to $1$--$2$\,mmag, per NISP detector, and in two disjoint color bins (SED bins) per filter. In case needed, we can fold in twice as many galaxies to improve the statistics. 

Should the monthly sampling turn out to be insufficient, e.g. in the presence of faster ZP variations due to material outgassing, then we can include photometry from sources appearing in the overlap areas of adjacent survey fields. \Euclid observes about 19 adjacent fields every
24 hours.

The monthly visits of the Self-cal field also provide spectroscopic observations of the WD standard, and of about 2000 field stars with $S/N\gtrsim10$ per spectral resolution element. Using statistical stacking of all spectra, we can detect chromatic changes per spectral resolution element on the order of 1\,mmag, globally for the full focal plane -- although not per detector as with the imaging observations. 

Note that these observations can only weakly distinguish which optical surface is losing transmission, or whether detectors degrade in QE, and due to which effects. Nonetheless, these calibrations put \Euclid in an excellent position to determine the net effect on the data, as encoded in $T_{\rm Evol}(\lambda,t_{\rm m})$, maintaining the NISP photometric system throughout \Euclid's lifetime.

\section{Summary and data products\label{sec:summary}}
\subsection{Context and main results}
\Euclid will observe 15\,000\,deg$^2$ of the darkest extragalactic sky to a median depth of $24.4$\,AB\,mag for $5\,\sigma$ point sources \citep{scaramella2021}, establishing a photometric reference data set with considerable legacy value. Using detailed filter transmission measurements and ray tracing of the optical system, we determine the edges of the spectral response curves for any position in the focal plane with a -- conservatively estimated -- accuracy of 0.8\,nm. The out-of-band blocking of the \ymagm, \jmagm and \hmagm passbands is excellent over the $0.3$--$3.0$\,\micron\ range, with out-of-band contributions of $0.2$\,mmag for typical power-law SEDs. The main passband properties, and the polynomial coefficients to compute the passband edges as a function of field position, are given in Tables~\ref{table:passbandproperties} and \ref{table:pbvar_coeffs}, respectively.

We also derived the photometric zeropoints in the AB mag system. The NISP passbands are about twice as wide as their ground-based counterparts that are constrained by the atmospheric transmission windows. We provide linear transformations for stars and galaxies separately, to convert between the NISP and the ground-based photometric systems. A {\tt Python} module is available for the computation of arbitrary transformations and magnitudes. 

We strongly recommend authors to use the `E' subscript -- as in \yband\footnote{Euclid papers use the following type-setting in LaTeX, e.g. for \ymagm: {\tt \$Y\_\{\textbackslash scriptscriptstyle\textbackslash rm E\}\$}}\;-- when referring to Euclid passbands and magnitudes, to avoid any confusion with the ground-based $YJH$ passbands that have only half the spectral width.

Overall, we designed and built a well-defined photometric system for NISP with great legacy value. Our rigorous in-flight calibration program puts \Euclid in an excellent position to maintain this photometric system throughout the mission. 

\subsection{Published data products and versioning\label{sec:data_products}}
The \ymagm, \jmagm and \hmagm spectral responses at the center of the NISP FPA are available at an ESA
server\footnote{\texttt{\url{https://euclid.esac.esa.int/msp/refdata/nisp/NISP-PHOTO-PASSBANDS-V1}}}. Included are the transmission for the filters, telescope, and NI-OA, as well as the mean detector QE, from which the total spectral response was computed; an excerpt is reproduced in Table~\ref{table:publicdata}. For convenience, the tables are available in both ASCII and FITS \citep{hanisch2001} format, and can be readily displayed with e.g. {\tt TOPCAT} \citep{taylor2005} and processed with {\tt astropy} \citep{robitaille2013,price-whelan2018}. 

We registered an Digital Object Identifier (DOI)\footnote{\texttt{\url{https://doi.org/10.5270/esa-kx8w57c}}} for the electronic response tables (version v1.0) published in this paper. In case updated response curves become available, a new DOI will be registered for them. The landing web page for the old DOI will inform the visitor that a new version is available. Older versions will be retained for reference.

\begin{table}[t]
\caption{Excerpt of the published NISP response and transmission curves (abridged, reduced numeric precision). The columns give the wavelength in nm, total system response, filter transmission, telescope transmission (mirrors and dichroic), transmission of the NI-OA, and detector QE. The response in column 2 is the product of columns 3 to 6.}
\smallskip
\label{table:publicdata}
\smallskip
\begin{tabular}{|l|r|r|r|r|r|}
\hline
  WAVE & T\_TOT & T\_FILT & T\_TEL & T\_NIOA & T\_QE \\
\hline
943 & 0.0276 &  0.0372 &  0.8670 &  0.9383 &  0.9121 \\ 
944 & 0.0442 &  0.0594 &  0.8680 &  0.9393 &  0.9126 \\ 
945 & 0.0619 &  0.0831 &  0.8676 &  0.9403 &  0.9132 \\ 
946 & 0.0834 &  0.1117 &  0.8686 &  0.9412 &  0.9137 \\ 
947 & 0.1244 &  0.1660 &  0.8698 &  0.9422 &  0.9142 \\ 
948 & 0.1913 &  0.2548 &  0.8704 &  0.9431 &  0.9147 \\ 
949 & 0.2969 &  0.3944 &  0.8716 &  0.9440 &  0.9151 \\ 
950 & 0.4310 &  0.5714 &  0.8719 &  0.9449 &  0.9156 \\ 
951 & 0.5617 &  0.7425 &  0.8730 &  0.9457 &  0.9162 \\ 
952 & 0.6626 &  0.8739 &  0.8739 &  0.9465 &  0.9167 \\ 
953 & 0.7178 &  0.9443 &  0.8748 &  0.9473 &  0.9173 \\ 
954 & 0.7362 &  0.9658 &  0.8760 &  0.9481 &  0.9178 \\ 
955 & 0.7411 &  0.9706 &  0.8763 &  0.9488 &  0.9183 \\ 
956 & 0.7425 &  0.9705 &  0.8768 &  0.9496 &  0.9189 \\ 
957 & 0.7438 &  0.9702 &  0.8776 &  0.9503 &  0.9194 \\ 
\hline
\end{tabular}
\end{table}

\begin{acknowledgements}
\AckEC\\
The first group of authors (up to and including M.~Weiler) worked directly on this paper. The other authors made substantial, multi-year contributions to the \Euclid\ project that enabled this paper in the first place.\\
The authors at MPIA acknowledge funding by the German Space Agency DLR under grant numbers 50~OR~1202 and 50~QE~2003. The work by J.M.~Carrasco and M.~Weiler was (partially) funded by the Spanish MICIN/AEI/10.13039/501100011033 and by ``ERDF A way of making Europe'' by the European Union through grant RTI2018-095076-B-C21, and the Institute of Cosmos Sciences University of Barcelona (ICCUB, Unidad de Excelencia ’Mar\'{\i}a de Maeztu’) through grant CEX2019-000918-M.\\
The authors thank Thomas Weber (OBJ; now Materion Balzers Optics) for the technical support with the transmission measurements of the NISP filter substrates, and the anonymous referee for their useful comments.\\
The plots in this publication were prepared with {\tt TOPCAT} \citep{taylor2005} and {\tt Matplotlib} \citep{hunter20007}.
\end{acknowledgements}
\bibliography{manuscript}
\begin{appendix}
\onecolumn
\section{Passband variations\label{apdx:pbshifts}}
\begin{table}[h]
\caption{Coefficients for Eq.~(\ref{eq:pbvar_polynomials}) to compute the blueshifted cut-on and cut-off wavelengths $\lambda_{50}^{\rm on/off}(z,y)$ of the filter flanks in the focal plane's {\tt R\_mosaic} coordinate system (Sect.~\ref{sec:passband_variations}). We use the e-notation to simplify a software implementation. The last column, `Test', can be used to verify computer code for $z=30$, $y=50$. The corresponding blueshift of the passbands is shown in Fig.~\ref{fig:pbvar_YJH}.}
\smallskip
\label{table:pbvar_coeffs}
\smallskip
\small
\begin{tabular}{|l|r|rrr|rrr||r|}
\hline
Flank & $a_0$ & $b_1$ & $b_2$ & $b_3$ & $c_1$ & $c_2$ & $c_3$ & Test\\
\hline
\ymagm cut-on  &  949.58 &  4.21895e$-$04 & $-$1.86149e$-$04 &  8.59708e-10 &  3.83384e$-$03 & $-$1.83650e$-$04 & $-$4.81958e$-$08 &  949.15 \\
\ymagm cut-off & 1212.22 & $-$1.14852e$-$03 & $-$2.32133e$-$04 & $-$6.27018e$-$09 &  4.13559e$-$03 & $-$2.28732e$-$04 & $-$5.85140e$-$08 & 1211.60 \\
\jmagm cut-on  & 1167.61 & $-$2.53327e$-$03 & $-$2.36385e$-$04 & $-$1.71028e$-$08 &  4.90696e$-$03 & $-$2.37536e$-$04 & $-$5.19171e$-$08 & 1166.97 \\
\jmagm cut-off & 1566.94 & $-$1.57482e$-$03 & $-$3.12973e$-$04 & $-$8.43209e$-$09 &  5.60578e$-$03 & $-$3.07454e$-$04 & $-$8.81555e$-$08 & 1566.11 \\
\hmagm cut-on  & 1521.51 &  6.95575e$-$05 & $-$3.00405e$-$04 & $-$4.34174e$-$09 &  8.25013e$-$03 & $-$2.83213e$-$04 & $-$7.85756e$-$08 & 1520.94 \\
\hmagm cut-off & 2021.30 & $-$8.28076e$-$05 & $-$3.91185e$-$04 & $-$7.98574e$-$09 &  1.14732e$-$02 & $-$3.67085e$-$04 & $-$1.02239e$-$07 & 2020.59 \\
\hline
\end{tabular}
\end{table}

\begin{figure}[h]
\centering
\includegraphics[angle=0,width=1.0\hsize]{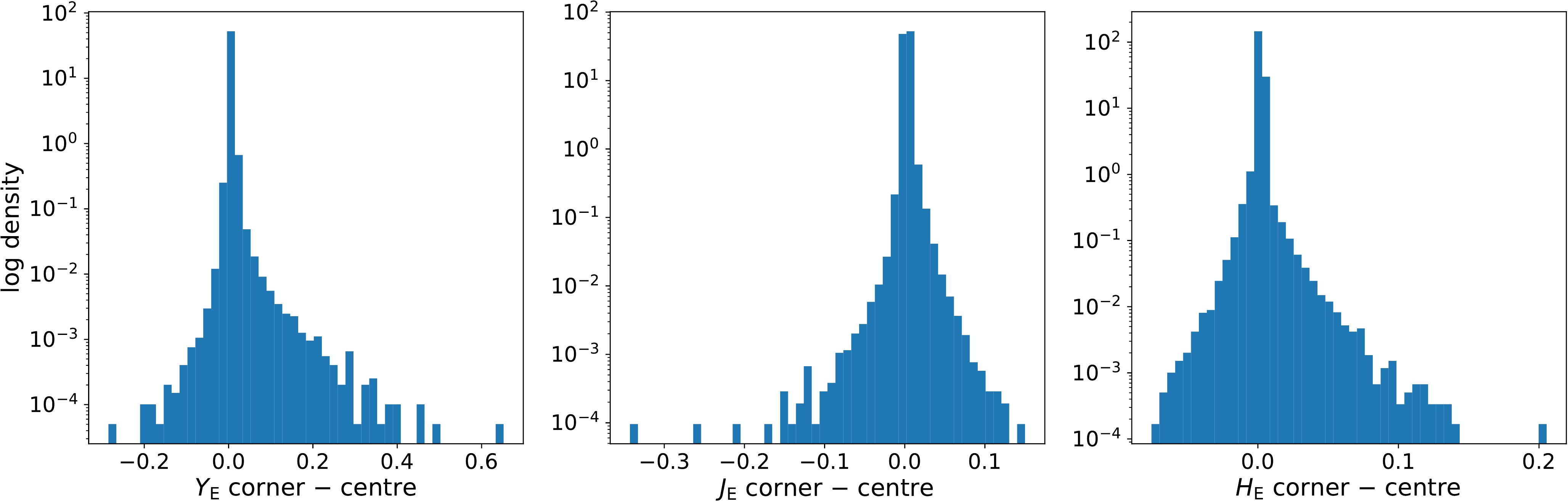}
\caption{Maximum magnitude difference {\bf for galaxies} in the Euclid SC8 catalog due to passband variations, when a source moves from the center of the FPA to the corner. Note the log-scaling of the y-axis, the effect is typically on the order of a few milli-mag (see Table~\ref{table:pbvar_photometry_residuals}).}
\label{fig:photometry_difference_galaxies}
\end{figure}

\begin{figure}[h]
\centering
\includegraphics[angle=0,width=1.0\hsize]{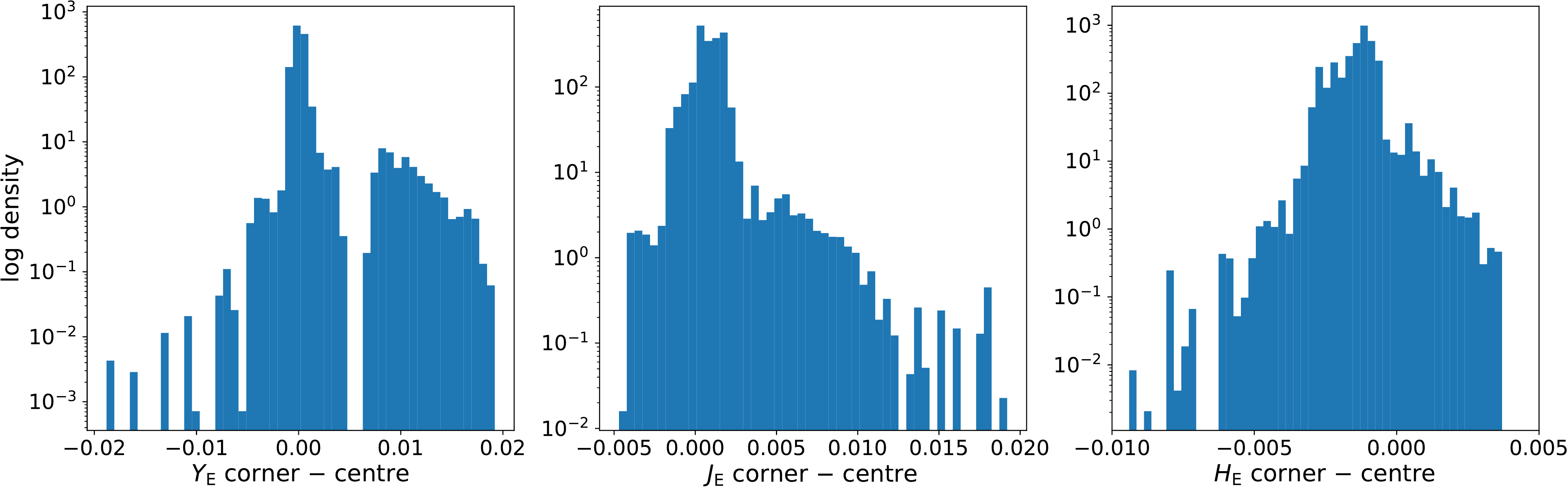}
\caption{Same as Fig.~\ref{fig:photometry_difference_galaxies}, but {\bf for stars}.}
\label{fig:photometry_difference_stars}
\end{figure}

\begin{table}[h]
\caption{Mean and RMS (in mag) of the distributions shown in Figs.~\ref{fig:photometry_difference_galaxies} and \ref{fig:photometry_difference_stars}.}
\smallskip
\label{table:pbvar_photometry_residuals}
\smallskip
\begin{tabular}{|l|rr|rr|}
\hline
& \multicolumn{2}{c|}{{\bf Galaxies}} & \multicolumn{2}{c|}{{\bf Stars}} \\
\hline
Filter & Mean & RMS & Mean & RMS\\
\hline
\ymagm & 0.0047 & 0.0052 & 0.0004 & 0.0020 \\
\jmagm & 0.0027 & 0.0028 & 0.0009 & 0.0013 \\
\hmagm & 0.0019 & 0.0024 & -0.00142 & 0.00074 \\
\hline
\end{tabular}
\end{table}

\newpage

\section{Transformation from and to the NISP photometric system for galaxies\label{apx:transformations_galaxies}}
Equations~(\ref{eq:nisp_estimates_galaxies_a}) through (\ref{eq:nisp_estimates_galaxies_o}) convert NISP AB magnitudes to 2MASS, MKO and VISTA AB magnitudes {\bf for galaxies}. Before using these transformations, read Sect.~\ref{sec:limitations}. 
See also Fig.~\ref{fig:nisp_photo_colors_galaxies}.
\begin{alignat}{4}
J_{\scriptscriptstyle\rm 2MASS} &= J_{\scriptscriptstyle\rm E} && +0.005 +0.314 \;(Y_{\scriptscriptstyle\rm E}-J_{\scriptscriptstyle\rm E}) && \quad(\sigma = 0.037)\label{eq:nisp_estimates_galaxies_a}\\
H_{\scriptscriptstyle\rm 2MASS} &= H_{\scriptscriptstyle\rm E} && -0.004 +0.243 \;(J_{\scriptscriptstyle\rm E}-H_{\scriptscriptstyle\rm E}) && \quad(\sigma = 0.047)\\
K_{\rm s,\scriptscriptstyle\rm 2MASS} &= H_{\scriptscriptstyle\rm E} && -0.001 -0.590 \;(J_{\scriptscriptstyle\rm E}-H_{\scriptscriptstyle\rm E}) && \quad(\sigma = 0.101)\\[2mm]
Z_{\scriptscriptstyle\rm MKO} &= J_{\scriptscriptstyle\rm E} && +0.070 +2.072 \;(Y_{\scriptscriptstyle\rm E}-J_{\scriptscriptstyle\rm E}) && \quad(\sigma = 0.204)\\
Y_{\scriptscriptstyle\rm MKO} &= J_{\scriptscriptstyle\rm E} && -0.002 +1.234 \;(Y_{\scriptscriptstyle\rm E}-J_{\scriptscriptstyle\rm E}) && \quad(\sigma = 0.068)\\
J_{\scriptscriptstyle\rm MKO} &= J_{\scriptscriptstyle\rm E} && +0.009 +0.286 \;(Y_{\scriptscriptstyle\rm E}-J_{\scriptscriptstyle\rm E}) && \quad(\sigma = 0.053)\\
H_{\scriptscriptstyle\rm MKO} &= H_{\scriptscriptstyle\rm E} && -0.005 +0.264 \;(J_{\scriptscriptstyle\rm E}-H_{\scriptscriptstyle\rm E}) && \quad(\sigma = 0.044)\\
K_{\scriptscriptstyle\rm MKO} &= H_{\scriptscriptstyle\rm E} && +0.004 -0.635 \;(J_{\scriptscriptstyle\rm E}-H_{\scriptscriptstyle\rm E}) && \quad(\sigma = 0.100)\\[2mm]
Z_{\scriptscriptstyle\rm VISTA} &= J_{\scriptscriptstyle\rm E} && +0.055 +2.224 \;(Y_{\scriptscriptstyle\rm E}-J_{\scriptscriptstyle\rm E}) && \quad(\sigma = 0.225)\\
Y_{\scriptscriptstyle\rm VISTA} &= J_{\scriptscriptstyle\rm E} && +0.000 +1.282 \;(Y_{\scriptscriptstyle\rm E}-J_{\scriptscriptstyle\rm E}) && \quad(\sigma = 0.073)\\
J_{\scriptscriptstyle\rm VISTA} &= J_{\scriptscriptstyle\rm E} && +0.008 +0.282 \;(Y_{\scriptscriptstyle\rm E}-J_{\scriptscriptstyle\rm E}) && \quad(\sigma = 0.051)\\
H_{\scriptscriptstyle\rm VISTA} &= H_{\scriptscriptstyle\rm E} && -0.004 +0.244 \;(J_{\scriptscriptstyle\rm E}-H_{\scriptscriptstyle\rm E}) && \quad(\sigma = 0.044)\\
K_{\rm s,\scriptscriptstyle\rm VISTA} &= H_{\scriptscriptstyle\rm E} && -0.001 -0.569 \;(J_{\scriptscriptstyle\rm E}-H_{\scriptscriptstyle\rm E}) && \quad(\sigma = 0.098)\label{eq:nisp_estimates_galaxies_o}
\end{alignat}
\\Equations~(\ref{eq:ext_estimates_galaxies_a}) through (\ref{eq:ext_estimates_galaxies_o}) convert 2MASS, MKO and VISTA AB magnitudes to NISP AB magnitudes {\bf for galaxies}. Before using these transformations, read Sect.~\ref{sec:limitations}.
See also Fig.~\ref{fig:ext_photo_colors_galaxies}.
\begin{alignat}{4}
Y_{\scriptscriptstyle\rm E} &= H_{\scriptscriptstyle\rm 2MASS} && +0.052 +1.479 \;(J_{\scriptscriptstyle\rm 2MASS}-H_{\scriptscriptstyle\rm 2MASS}) && \quad(\sigma = 0.105)\label{eq:ext_estimates_galaxies_a}\\
Y_{\scriptscriptstyle\rm E} &= J_{\scriptscriptstyle\rm MKO} && +0.014 +0.703 \;(Y_{\scriptscriptstyle\rm MKO}-J_{\scriptscriptstyle\rm MKO}) && \quad(\sigma = 0.047)\\
Y_{\scriptscriptstyle\rm E} &= J_{\scriptscriptstyle\rm VISTA} && +0.015 +0.659 \;(Y_{\scriptscriptstyle\rm VISTA}-J_{\scriptscriptstyle\rm VISTA}) && \quad(\sigma = 0.049)\\[2mm]
J_{\scriptscriptstyle\rm E} &= H_{\scriptscriptstyle\rm 2MASS} && -0.005 +0.676 \;(J_{\scriptscriptstyle\rm 2MASS}-H_{\scriptscriptstyle\rm 2MASS}) && \quad(\sigma = 0.037)\\
J_{\scriptscriptstyle\rm E} &= H_{\scriptscriptstyle\rm MKO} && +0.005 +0.652 \;(J_{\scriptscriptstyle\rm MKO}-H_{\scriptscriptstyle\rm MKO}) && \quad(\sigma = 0.042)\\
J_{\scriptscriptstyle\rm E} &= H_{\scriptscriptstyle\rm VISTA} && +0.004 +0.665 \;(J_{\scriptscriptstyle\rm VISTA}-H_{\scriptscriptstyle\rm VISTA}) && \quad(\sigma = 0.042)\\[2mm]
H_{\scriptscriptstyle\rm E} &= H_{\scriptscriptstyle\rm 2MASS} && +0.000 -0.192 \;(J_{\scriptscriptstyle\rm 2MASS}-H_{\scriptscriptstyle\rm 2MASS}) && \quad(\sigma = 0.057)\\
H_{\scriptscriptstyle\rm E} &= H_{\scriptscriptstyle\rm MKO} && -0.002 -0.207 \;(J_{\scriptscriptstyle\rm MKO}-H_{\scriptscriptstyle\rm MKO}) && \quad(\sigma = 0.055)\\
H_{\scriptscriptstyle\rm E} &= H_{\scriptscriptstyle\rm VISTA} && -0.002 -0.192 \;(J_{\scriptscriptstyle\rm VISTA}-H_{\scriptscriptstyle\rm VISTA}) && \quad(\sigma = 0.054)\label{eq:ext_estimates_galaxies_o}
\end{alignat}

\begin{figure}[t]
\centering
\includegraphics[angle=0,width=0.8\hsize]{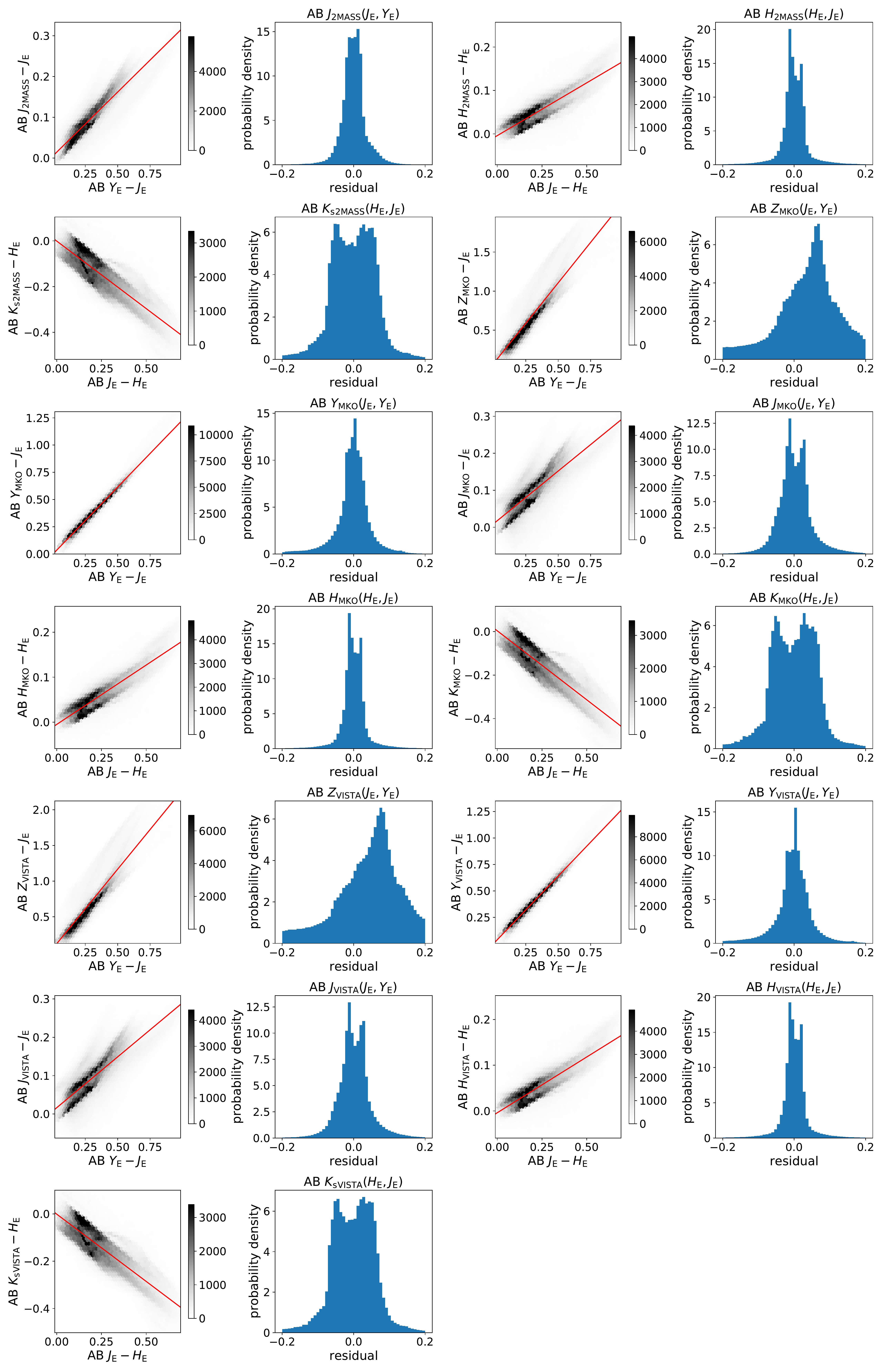}
\caption{Magnitude differences {\bf for galaxies} between external systems (2MASS, MKO, VISTA) and NISP, as a function of NISP color. The red lines show the linear fits given in Eqs.~(\ref{eq:nisp_estimates_galaxies_a}) through (\ref{eq:nisp_estimates_galaxies_o}).}
\label{fig:nisp_photo_colors_galaxies}
\end{figure}

\begin{figure}[t]
\centering
\includegraphics[angle=0,width=1.0\hsize]{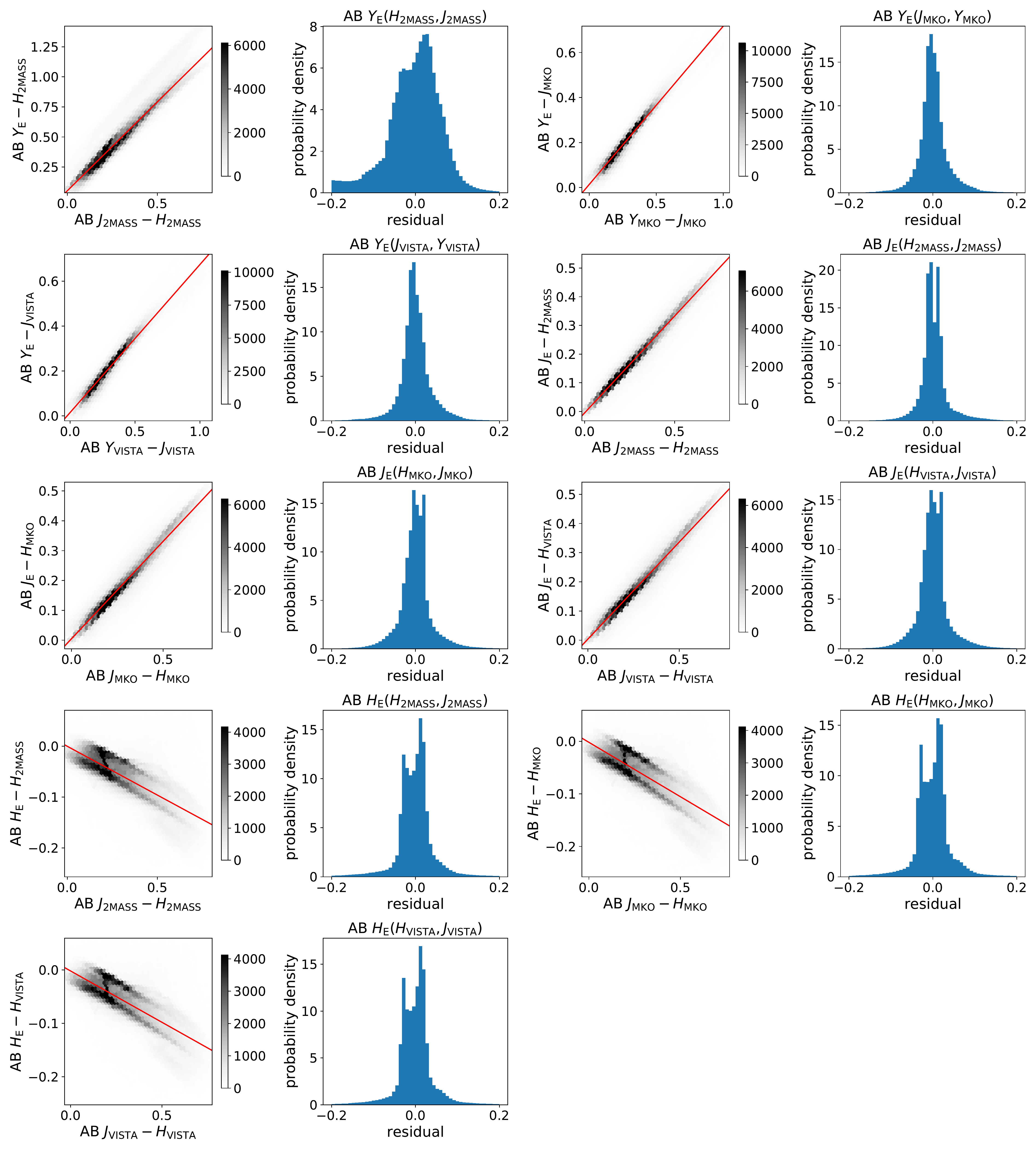}
\caption{Magnitude differences {\bf for galaxies} between NISP and external systems (2MASS, MKO, VISTA), as a function of color in an external system. The red lines show the linear fits given in Eqs.~(\ref{eq:ext_estimates_galaxies_a}) through (\ref{eq:ext_estimates_galaxies_o}).}
\label{fig:ext_photo_colors_galaxies}
\end{figure}

\subsection{About the substructure in Figs.~\ref{fig:nisp_photo_colors_galaxies} and \ref{fig:ext_photo_colors_galaxies}}
Some of the color-color plots in Figs.~\ref{fig:nisp_photo_colors_galaxies} and \ref{fig:ext_photo_colors_galaxies} show substructure that appear artificial. Remember that these are not real observations, but the SC8 synthetic SED templates, redshifted between $z=0.0$ and $2.3$. Some of the redshifted templates distribute fairly widely over the color-color space, whereas others -- mostly star-forming galaxies -- remain more concentrated. The latter then visually dominate the appearance of the linearly scaled color-color density maps. We choose a linear scaling so that the causal connection between the data and the linear fits (red lines) becomes evident. A logarithmic scaling is misleading in most cases, insinuating a poor fit quality.

\begin{figure}[t]
\centering
\includegraphics[angle=0,width=1.0\hsize]{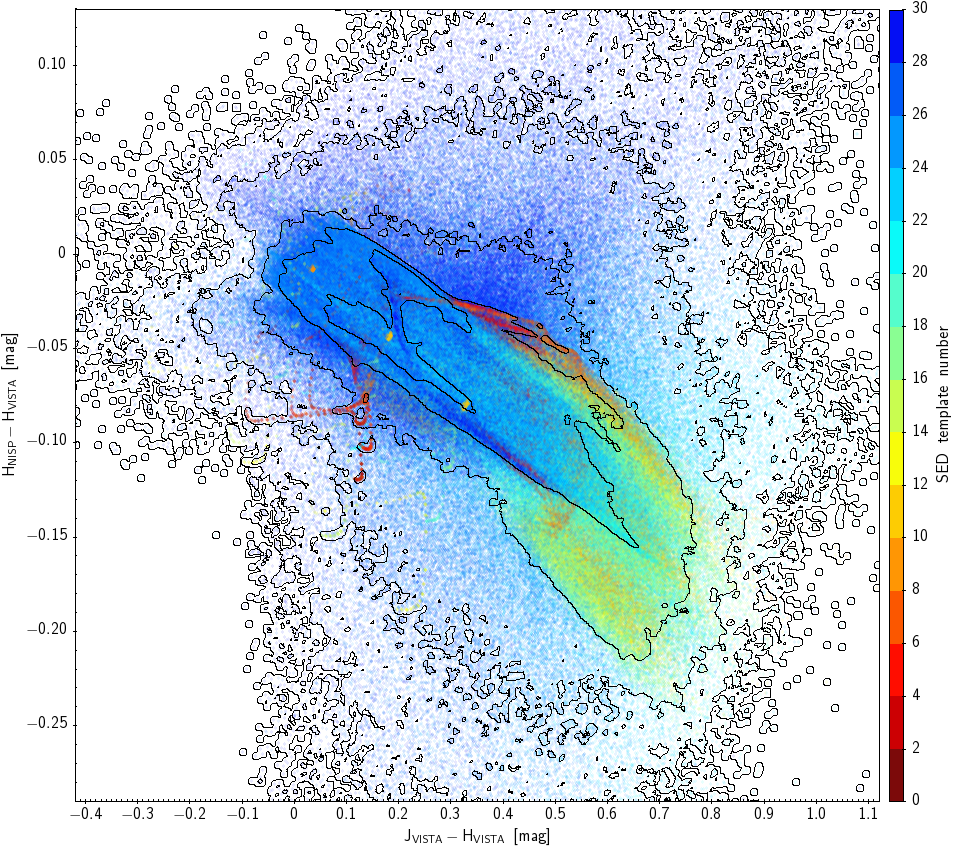}
\caption{Zoom into the lower left panel of Fig.~\ref{fig:ext_photo_colors_galaxies}, showing more data and details about how the redshifted \Euclid SC8 galaxy SEDs populate this particular color-color space. The contours trace the point density with a logarithmic spacing. Template numbers $0$--$7$ correspond to elliptical and lenticular types, templates $8$--$11$ to Hubble types Sa and Sb, templates $12$--$18$ to types Sc and Sd, and templates $18$-$30$ to starbursts of various age and metallicity.}
\label{fig:debug}
\end{figure}

As an example for substructure, consider the bottom two rows of Fig.~\ref{fig:ext_photo_colors_galaxies}, showing the difference in the NISP \hband and VISTA $H$-band magnitudes on the y-axis. Two parallel main branches are seen, mostly from star-forming galaxies. This is a consequence of the NISP \hband being nearly twice as wide as the VISTA $H$-band (see Fig.~\ref{fig:passband_comparison}). The two bands respond differently to spectra with specific continuum shapes and emission lines, also depending on redshift. The small bridge connecting the two branches represents starbursts with a specific metallicity and extinction evolving from redshift $z=0.25$ to $0.55$. The true complexity of this specific color-color space is shown in Fig.~\ref{fig:debug}. This also illustrates that the linear transformation equations may fail catastrophically for individual galaxies. 

\clearpage

\section{Transformation from and to the NISP photometric system for stars (spectral types O to K)\label{apx:transformations_stellarMAIN}}
\begin{figure}[t]
\centering
\includegraphics[angle=0,width=0.8\hsize]{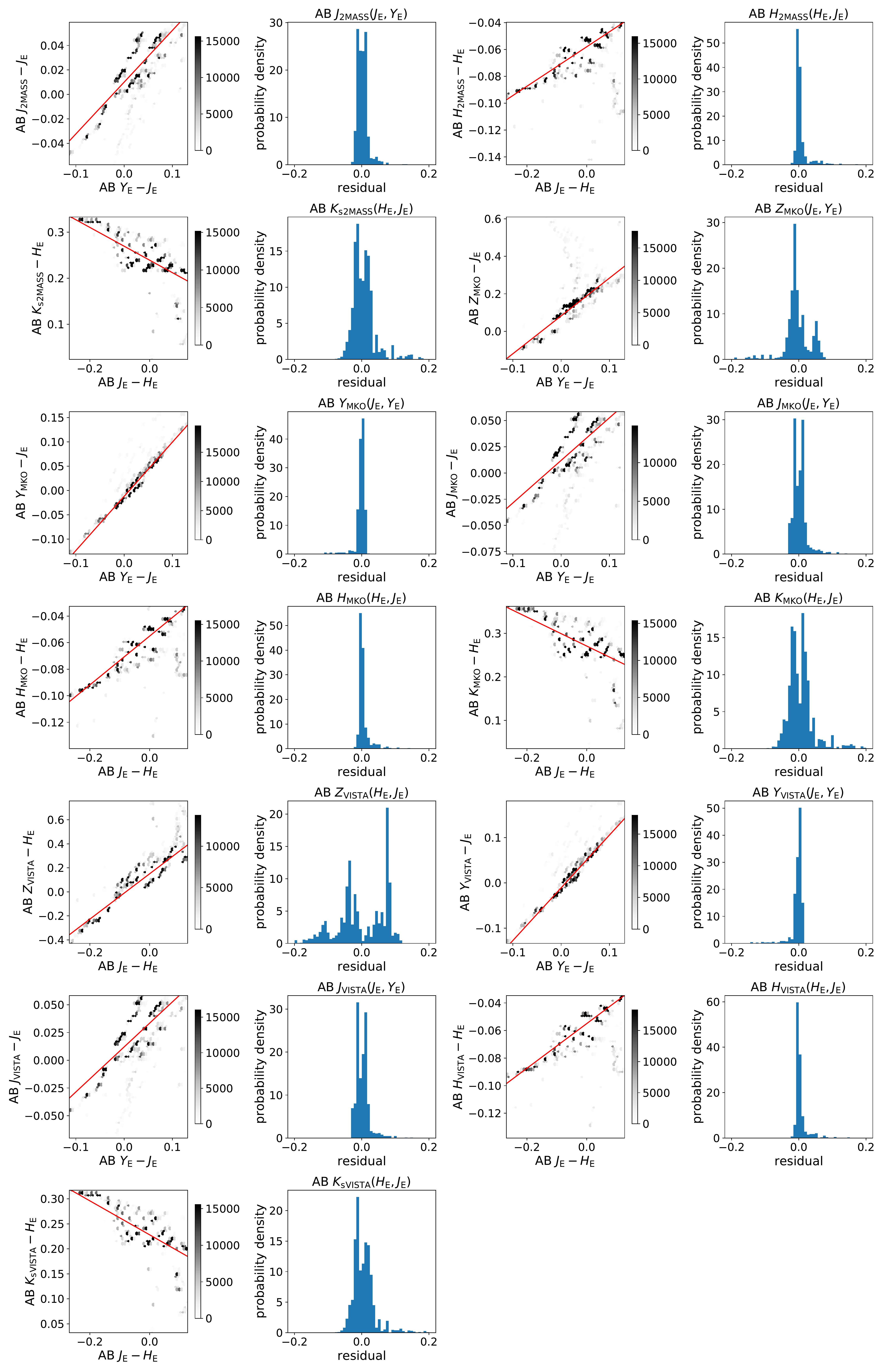}
\caption{Magnitude differences {\bf for stars of spectral types O to K} between external systems (2MASS, MKO, VISTA) and NISP, as a function of NISP color. Data are taken from the Euclid SC8 simulation pilot catalogs. The red lines show the linear fits given in Eqs.~(\ref{eq:nisp_estimates_stellarMAIN_a}) through (\ref{eq:nisp_estimates_stellarMAIN_o}).}
\label{fig:nisp_photo_colors_starsMAIN}
\end{figure}

\begin{figure}[t]
\centering
\includegraphics[angle=0,width=1.0\hsize]{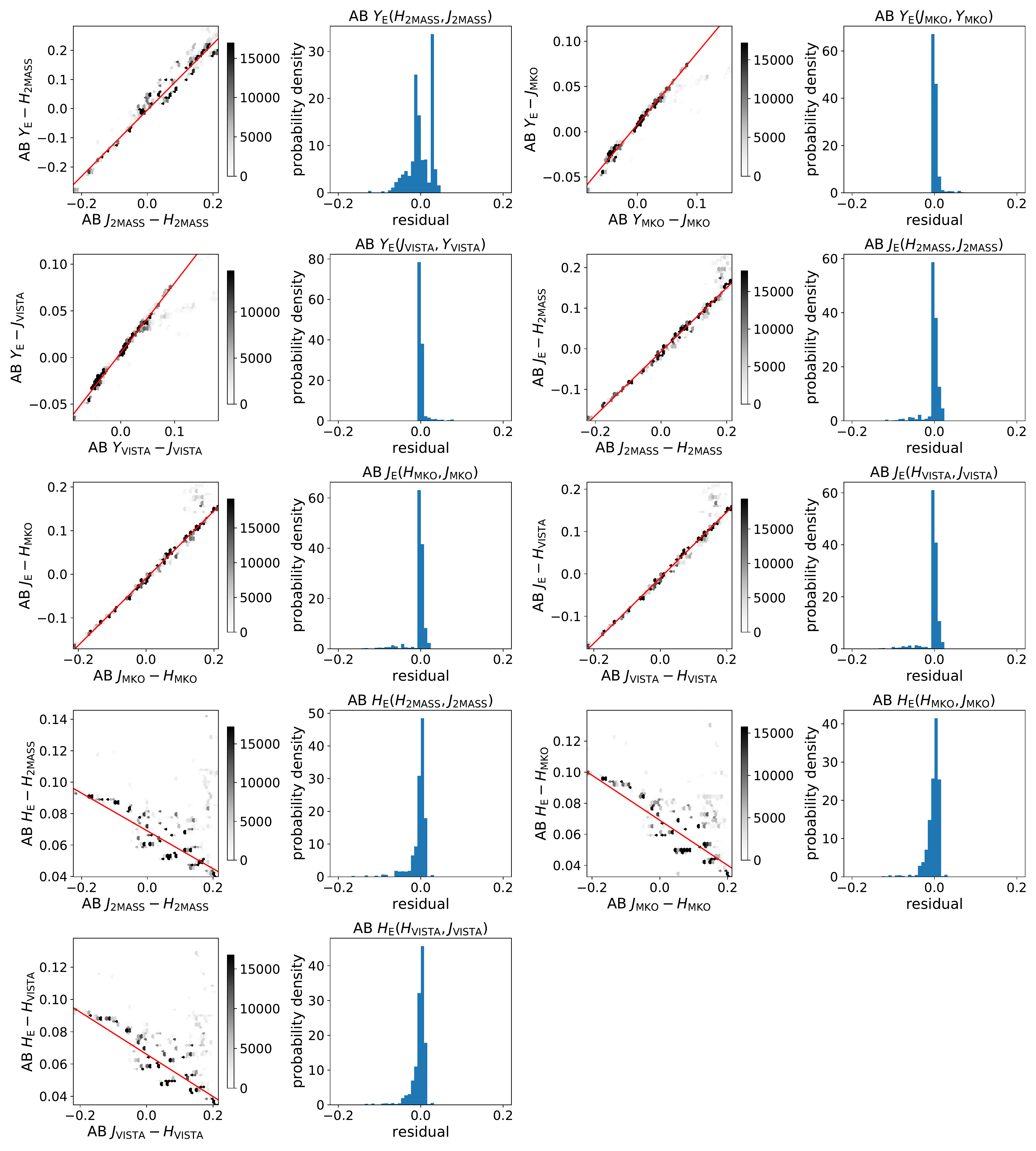}
\caption{Magnitude differences {\bf for stars of spectral types O to K} between NISP and external systems (2MASS, MKO, VISTA), as a function of color in an external system. The red lines show the linear fits given in Eqs.~(\ref{eq:ext_estimates_stellarMAIN_a}) through (\ref{eq:ext_estimates_stellarMAIN_o}).}
\label{fig:ext_photo_colors_starsMAIN}
\end{figure}

Equations~(\ref{eq:nisp_estimates_stellarMAIN_a}) through (\ref{eq:nisp_estimates_stellarMAIN_o}) convert NISP AB magnitudes to 2MASS, MKO and VISTA AB magnitudes {\bf for stars of spectral types O to K}. The $1\,\sigma$ residuals are given in parentheses. Before using these transformations, read Sect.~\ref{sec:limitations}. See also Fig.~\ref{fig:nisp_photo_colors_starsMAIN}.
\begin{alignat}{4}
J_{\scriptscriptstyle\rm 2MASS} &= J_{\scriptscriptstyle\rm E} && +0.010 +0.423 \;(Y_{\scriptscriptstyle\rm E}-J_{\scriptscriptstyle\rm E}) && \quad(\sigma = 0.016)\label{eq:nisp_estimates_stellarMAIN_a}\\
H_{\scriptscriptstyle\rm 2MASS} &= H_{\scriptscriptstyle\rm E} && -0.058 +0.147 \;(J_{\scriptscriptstyle\rm E}-H_{\scriptscriptstyle\rm E}) && \quad(\sigma = 0.021)\\
K_{\rm s,\scriptscriptstyle\rm 2MASS} &= H_{\scriptscriptstyle\rm E} && +0.239 -0.356 \;(J_{\scriptscriptstyle\rm E}-H_{\scriptscriptstyle\rm E}) && \quad(\sigma = 0.041)\\[2mm]
Z_{\scriptscriptstyle\rm MKO} &= J_{\scriptscriptstyle\rm E} && +0.082 +2.018 \;(Y_{\scriptscriptstyle\rm E}-J_{\scriptscriptstyle\rm E}) && \quad(\sigma = 0.107)\\
Y_{\scriptscriptstyle\rm MKO} &= J_{\scriptscriptstyle\rm E} && -0.012 +1.115 \;(Y_{\scriptscriptstyle\rm E}-J_{\scriptscriptstyle\rm E}) && \quad(\sigma = 0.020)\\
J_{\scriptscriptstyle\rm MKO} &= J_{\scriptscriptstyle\rm E} && +0.012 +0.406 \;(Y_{\scriptscriptstyle\rm E}-J_{\scriptscriptstyle\rm E}) && \quad(\sigma = 0.021)\\
H_{\scriptscriptstyle\rm MKO} &= H_{\scriptscriptstyle\rm E} && -0.055 +0.184 \;(J_{\scriptscriptstyle\rm E}-H_{\scriptscriptstyle\rm E}) && \quad(\sigma = 0.018)\\
K_{\scriptscriptstyle\rm MKO} &= H_{\scriptscriptstyle\rm E} && +0.271 -0.333 \;(J_{\scriptscriptstyle\rm E}-H_{\scriptscriptstyle\rm E}) && \quad(\sigma = 0.045)\\[2mm]
Z_{\scriptscriptstyle\rm VISTA} &= H_{\scriptscriptstyle\rm E} && +0.151 +1.891 \;(J_{\scriptscriptstyle\rm E}-H_{\scriptscriptstyle\rm E}) && \quad(\sigma = 0.113)\\
Y_{\scriptscriptstyle\rm VISTA} &= J_{\scriptscriptstyle\rm E} && -0.011 +1.165 \;(Y_{\scriptscriptstyle\rm E}-J_{\scriptscriptstyle\rm E}) && \quad(\sigma = 0.025)\\
J_{\scriptscriptstyle\rm VISTA} &= J_{\scriptscriptstyle\rm E} && +0.012 +0.406 \;(Y_{\scriptscriptstyle\rm E}-J_{\scriptscriptstyle\rm E}) && \quad(\sigma = 0.020)\\
H_{\scriptscriptstyle\rm VISTA} &= H_{\scriptscriptstyle\rm E} && -0.055 +0.162 \;(J_{\scriptscriptstyle\rm E}-H_{\scriptscriptstyle\rm E}) && \quad(\sigma = 0.018)\\
K_{\rm s,\scriptscriptstyle\rm VISTA} &= H_{\scriptscriptstyle\rm E} && +0.228 -0.340 \;(J_{\scriptscriptstyle\rm E}-H_{\scriptscriptstyle\rm E}) && \quad(\sigma = 0.036)\label{eq:nisp_estimates_stellarMAIN_o}
\end{alignat}
\\Equations~(\ref{eq:ext_estimates_stellarMAIN_a}) through (\ref{eq:ext_estimates_stellarMAIN_o}) convert from 2MASS, MKO and VISTA AB magnitudes to NISP AB magnitudes {\bf for stars of spectral types O to K}. Before using these transformations, read Sect.~\ref{sec:limitations}. 
See also Fig.~\ref{fig:ext_photo_colors_starsMAIN}.
\begin{alignat}{4}
Y_{\scriptscriptstyle\rm E} &= H_{\scriptscriptstyle\rm 2MASS} && -0.005 +1.134 \;(J_{\scriptscriptstyle\rm 2MASS}-H_{\scriptscriptstyle\rm 2MASS}) && \quad(\sigma = 0.029)\label{eq:ext_estimates_stellarMAIN_a}\\
Y_{\scriptscriptstyle\rm E} &= J_{\scriptscriptstyle\rm MKO} && +0.008 +0.789 \;(Y_{\scriptscriptstyle\rm MKO}-J_{\scriptscriptstyle\rm MKO}) && \quad(\sigma = 0.013)\\
Y_{\scriptscriptstyle\rm E} &= J_{\scriptscriptstyle\rm VISTA} && +0.005 +0.748 \;(Y_{\scriptscriptstyle\rm VISTA}-J_{\scriptscriptstyle\rm VISTA}) && \quad(\sigma = 0.015)\\[2mm]
J_{\scriptscriptstyle\rm E} &= H_{\scriptscriptstyle\rm 2MASS} && -0.007 +0.786 \;(J_{\scriptscriptstyle\rm 2MASS}-H_{\scriptscriptstyle\rm 2MASS}) && \quad(\sigma = 0.020)\\
J_{\scriptscriptstyle\rm E} &= H_{\scriptscriptstyle\rm MKO} && -0.009 +0.771 \;(J_{\scriptscriptstyle\rm MKO}-H_{\scriptscriptstyle\rm MKO}) && \quad(\sigma = 0.023)\\
J_{\scriptscriptstyle\rm E} &= H_{\scriptscriptstyle\rm VISTA} && -0.009 +0.775 \;(J_{\scriptscriptstyle\rm VISTA}-H_{\scriptscriptstyle\rm VISTA}) && \quad(\sigma = 0.022)\\[2mm]
H_{\scriptscriptstyle\rm E} &= H_{\scriptscriptstyle\rm 2MASS} && +0.069 -0.120 \;(J_{\scriptscriptstyle\rm 2MASS}-H_{\scriptscriptstyle\rm 2MASS}) && \quad(\sigma = 0.021)\\
H_{\scriptscriptstyle\rm E} &= H_{\scriptscriptstyle\rm MKO} && +0.069 -0.145 \;(J_{\scriptscriptstyle\rm MKO}-H_{\scriptscriptstyle\rm MKO}) && \quad(\sigma = 0.017)\\
H_{\scriptscriptstyle\rm E} &= H_{\scriptscriptstyle\rm VISTA} && +0.066 -0.131 \;(J_{\scriptscriptstyle\rm VISTA}-H_{\scriptscriptstyle\rm VISTA}) && \quad(\sigma = 0.018)\label{eq:ext_estimates_stellarMAIN_o}
\end{alignat}

\clearpage

\clearpage

\section{Transformation from and to the NISP photometric system for stars (MLT types)\label{apx:transformations_stellarMLT}}
\begin{figure}[t]
\centering
\includegraphics[angle=0,width=0.8\hsize]{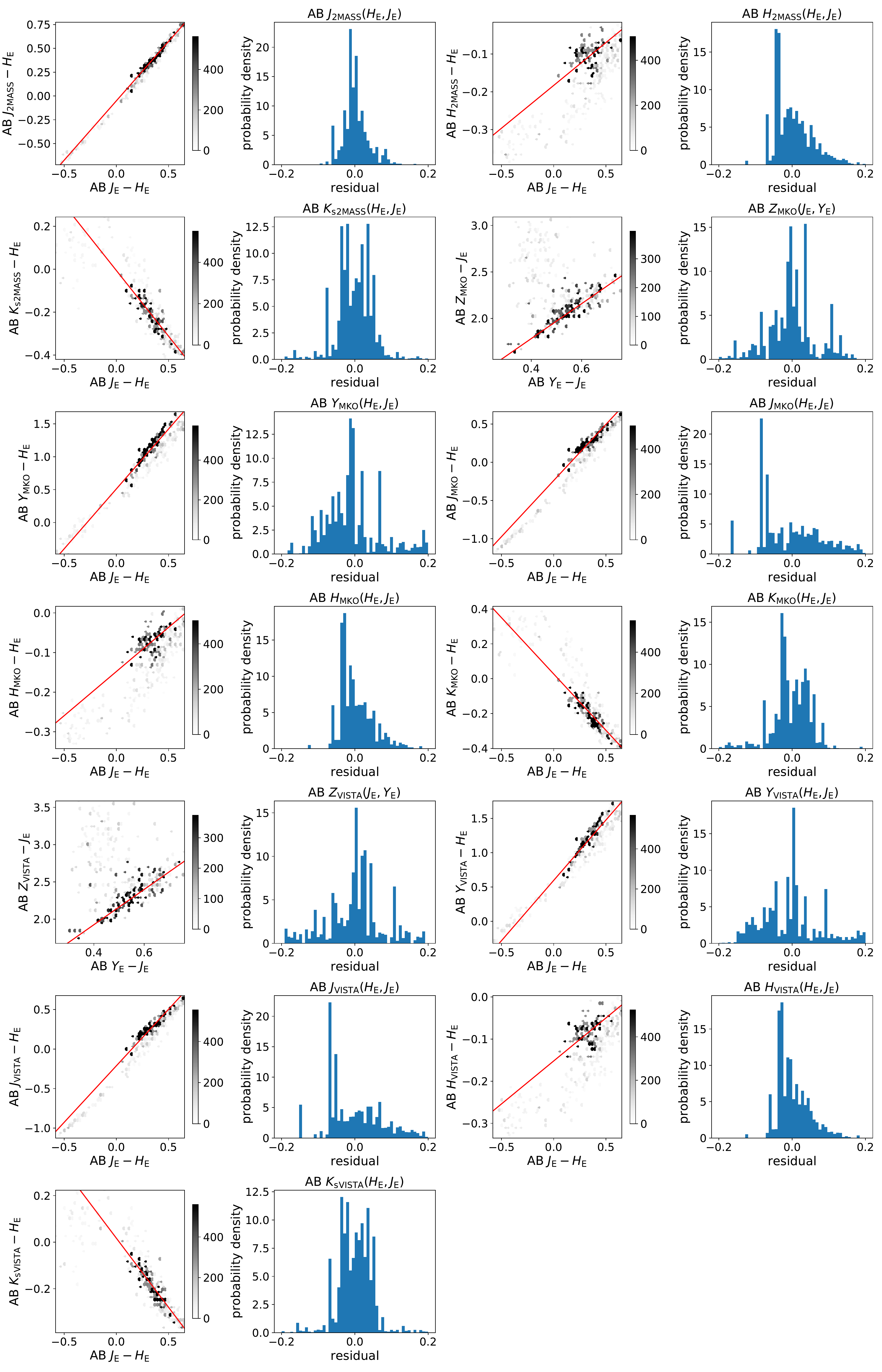}
\caption{Magnitude differences {\bf for MLT types} between external systems (2MASS, MKO, VISTA) and NISP, as a function of NISP color. Data are taken from the Euclid SC8 simulation pilot catalogs. The red lines show the linear fits given in Eqs.~(\ref{eq:nisp_estimates_stellarMLT_a}) through (\ref{eq:nisp_estimates_stellarMLT_o}).}
\label{fig:nisp_photo_colors_starsMLT}
\end{figure}

\begin{figure}[t]
\centering
\includegraphics[angle=0,width=1.0\hsize]{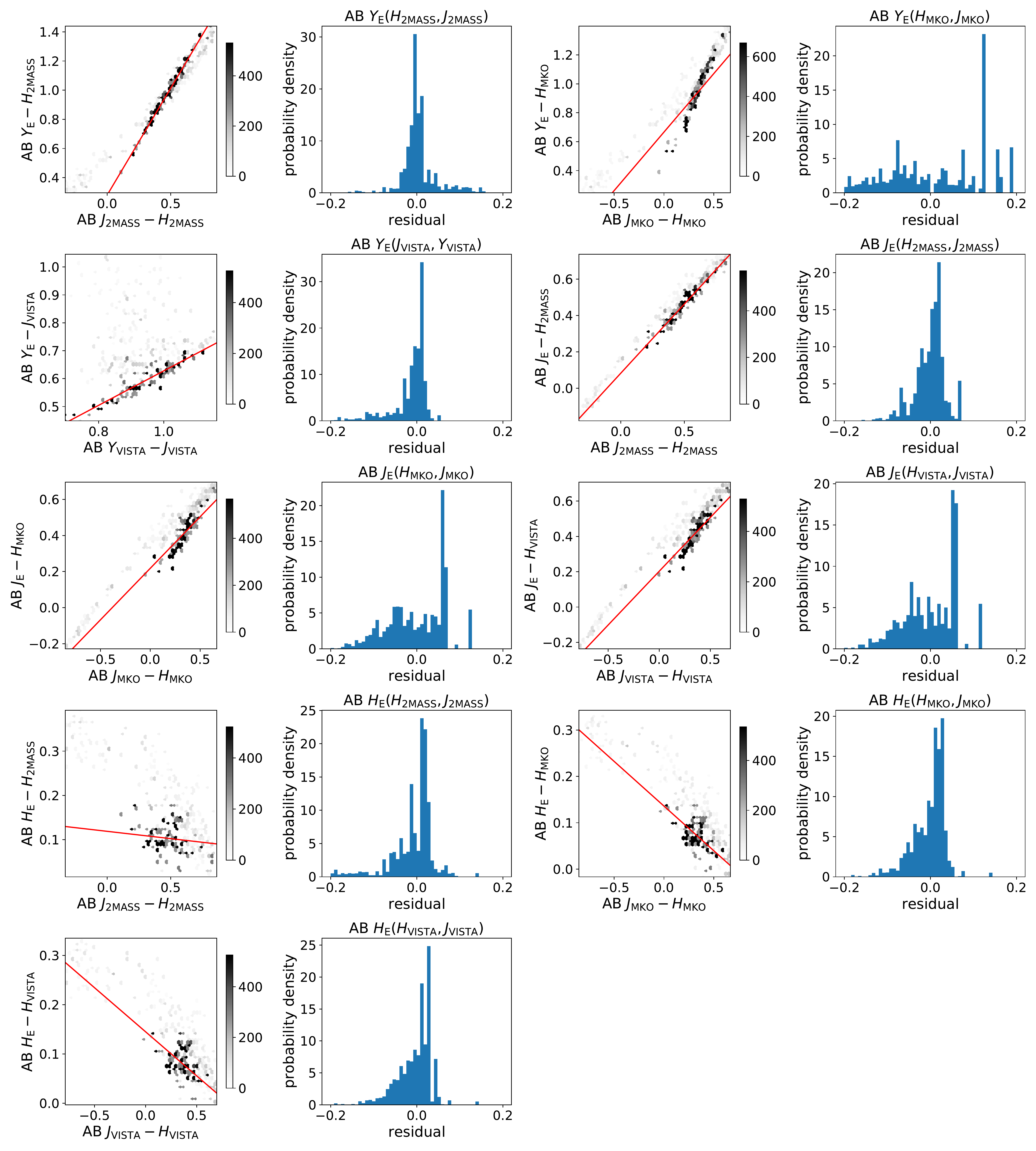}
\caption{Magnitude differences {\bf for MLT types} between NISP and external systems (2MASS, MKO, VISTA), as a function of color in an external system. The red lines show the linear fits given in Eqs.~(\ref{eq:ext_estimates_stellarMLT_a}) through (\ref{eq:ext_estimates_stellarMLT_o}).}
\label{fig:ext_photo_colors_starsMLT}
\end{figure}

Equations~(\ref{eq:nisp_estimates_stellarMLT_a}) through (\ref{eq:nisp_estimates_stellarMLT_o}) convert NISP AB magnitudes to 2MASS, MKO and VISTA AB magnitudes {\bf for MLT types}. The $1\,\sigma$ residuals are given in parentheses. Before using these transformations, read Sect.~\ref{sec:limitations}. See also Fig.~\ref{fig:nisp_photo_colors_starsMLT}.
\begin{alignat}{4}
J_{\scriptscriptstyle\rm 2MASS} &= H_{\scriptscriptstyle\rm E} && -0.051 +1.248 \;(J_{\scriptscriptstyle\rm E}-H_{\scriptscriptstyle\rm E}) && \quad(\sigma = 0.034)\label{eq:nisp_estimates_stellarMLT_a}\\
H_{\scriptscriptstyle\rm 2MASS} &= H_{\scriptscriptstyle\rm E} && -0.183 +0.226 \;(J_{\scriptscriptstyle\rm E}-H_{\scriptscriptstyle\rm E}) && \quad(\sigma = 0.051)\\
K_{\rm s,\scriptscriptstyle\rm 2MASS} &= H_{\scriptscriptstyle\rm E} && -0.006 -0.612 \;(J_{\scriptscriptstyle\rm E}-H_{\scriptscriptstyle\rm E}) && \quad(\sigma = 0.082)\\[2mm]
Z_{\scriptscriptstyle\rm MKO} &= J_{\scriptscriptstyle\rm E} && +1.032 +1.877 \;(Y_{\scriptscriptstyle\rm E}-J_{\scriptscriptstyle\rm E}) && \quad(\sigma = 0.289)\\
Y_{\scriptscriptstyle\rm MKO} &= H_{\scriptscriptstyle\rm E} && +0.511 +1.821 \;(J_{\scriptscriptstyle\rm E}-H_{\scriptscriptstyle\rm E}) && \quad(\sigma = 0.096)\\
J_{\scriptscriptstyle\rm MKO} &= H_{\scriptscriptstyle\rm E} && -0.239 +1.469 \;(J_{\scriptscriptstyle\rm E}-H_{\scriptscriptstyle\rm E}) && \quad(\sigma = 0.092)\\
H_{\scriptscriptstyle\rm MKO} &= H_{\scriptscriptstyle\rm E} && -0.148 +0.223 \;(J_{\scriptscriptstyle\rm E}-H_{\scriptscriptstyle\rm E}) && \quad(\sigma = 0.044)\\
K_{\scriptscriptstyle\rm MKO} &= H_{\scriptscriptstyle\rm E} && +0.027 -0.646 \;(J_{\scriptscriptstyle\rm E}-H_{\scriptscriptstyle\rm E}) && \quad(\sigma = 0.085)\\[2mm]
Z_{\scriptscriptstyle\rm VISTA} &= J_{\scriptscriptstyle\rm E} && +0.979 +2.368 \;(Y_{\scriptscriptstyle\rm E}-J_{\scriptscriptstyle\rm E}) && \quad(\sigma = 0.376)\\
Y_{\scriptscriptstyle\rm VISTA} &= H_{\scriptscriptstyle\rm E} && +0.596 +1.758 \;(J_{\scriptscriptstyle\rm E}-H_{\scriptscriptstyle\rm E}) && \quad(\sigma = 0.096)\\
J_{\scriptscriptstyle\rm VISTA} &= H_{\scriptscriptstyle\rm E} && -0.212 +1.429 \;(J_{\scriptscriptstyle\rm E}-H_{\scriptscriptstyle\rm E}) && \quad(\sigma = 0.081)\\
H_{\scriptscriptstyle\rm VISTA} &= H_{\scriptscriptstyle\rm E} && -0.152 +0.204 \;(J_{\scriptscriptstyle\rm E}-H_{\scriptscriptstyle\rm E}) && \quad(\sigma = 0.043)\\
K_{\rm s,\scriptscriptstyle\rm VISTA} &= H_{\scriptscriptstyle\rm E} && +0.017 -0.594 \;(J_{\scriptscriptstyle\rm E}-H_{\scriptscriptstyle\rm E}) && \quad(\sigma = 0.083)\label{eq:nisp_estimates_stellarMLT_o}
\end{alignat}
\\Equations~(\ref{eq:ext_estimates_stellarMLT_a}) through (\ref{eq:ext_estimates_stellarMLT_o}) convert from 2MASS, MKO and VISTA AB magnitudes to NISP AB magnitudes {\bf for MLT types}. Before using these transformations, read Sect.~\ref{sec:limitations}. 
See also Fig.~\ref{fig:ext_photo_colors_starsMLT}.
\begin{alignat}{4}
Y_{\scriptscriptstyle\rm E} &= H_{\scriptscriptstyle\rm 2MASS} && +0.280 +1.467 \;(J_{\scriptscriptstyle\rm 2MASS}-H_{\scriptscriptstyle\rm 2MASS}) && \quad(\sigma = 0.104)\label{eq:ext_estimates_stellarMLT_a}\\
Y_{\scriptscriptstyle\rm E} &= H_{\scriptscriptstyle\rm MKO} && +0.666 +0.810 \;(J_{\scriptscriptstyle\rm MKO}-H_{\scriptscriptstyle\rm MKO}) && \quad(\sigma = 0.125)\\
Y_{\scriptscriptstyle\rm E} &= J_{\scriptscriptstyle\rm VISTA} && +0.009 +0.619 \;(Y_{\scriptscriptstyle\rm VISTA}-J_{\scriptscriptstyle\rm VISTA}) && \quad(\sigma = 0.110)\\[2mm]
J_{\scriptscriptstyle\rm E} &= H_{\scriptscriptstyle\rm 2MASS} && +0.083 +0.759 \;(J_{\scriptscriptstyle\rm 2MASS}-H_{\scriptscriptstyle\rm 2MASS}) && \quad(\sigma = 0.034)\\
J_{\scriptscriptstyle\rm E} &= H_{\scriptscriptstyle\rm MKO} && +0.218 +0.572 \;(J_{\scriptscriptstyle\rm MKO}-H_{\scriptscriptstyle\rm MKO}) && \quad(\sigma = 0.070)\\
J_{\scriptscriptstyle\rm E} &= H_{\scriptscriptstyle\rm VISTA} && +0.203 +0.609 \;(J_{\scriptscriptstyle\rm VISTA}-H_{\scriptscriptstyle\rm VISTA}) && \quad(\sigma = 0.065)\\[2mm]
H_{\scriptscriptstyle\rm E} &= H_{\scriptscriptstyle\rm 2MASS} && +0.119 -0.033 \;(J_{\scriptscriptstyle\rm 2MASS}-H_{\scriptscriptstyle\rm 2MASS}) && \quad(\sigma = 0.070)\\
H_{\scriptscriptstyle\rm E} &= H_{\scriptscriptstyle\rm MKO} && +0.136 -0.193 \;(J_{\scriptscriptstyle\rm MKO}-H_{\scriptscriptstyle\rm MKO}) && \quad(\sigma = 0.038)\\
H_{\scriptscriptstyle\rm E} &= H_{\scriptscriptstyle\rm VISTA} && +0.145 -0.179 \;(J_{\scriptscriptstyle\rm VISTA}-H_{\scriptscriptstyle\rm VISTA}) && \quad(\sigma = 0.039)\label{eq:ext_estimates_stellarMLT_o}
\end{alignat}
\clearpage

\clearpage

\section{Transformation from and to the NISP photometric system for stars (all spectral types)\label{apx:transformations_stellar}}
\begin{figure}[t]
\centering
\includegraphics[angle=0,width=0.8\hsize]{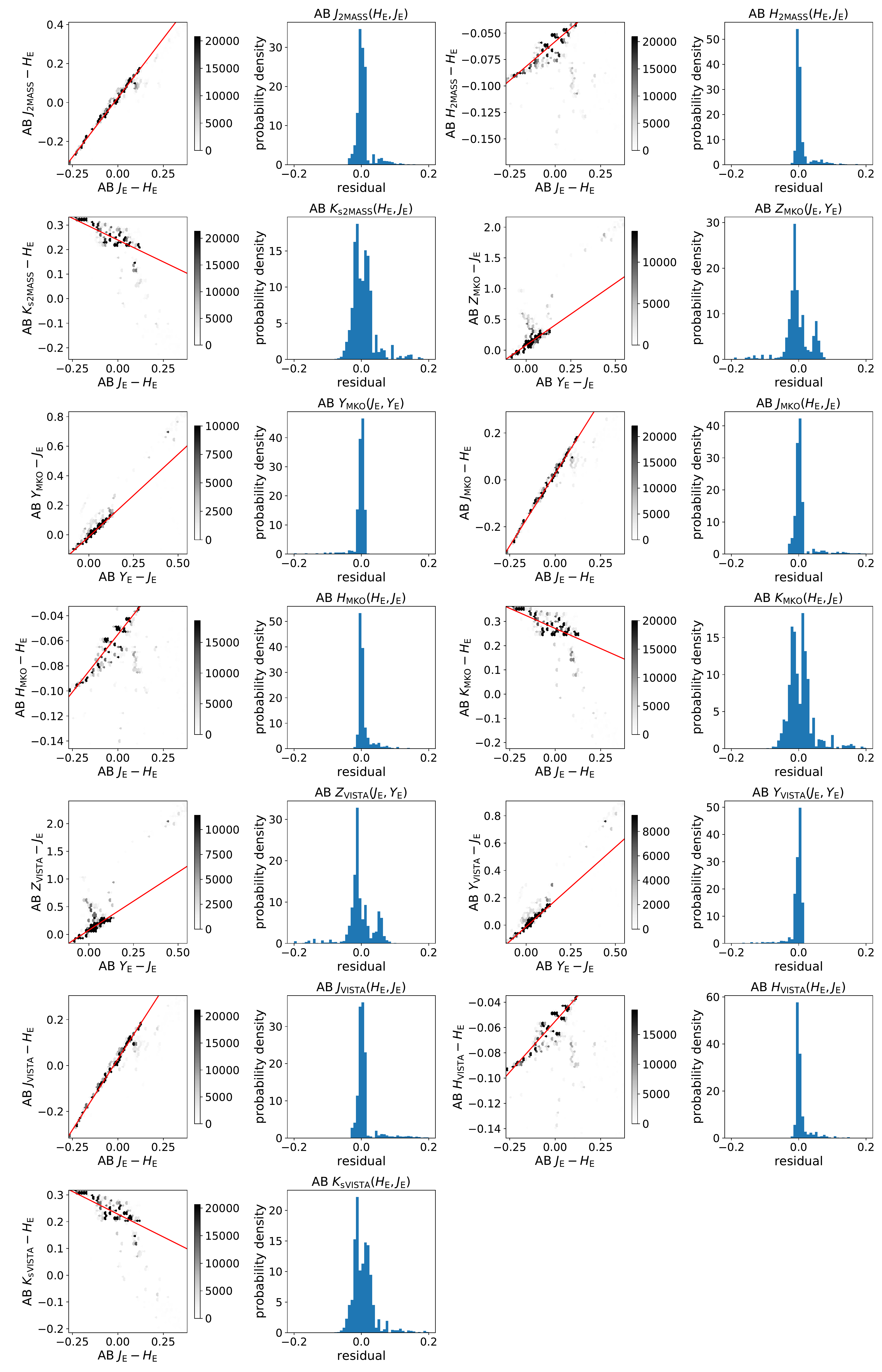}
\caption{Magnitude differences {\bf for stars of all spectral types} between external systems (2MASS, MKO, VISTA) and NISP, as a function of NISP color. Data are taken from the Euclid SC8 simulation pilot catalogs. The red lines show the linear fits given in Eqs.~(\ref{eq:nisp_estimates_stellar_a}) through (\ref{eq:nisp_estimates_stellar_o}).}
\label{fig:nisp_photo_colors_stars}
\end{figure}

\begin{figure}[t]
\centering
\includegraphics[angle=0,width=1.0\hsize]{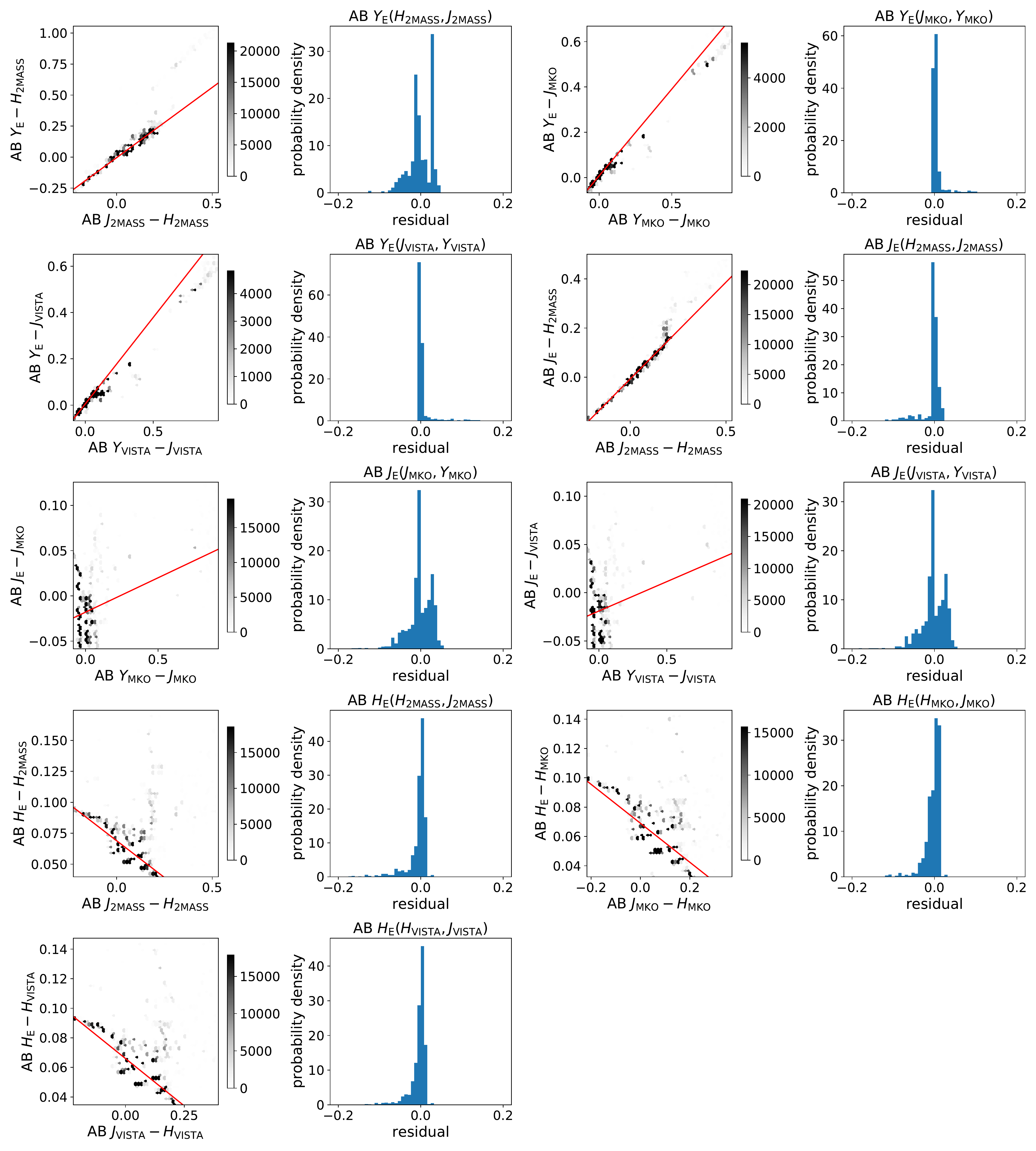}
\caption{Magnitude differences {\bf for stars of all spectral types} between NISP and external systems (2MASS, MKO, VISTA), as a function of color in an external system. The red lines show the linear fits given in Eqs.~(\ref{eq:ext_estimates_stellar_a}) through (\ref{eq:ext_estimates_stellar_o}). {\bf Note that the two fits displayed in the third row} are skewed by a number of very red sources that are not well visible in these linear gray scale maps. This shows that a simple, 'one-fits-all' model is not always an appropriate way to transform between photometric systems.}
\label{fig:ext_photo_colors_stars}
\end{figure}

Equations~(\ref{eq:nisp_estimates_stellar_a}) through (\ref{eq:nisp_estimates_stellar_o}) convert NISP AB magnitudes to 2MASS, MKO and VISTA AB magnitudes {\bf for stars of all spectral types}. The $1\,\sigma$ residuals are given in parentheses. Before using these transformations, read Sect.~\ref{sec:limitations}. See also Fig.~\ref{fig:nisp_photo_colors_stars}.
\begin{alignat}{4}
J_{\scriptscriptstyle\rm 2MASS} &= H_{\scriptscriptstyle\rm E} && +0.024 +1.220 \;(J_{\scriptscriptstyle\rm E}-H_{\scriptscriptstyle\rm E}) && \quad(\sigma = 0.026)\label{eq:nisp_estimates_stellar_a}\\
H_{\scriptscriptstyle\rm 2MASS} &= H_{\scriptscriptstyle\rm E} && -0.058 +0.147 \;(J_{\scriptscriptstyle\rm E}-H_{\scriptscriptstyle\rm E}) && \quad(\sigma = 0.030)\\
K_{\rm s,\scriptscriptstyle\rm 2MASS} &= H_{\scriptscriptstyle\rm E} && +0.239 -0.356 \;(J_{\scriptscriptstyle\rm E}-H_{\scriptscriptstyle\rm E}) && \quad(\sigma = 0.072)\\[2mm]
Z_{\scriptscriptstyle\rm MKO} &= J_{\scriptscriptstyle\rm E} && +0.082 +2.018 \;(Y_{\scriptscriptstyle\rm E}-J_{\scriptscriptstyle\rm E}) && \quad(\sigma = 0.212)\\
Y_{\scriptscriptstyle\rm MKO} &= J_{\scriptscriptstyle\rm E} && -0.012 +1.115 \;(Y_{\scriptscriptstyle\rm E}-J_{\scriptscriptstyle\rm E}) && \quad(\sigma = 0.045)\\
J_{\scriptscriptstyle\rm MKO} &= H_{\scriptscriptstyle\rm E} && +0.027 +1.234 \;(J_{\scriptscriptstyle\rm E}-H_{\scriptscriptstyle\rm E}) && \quad(\sigma = 0.048)\\
H_{\scriptscriptstyle\rm MKO} &= H_{\scriptscriptstyle\rm E} && -0.055 +0.184 \;(J_{\scriptscriptstyle\rm E}-H_{\scriptscriptstyle\rm E}) && \quad(\sigma = 0.025)\\
K_{\scriptscriptstyle\rm MKO} &= H_{\scriptscriptstyle\rm E} && +0.271 -0.332 \;(J_{\scriptscriptstyle\rm E}-H_{\scriptscriptstyle\rm E}) && \quad(\sigma = 0.077)\\[2mm]
Z_{\scriptscriptstyle\rm VISTA} &= J_{\scriptscriptstyle\rm E} && +0.077 +2.091 \;(Y_{\scriptscriptstyle\rm E}-J_{\scriptscriptstyle\rm E}) && \quad(\sigma = 0.248)\\
Y_{\scriptscriptstyle\rm VISTA} &= J_{\scriptscriptstyle\rm E} && -0.011 +1.165 \;(Y_{\scriptscriptstyle\rm E}-J_{\scriptscriptstyle\rm E}) && \quad(\sigma = 0.052)\\
J_{\scriptscriptstyle\rm VISTA} &= H_{\scriptscriptstyle\rm E} && +0.028 +1.234 \;(J_{\scriptscriptstyle\rm E}-H_{\scriptscriptstyle\rm E}) && \quad(\sigma = 0.045)\\
H_{\scriptscriptstyle\rm VISTA} &= H_{\scriptscriptstyle\rm E} && -0.055 +0.161 \;(J_{\scriptscriptstyle\rm E}-H_{\scriptscriptstyle\rm E}) && \quad(\sigma = 0.025)\\
K_{\rm s,\scriptscriptstyle\rm VISTA} &= H_{\scriptscriptstyle\rm E} && +0.228 -0.340 \;(J_{\scriptscriptstyle\rm E}-H_{\scriptscriptstyle\rm E}) && \quad(\sigma = 0.065)\label{eq:nisp_estimates_stellar_o}
\end{alignat}
\\Equations~(\ref{eq:ext_estimates_stellar_a}) through (\ref{eq:ext_estimates_stellar_o}) convert from 2MASS, MKO and VISTA AB magnitudes to NISP AB magnitudes {\bf for stars of all spectral types}. Before using these transformations, read Sect.~\ref{sec:limitations}. 
See also Fig.~\ref{fig:ext_photo_colors_stars}.
\begin{alignat}{4}
Y_{\scriptscriptstyle\rm E} &= H_{\scriptscriptstyle\rm 2MASS} && -0.005 +1.134 \;(J_{\scriptscriptstyle\rm 2MASS}-H_{\scriptscriptstyle\rm 2MASS}) && \quad(\sigma = 0.086)\label{eq:ext_estimates_stellar_a}\\
Y_{\scriptscriptstyle\rm E} &= J_{\scriptscriptstyle\rm MKO} && +0.008 +0.765 \;(Y_{\scriptscriptstyle\rm MKO}-J_{\scriptscriptstyle\rm MKO}) && \quad(\sigma = 0.026)\\
Y_{\scriptscriptstyle\rm E} &= J_{\scriptscriptstyle\rm VISTA} && +0.005 +0.747 \;(Y_{\scriptscriptstyle\rm VISTA}-J_{\scriptscriptstyle\rm VISTA}) && \quad(\sigma = 0.028)\\[2mm]
J_{\scriptscriptstyle\rm E} &= H_{\scriptscriptstyle\rm 2MASS} && -0.007 +0.787 \;(J_{\scriptscriptstyle\rm 2MASS}-H_{\scriptscriptstyle\rm 2MASS}) && \quad(\sigma = 0.026)\\
J_{\scriptscriptstyle\rm E} &= J_{\scriptscriptstyle\rm MKO} && -0.018 +0.076 \;(Y_{\scriptscriptstyle\rm MKO}-J_{\scriptscriptstyle\rm MKO}) && \quad(\sigma = 0.040)\\
J_{\scriptscriptstyle\rm E} &= J_{\scriptscriptstyle\rm VISTA} && -0.019 +0.061 \;(Y_{\scriptscriptstyle\rm VISTA}-J_{\scriptscriptstyle\rm VISTA}) && \quad(\sigma = 0.037)\\[2mm]
H_{\scriptscriptstyle\rm E} &= H_{\scriptscriptstyle\rm 2MASS} && +0.069 -0.119 \;(J_{\scriptscriptstyle\rm 2MASS}-H_{\scriptscriptstyle\rm 2MASS}) && \quad(\sigma = 0.030)\\
H_{\scriptscriptstyle\rm E} &= H_{\scriptscriptstyle\rm MKO} && +0.069 -0.133 \;(J_{\scriptscriptstyle\rm MKO}-H_{\scriptscriptstyle\rm MKO}) && \quad(\sigma = 0.021)\\
H_{\scriptscriptstyle\rm E} &= H_{\scriptscriptstyle\rm VISTA} && +0.066 -0.128 \;(J_{\scriptscriptstyle\rm VISTA}-H_{\scriptscriptstyle\rm VISTA}) && \quad(\sigma = 0.023)\label{eq:ext_estimates_stellar_o}
\end{alignat}
\clearpage

\clearpage

\end{appendix}

\end{document}